\documentclass[a4paper,11pt]{article}
\usepackage{setspace}
\pdfoutput=1 

\usepackage{jcappub} 

\usepackage[T1]{fontenc} 
\usepackage{hyperref}

\usepackage{xfrac}
\usepackage{graphicx}
\usepackage{rotating}
\usepackage{physics}
\usepackage{multirow}
\usepackage[normalem]{ulem}
\usepackage{amssymb,amsmath}
\usepackage[dvipsnames]{xcolor}
\usepackage{tikz}
\usetikzlibrary{shapes,arrows}
\usepackage{pdflscape} 
\usepackage{array}
\usepackage{xspace}
\newcolumntype{L}{>{$}l<{$}}
\newcolumntype{C}{>{$}c<{$}}

\newcolumntype{P}[1]{>{\centering\arraybackslash}p{#1}}

\newcommand{\SF}{\textit{ShapeFit}\xspace}

\newcommand{\Om}{\Omega_\mathrm{m}}
\newcommand{\Ob}{\Omega_\mathrm{b}}
\newcommand{\Oc}{\Omega_\mathrm{c}}
\newcommand{\rd}{r_\mathrm{d}}
\newcommand{\keq}{k_\mathrm{eq}}

\newcommand{\Mpcoverhs}{\rd^\mathrm{fid} \rd^{-1} h^{-1}\mathrm{Mpc}}

\newcommand{\hoverMpc}{\left[ h\mathrm{Mpc}^{-1} \right]}
\newcommand{\bnl}{b_{3\mathrm{nl}}}
\newcommand{\bs}{b_{s^2}}

\title{\boldmath  A tale of two (or more) $h$'s }

\author[1,2]{Samuel Brieden}
\author[2]{H\'ector Gil-Marín}
\author[2,3]{Licia Verde}

\affiliation[1]{Institute for Astronomy,  University of Edinburgh, Blackford Hill, Edinburgh, EH9 3HJ, UK}
\affiliation[2]{ICC, University of Barcelona, IEEC-UB, Mart\'i i Franqu\`es, 1, E-08028 Barcelona, Spain}
\affiliation[3]{ICREA, Pg. Llu\'is Companys 23, Barcelona, E-08010, Spain}

\emailAdd{samuel.brieden@ed.ac.uk}
\emailAdd{hectorgil@icc.ub.edu}
\emailAdd{liciaverde@icc.ub.edu }

\abstract{We use the large-scale structure galaxy data (LSS) from the BOSS and eBOSS surveys, in combination with abundances information from Big Bang Nucleosynthesis (BBN), to measure two values of the Hubble expansion rate, $H_0=100h\,[{\rm km}\, {\rm s}^{-1}\,{\rm Mpc}^{-1}]$, each of them based on very different physical processes. One is a (traditional) late-time-background measurement, based on determining the BAO scale and using BBN abundances on baryons for calibrating its absolute size (BAO+BBN). This method anchors $H_0$ to the (standard) physics of the sound horizon scale at pre-recombination times. The other is a newer, early-time based measurement, associated with the broadband shape of the power spectrum. This second method anchors $H_0$ to the physics of the matter-radiation equality scale, which also needs BBN information for determining the suppression of baryons in the power spectrum shape (shape+BBN). Within the $\Lambda$CDM model, we find very good consistency among these two $H_0$'s: BAO+BBN 
(+growth) 
delivers $H_0=67.42_{-0.94}^{+0.88}$ $(67.37_{-0.95}^{+0.86})$ km s$^{-1}$Mpc$^{-1}$
, whereas the shape+BBN 
(+growth) 
delivers $H_0 = 70.1_{-2.1}^{+2.1}$ $(70.1_{-2.1}^{+1.9})$ km s$^{-1}$ Mpc$^{-1}$,
where `growth' stands for information from the late-time-perturbations captured by the growth of structure parameter. 
These are the tightest sound-horizon free $H_0$ constraints from LSS data to date.

As a consequence, to be viable any $\Lambda$CDM extension proposed to address the so-called ``Hubble tension'' needs to modify consistently not only the sound horizon scale physics, but also the matter-radiation equality scale, in such a way that
 both late- and early-based $H_0$'s return results mutually consistent and consistent with the high $H_0$ value recovered by the standard cosmic distance ladder (distance-redshift relation) determinations.}

\begin{document}
\maketitle
\flushbottom
\section{Introduction}
\label{sec:intro}
Over the last two decades cosmology has become a precision science; cosmological observations have consolidated the standard model of cosmology, the $\Lambda$ cold dark matter ($\Lambda$CDM) model: an effective model with only six parameters, now measured with unprecedented precision. In the near future  new surveys e.g., \cite{desi,euclid,lsst,wfirst,SKA, SimonsObs} are expected to provide further improvements.  The model's  parameters  can, and have been,  measured with errors of few percent or sometimes  sub-percent using different approaches and different observables that rely on different physics.  The resounding success of the $\Lambda$CDM model rests on the  broad agreement of  these different measurements. 

Over the last decade or so however,  this agreement has worsened progressively and significantly for the Hubble expansion rate today, the parameter $H_0$, also expressed through the dimensionless Hubble parameter $h\equiv H_0/(100\,{\rm km\, s}^{-1}{\rm Mpc}^{-1})$. The discrepancy is now at the $\sim5\sigma$ level ~\cite{Riess:2021jrx} and deemed a `tension'~\cite{VTR_2019}.
Direct, local measurements of $H_0$, based on the traditional cosmic distance ladder and relying on type Ia supernovae (SNe) calibrated through Cepheid stars, report a `high-$H_0$' value of $H_0=73.0\pm1.0\,{\rm km/s/Mpc }$ \cite{Riess:2021jrx}. But indirect measurements of $H_0$, based on the interpretation of Cosmic Microwave Background (CMB) anisotropies within the standard $\Lambda$CDM model, yield a `low-$H_0$' value: $H_0=67.4\pm0.5\, {\rm km/s/Mpc}$ \cite{Aghanim:2018eyx}. Other techniques have complemented these two type of measurements: e.g,  direct measurements of type Ia SNe calibrated using the tip of Red Giant Branch (TRGB) \cite{wendy}, time-delays in strong lensing systems \cite{holicow}, and indirect measurements from large scale structure data (LSS), which rely on the Baryon Acoustic Oscillation (BAO) feature imprinted on late-time dark matter tracers \cite{alamdr12,eboss_collaboration_dr16,wigglez}. 
Observations seem to segregate in two `camps' \cite{DiValentino:2021izs}: 
local, direct measurements, which cluster around the  high-$H_0$ value, and indirect measurements which fall around the low-$H_0$ value \cite{VTR_2019}. Local direct measurements are based on the (direct) cosmic distance ladder approach; indirect measurement broadly fall in the inverse cosmic distance ladder approach \cite{Cuesta:2014asa,Aubourgetal15}. As, after almost a decade of checks by the community, experimental systematics seem to be  an increasingly unlikely  explanation for the tension, a suite of exotic extensions of the $\Lambda$CDM model has been investigated to alleviate this tension e.g., \cite{H0olympics,Kamion_Riess22} and references therein.

No measurement is completely  free of assumptions; relevant to this paper is the fact that indirect determinations of $H_0$ are particularly reliant on the adopted cosmological model.  If the $H_0$ tension represents a shortcoming of the standard model, indicating that  the model is in need of an extension, it is of paramount importance to analyze the data, especially those yielding indirect measurements of $H_0$, in the most model-agnostic possible way, making the least possible number of assumptions, and spelling them out clearly. 
Then, suitable combinations of such measurements can be constructed to design {\it diagnostic tests} of the $\Lambda$CDM model or of  proposed extensions. 

The recently-developed  \SF compression technique \cite{ShapeFit} is particularly well-suited to develop such diagnostic tests for the analysis of  galaxy spectroscopic surveys data (see \cite{Brieden:2022lsd,ShapeFit:data}).
\SF compresses the information enclosed in the full power spectrum anisotropic signal in few physical variables per redshift bin, which are essentially model-independent\footnote{The variables are model-independent, their interpretation however is not.}.
These variables represent the position of the BAO peak along and across the line-of-sight (LOS) in units of the sound horizon scale,  the growth rate of structure, and the slope of the smoothed broadband of the power spectrum pivoting at a certain large reference scale. 
They effectively parameterise changes of different physical features relative to a linear matter power spectrum template, which is why it is also known as `fixed-template' approach.
The four physical  variables per redshift bin can be later interpreted in the light of a model, as  to extract information about the parameters of that model. 

This model-agnostic approach is qualitatively different from other approaches --often referred to 
as `direct fit' or `full-modeling'-- which infer the parameters of a given model directly from the full power spectrum signal and  simultaneously at all available scales and  redshifts (e.g.,  \cite{SDSS:2003tbn,DAmico:2019fhj,Ivanov:2019pdj,Trosteretal:2020}).
Within the `full-modeling' approach it is not straightforward to unambiguously 
disentangle the information relative to different physical processes, and thus it is very difficult to run diagnostic tests of $\Lambda$CDM, such as whether the late- and early-time signals are consistent. 
These diagnostic tests, on the other hand,  are enabled by the \SF approach:
as we illustrate in this paper, by playing with the variables considered during the cosmology inference step, one can turn off and on  different assumptions of the model, as the different compressed variables represent early- or late-time physical processes within the model, or express background vs. perturbation physics (see also \cite{Hamann_shape}). 
One possible initial approach to design a diagnostic test, which we attempt here, is to consider a `piecewise' $\Lambda$CDM model:  for each relevant epoch or physical process we assume that $\Lambda$CDM is an effective model, but we do not necessarily force  the inferred parameters of the model to be coherent across epochs or physical signatures.

These consistency checks are of particular relevance when it comes to indirect (or model-dependent) measurements of $H_0$.
In a similar way as
(direct) distance ladder-based $H_0$ measurements 
need  low-$z$ `anchors', indirect inverse distance ladder-based $H_0$ determinations need high-$z$ `anchors', whose notion usually relies on a model. The traditional and best calibrated anchor is the size of the sound horizon at radiation drag, which is usually provided by the analysis of CMB data. The robustness of the direct distance ladder approach  relies on having several different `anchors' and calibrators; it is thus important to  provide several independent and different anchors also for the inverse ladder.

This paper is one attempt in this direction. We consider and compare different ways of measuring the Hubble expansion rate from large-scale structure data, where each approach relies on different assumptions and physical processes that are in-built in the $\Lambda$CDM model.
One way is closely related to the traditional, BAOofLSS,
approach: measurements of the BAO signal imprinted in the distribution of dark-matter tracers (galaxies, quasars and Lyman-$\alpha$) at different redshifts, which are calibrated with the absolute size of the sound horizon scale at radiation drag $r_{\rm d}$. Instead of being given by the analysis of CMB data,  the adopted $r_{\rm d}$ value is obtained by assuming standard early-time physics and big bang nucleosynthesis (BBN) \cite{Pisanti:2007hk} combined with measurements of abundances of light elements \cite{Adelberger:2010qa}.
This method, referred to as BAO+BBN, relies on assuming  standard physics at pre-recombination times and provides an indirect measurement of $H_0$ which is independent of CMB anisotropies \cite{BAOBBN,Cuceu20,NilsBBN,NilsBBN2}. 

An alternative route to $H_0$ from LSS data is to use  the  matter-radiation equality scale, $k_{\rm eq}$, for example employing it as a standard ruler (calibrated also at early times), just as done with the sound horizon scale (e.g., \cite{cunnington22} and references therein, also \cite{2005MNRAS.363.1329B}).
Unlike the sound horizon $r_{\rm d}$, this scale imprints a signal at much larger scales (the matter power spectrum turn-around), and this feature also includes the suppression of power on nearby smaller scales, which is related to how fast modes re-enter the horizon just before matter-radiation equality. One drawback of this approach is that this turn-around feature is not well measured in current LSS surveys: partly because of the limitation of the number of large-scale modes available, and partly because at these scales imaging systematics tend to contaminate the signal of the power spectrum significantly. 
However, information can still be accessed in the broadband shape of the power spectrum at scales just below the turn-around, typically $0.03<k\,\hoverMpc<0.10$ \cite{Brieden:2022lsd,Philcox_Sherwin_Farren_Baxter_21,Farren:2021grl,Philcox:2022sgj,Smith:2022iax}, which are also sensitive to the physics around matter-radiation equality times. From the \SF perspective this information is encoded by the slope of the smoothed broadband of the power spectrum.

Having two distinct measurements of the Hubble parameter, informed by very different physical processes (right before recombination or at matter-radiation equality epochs), is very important to shed light on the current $H_0$-tension. Any proposed solution to the Hubble tension should improve the performance of $\Lambda$CDM in terms of agreement of all the different $H_0$'s, and also improve, or at least not worsen, other emerging tensions such as the $\sigma_8$ tension (see e.g. \cite{Amon_Efstathiou22} and references therein).

The main goal of this paper is then twofold:
\begin{enumerate} 
\item to obtain the tightest constraints on $H_0$ from LSS spectroscopic data to date which are independent from the sound horizon at radiation drag and based on $\Lambda$CDM matter-radiation equality physics,
\item  to provide an array of internal consistency checks of the cosmological model, assuming  that systematic effects in the observations are under control (alternatively it can provide an internal consistency test of the observations themselves, but we will not consider this case here). In particular, the goal is to determine whether this measurement is in agreement with that provided by sound-horizon scale BAOofLSS-type of analyses, where the sound horizon ruler is calibrated assuming $\Lambda$CDM pre-recombination physics. 
\end{enumerate}
These type of checks can then provide diagnostic tests for the $
\Lambda$CDM model and `guardrails' for new physics beyond $\Lambda$CDM.

The rest of the  paper is structured as follows. 
In section~\ref{sec:standard_ruler} we review  very pedagogically the assumptions that a ruler must fulfill to be called standard, and which of these conditions will be assumed later when performing the analysis. In section~\ref{sec:SF} we summarise some notions and dependencies we expect to obtain when we infer constraints on parameters of $\Lambda$CDM type of models from the same type of variables as used in \SF, and which assumptions are being made when inferring the Hubble expansion rate parameters from different physical variables.
Our main results, consisting of the different types of Hubble parameter measurements obtained from the publicly available BOSS and eBOSS compressed physical variables and their physical implications, are presented in section~\ref{sec:results}. 
In section~\ref{sec:keq} we explore in details the differences between the approach of  using the matter-radiation equality scale as a standard ruler,  and that of using instead the slope of the broadband shape as a feature to infer Hubble parameter measurements. Finally  we conclude in  section~\ref{sec:conclusions}. 

\section{Standard rulers and cosmological distances}\label{sec:standard_ruler}

Measuring distances in cosmology is one of the most challenging tasks the field  routinely faces. Beside standard candles, standard rulers make this possible. 
In order to  convert the observed redshifts and object's positions in the sky  into comoving distances, we can employ the physical size of a certain object or feature (a ruler). The ruler length can be derived from first principles or assumed to be well known. The BAO feature on low redshift clustering of large-scale structure is such a standard ruler, which enables us to employ LSS data to derive geometric constraints on the Universe's expansion history.

In this work we assume (as it is customary to do but not always spelled out explicitly)  that a standard ruler must satisfy the following two key conditions,
\begin{itemize}
    \item[\textit{i)}] it is isotropic, 
    \item[\textit{ii)}] its length is fixed, it does not evolve with redshift.
\end{itemize}
 A standard ruler is particularly useful if also 
\begin{itemize}
    \item[\textit{iii)}] the length of the ruler is known.
   \end{itemize}

In particular, the sound horizon scale at radiation drag epoch is set by early-time physics and imprints a characteristic scale on the matter clustering  early on, which can thus be assumed to be a ruler. This has been key to the spectacular success of BAOofLSS for cosmology. In fact, condition \textit{i)} is very conservative: violating isotropy for rulers set by early-time physics would violate 
one of the pillars of the Cosmological Principle \cite{Maartens:2011yx}.

For the BAO feature imprinted in late-time dark matter tracers, assumption \textit{ii)} is also very conservative.   While there are models predicting a tiny redshift dependence of the feature connected to the sound horizon due to non-linear couplings \cite{Blas2016_BAOIR}, these effects are usually taken into account in the standard approach either by modifying the fixed template accordingly or via reconstruction \cite{Eis2007,burden_reconstruction_2015,Sherwin_2019}. Beyond that, no cosmological models exist (yet) that would predict a running of the BAO scale with redshift. 
As such a scenario lacks of physical motivation, assumption \textit{ii)} is usually considered true, such that the sound horizon can indeed be treated as a standard ruler.

Condition \textit{iii)}, however, requires that the BAO scale, as seen in the late-time clustering, can be uniquely identified with the sound horizon (i.e., no mismatch) and requires a model of the early Universe (although not necessarily early Universe data). 
Although such a mismatch between the physical sound horizon inferred from the CMB and the length of the standard ruler observed in the BAO has been quantified already following perturbation theory arguments \cite{Blazek:2015ula,Slepian:2016nfb,Hirata:2017ivs}, this shift is small enough not to be statistically relevant (yet) and could easily be incorporated in the modeling.
Then, for a $\Lambda$CDM model for example,  it can be set by late-Universe observations of baryon abundances when the physics of specific  early-Universe epochs is assumed to be known.

In the case of standard candles, the magnitude of supernovae can be known to be fixed only under certain conditions or assumptions. For this reason sometimes supernovae Ia are called ``standardizable'' candles. Similarly, the length of a standard ruler could be known, or known to be fixed, only under certain assumptions. One could then also  use  the word ``standardizable'' rulers. Here we spell out clearly the assumptions under which a given ruler can be considered standard.
Along this paper we will comment on which of the above 3 assumptions about the properties of a ruler are used when deriving constraints on the Hubble parameter.

\section{ShapeFit compression and cosmological interpretation} \label{sec:SF}
This work makes use of the \SF methodology introduced in \cite{ShapeFit}, which has passed blind high precision validation tests on the large volume of the PT challenge simulation \cite{Nishimichi:2020tvu, PTchallenge:data} in \cite{ShapeFitPT}. Recently in \cite{Brieden:2022lsd}, \SF has been applied to the full BOSS+eBOSS Luminous Red Galaxy (LRG) and Quasar (QSO) samples \cite{Reid:2015gra,ebossLRG_catalogue,ebossQSO_catalogue}, yielding constraints on cosmological parameters with unprecedented precision from LSS data alone. 

Compared to classic analysis approaches relying on the BAO and redshift space distortion (RSD) signals, the additional constraining power of \SF mainly comes from the large-scale shape of the linear matter power spectrum, $k<0.1\,h\,{\rm Mpc}^{-1}$, encapsulated by a new degree of freedom of the template-fit, the shape parameter $m$ \cite{BriedenPRL21}.
This has been subsequently corroborated by  \cite{Simon:2022lde,Chenetal21} who  showed that even for direct  model fits which include smaller scales pushing into the non-linear regime,  most of the  robust cosmological signal is encoded in the large scales. 

Here we build on these findings by providing a modular cosmological analysis of each of the physical features captured by \SF and focusing on their individual relation with the Hubble parameter. For that, we use the measured compressed variables $\mathbf{\Theta}(z)$ of BOSS DR12 LRG (the first two overlapping redshift bins, $0.2<z<0.5$ and $0.4<z<0.6$), eBOSS DR16 LRG ($0.6<z<1.0$) and eBOSS DR16 QSO ($0.8<z<2.2$) samples provided in appendix E of \cite{Brieden:2022lsd},
\begin{align} \label{eq:theta}
    \mathbf{\Theta}(z) = \left\lbrace D_H(z)/\rd, D_M(z)/\rd, f(z)\sigma_{s8}(z), m(z) \right\rbrace ,
\end{align}
where $D_H(z)$ and $D_M(z)$ represent the distance along and across the LOS, respectively, $\rd$ is the sound horizon scale at baryon drag used as standard ruler in the fixed-template method, $f(z)$ is the logarithmic growth rate of structure,  $\sigma_{s8}(z)$ is the linear matter fluctuation amplitude filtered on the scale of $8\, \Mpcoverhs$, and $m(z)$ (or the shape parameter) is the linear power spectrum slope at a pivot scale of $k_p=\pi/\rd$.

We now summarize  the physical signal each of these quantities encode, focusing on their cosmological implications and in particular how they can be used to obtain qualitatively different measurements of $h$. Subsections \ref{sec:SF-geom}, \ref{sec:SF-shape} and \ref{sec:2h} are somewhat pedagogical, but provide key concepts to highlight the motivation of this work.

\subsection{BAO and geometry information - \texorpdfstring{$h$}{} from the sound horizon scale}

\label{sec:SF-geom}

The first two quantities, $D_H(z)/\rd$ and $ D_M(z)/\rd$, contain purely geometrical information. The numerators, $D_H(z)$ and $D_M(z)$ are determined by the late-time (post-recombination) geometry, setting the \textit{coordinate} system of the universe. The denominator, the sound horizon scale $\rd$, is set by early-time (pre-recombination) physics; by design
it is a standard ruler and yields the natural \textit{unit} system for the  BAO-based geometric probe.
By construction, fixed-template fits such as the standard BAO, RSD and \textit{ShapeFit} analyses yield distance measurements {\it in units of the standard ruler} which are independent of the adopted template.
In such analyses the ruler length is unknown \textit{a priori} and its absolute value can only be inferred (indirectly) under the assumption of a model of the early-universe.

Now, it is interesting to take a closer look at the very different cosmological implications of the early- and late-time quantities separately, in particular how these relate to the standard ruler assumptions \textit{i)} - \textit{iii)} (see section~\ref{sec:standard_ruler}). For this purpose, it is useful to  consider the derived basis, where  the signal is explicitly  decomposed  in an  isotropic  component and the classic Alcock-Paczynski (AP) parameter, $\left\lbrace D_V(z)/\rd,\, F_\mathrm{AP}(z) \right\rbrace$, respectively, defined as,
\begin{align} \label{eq:DVrd}
    D_V(z)/\rd &\equiv \left[ z \left(D_H(z)/\rd\right)  \left(D_M(z)/\rd\right)^2 \right] ^{1/3} = \left[ z D_H(z)  D_M^2(z) \right] ^{1/3}/\rd ~, \\ \label{eq:FAP} F_\mathrm{AP}(z) &\equiv \frac{D_M(z)/\rd}{D_H(z)/\rd} = D_M(z)/D_H(z)~,
\end{align}
which makes it easier to  disentangle the early- and late-time effects. 
The spherically-averaged distance scaling, $D_V(z)/\rd$, represents the scaling of the  late-time isotropic distance $D_V(z)$ versus the early-time sound horizon $\rd$ (assumed isotropic in {\it i)}) . 

The AP parameter, $F_\mathrm{AP}(z)$, is a pure late-time variable which captures the apparent distortion of an isotropic feature (here $r_{\rm d}$, but any scale can be used as a reference) along vs. across the LOS.
Note that for $F_{\rm AP}(z)$ the only early-time assumption entering this quantity is standard ruler condition  \textit{i)},  not  \textit{ii)}: eq.~\eqref{eq:FAP} remains true even if the ruler (the sound horizon)
were to change with redshift. 

It is possible to separate early- from late-time information in the interpretation of the compressed variables as follows. The redshift dependence of $D_V(z)/r_{\rm d}$,  regardless of the signal amplitude,  only encloses information about  late-time  kinematics. 
In this case one only uses the relative $D_V(z)/r_{\rm d}$ information across different redshifts;
we refer to this as uncalibrated (do not use of the absolute value of $r_{\rm d}$) and unnormalized (the overall amplitude of $D_V$ is irrelevant) isotropic distance, which we note as `$\mathfrak{D}_V(z)$'.

The AP distortion information (fully described by $F_\mathrm{AP}$) combined with the isotropic late-time part, `$\mathfrak{D}_V+F_\mathrm{AP}$', represents the full `unnormalized and uncalibrated BAO' information. 
This information is particularly relevant for model building, since it is associated to the cosmological background only.
In fact, its measurement only relies on the assumption of the cosmological principle (e.g., that the universe is homogeneous and isotropic at any given time, without the need of assuming General Relativity (GR)\footnote{A small caveat is due: in practice the BAO compressed variables are often  extracted from the data after the so-called reconstruction step. Reconstruction is designed to undo the linear bulk-flows and non-linear peculiar motion, reduce statistical and systematic errors associated with non-linearities and hence sharpen the BAO signal. 
Reconstruction assumes a model for the growth of structure and thus GR. However these are small corrections and apply to small enough scales where Newtonian gravity holds and GR effects are unimportant.) see e.g., \cite{Sherwin_2019} and references therein.} and without early-time assumptions (apart from the existence of a standard ruler) and can hence be compared to the uncalibrated SNe Pantheon $\Om$
measurement. For a Friedmann-Lema\^itre-Robertson-Walker universe (and for any theory of gravity) in combination with SNe, BAOofLSS can be used to constrain the Etherington relation and global curvature. It is well known that the AP test, within the flat $\Lambda$CDM model (at late-time), constrains $\Omega_\mathrm{m}$; within a more general dark energy model it can also constrain its associated equation of state parameter(s). 

There is also an intermediate step,  whereby 
the absolute value of $D_V$ (i.e., normalized) but uncalibrated ruler (so condition {\it iii)} still does not apply), $D_V/r_{\rm d}$,  is considered. This is what gives the usual combination $h r_{\rm d}$ that standard BAOofLSS analyses retrieve. Then, the final step is to get a value or constraint on $r_{\rm d}$ (i.e., now {\it iii)} does apply) to finally obtain $h$ \cite{bernal_trouble_2016, 2014PhRvL.113x1302H, Standards2}.  
Hence, the BAOofLSS information alone (in terms of angles and redshifts) is not capable of constraining $h$ unless a certain model is assumed about the early-time physics.  Recall that the sound horizon scale is given by,\footnote{Alternative scaling reported by \citep{Aubourgetal15} is $r_{\rm d}\simeq \frac{55.154 e^{-72.3(w_\nu+0.0006)^2}}{(w_c+w_b)^{0.25351}w_b^{0.12807}}$, which we report here just for reference.} 
\begin{equation}
\label{eq:rw_lcdm} r_{\rm d}=\int^{z_d}_{\infty} \frac{c_s(z)}{H(z)}dz \simeq \mid_{\Lambda {\rm CDM}} \frac{147.05}{\rm Mpc} \left(\frac{\Omega_m h^2}{0.1432}\right)^{-0.23}\left(\frac{N_{\rm eff}}{3.04}\right)^{-0.1} \left(\frac{\Omega_{\rm b} h^2}{0.02236}\right)^{-0.13},
\end{equation} 
  where $z_d$ denotes the redshift at radiation drag, and $c_s$ is the sound speed in the photon-baryon fluid (and we have used eq.~2.4 of \cite{NilsBBN2} on the RHS).
For example, within a flat $\Lambda$CDM model (shortly) before recombination, with fixed CMB temperature, $\rd$ is a function of the physical matter density $\omega_m\equiv\Omega_\mathrm{m}h^2$ and the physical baryon density $\omega_{\rm b}\equiv\Omega_\mathrm{b}h^2$ only. Note that for more exotic models, such as those with dark radiation or with early dark energy, this is no longer the case. But for standard early-time physics, the BAO measurement of $h\rd$ and $\Omega_\mathrm{m}$ can be assisted by a BBN prior on $\Omega_\mathrm{b}h^2$ to provide a measurement of $h$. Invoking such a prior on the physical baryon density is the methodology chosen in this paper, as further specified in section \ref{sec:results-data}. For a more sophisticated treatment we refer to \cite{NilsBBN2}, where \SF is combined with the full BBN likelihood operating with the exact measurements of the relevant nuclear reaction rates. All the results obtained from BOSS+eBOSS for the respective cases mentioned here are shown in section \ref{sec:results}.

As noted before, exotic dark-radiation models aim to explain the Hubble tension as an effect of a dark-radiation component on the expansion rate at pre-recombination times, which modifies the sound horizon scale. As noted by \cite{Knox_Millea20}, an increased expansion rate at early-times leads to a decrease of $r_{\rm d}$, essentially because the time required to reach the threshold temperature for decoupling is lower. However, this prior-to-recombination change in the shape of $H(z)$ may also impact other relevant scales such as the sound horizon at matter-radiation equality and the scale of photon diffusion at recombination. Current LSS and CMB data yield tight constraints on these scales, therefore naive dark-radiation models which aim to solve the Hubble tension may fail to  provide a good description of other observables.

To understand how internal tensions in the LSS data
might hint to new physics beyond the standard, vanilla, $\Lambda$CDM model, we consider the isotropic BAO distances given by the Planck best-fit cosmology under the assumption of a $\Lambda$CDM model at different redshifts of interest: 
\begin{eqnarray}
\label{dv_first} D_V(z=0.38)&\simeq\mid_{\Lambda {\rm CDM}}& \frac{1477.07}{[{\rm Mpc}]} \left(\frac{\Omega_m}{0.316}\right)^{-0.11} \left(\frac{h}{0.674}\right)^{-1}; \\
D_V(z=0.70)&\simeq\mid_{\Lambda {\rm CDM}}& \frac{2375.74}{[{\rm Mpc}]} \left(\frac{\Omega_m}{0.316}\right)^{-0.18} \left(\frac{h}{0.674}\right)^{-1}; \\ 
D_V(z=1.48)&\simeq\mid_{\Lambda {\rm CDM}}&\frac{3808.19}{[{\rm Mpc}]} \left(\frac{\Omega_m}{0.316}\right)^{-0.29} \left(\frac{h}{0.674}\right)^{-1};\\
\label{dv_last} D_V(z=2.44)&\simeq\mid_{\Lambda {\rm CDM}}&\frac{4603.25}{[{\rm Mpc}]}\left(\frac{\Omega_m}{0.316}\right)^{-0.34} \left(\frac{h}{0.674}\right)^{-1}.
\end{eqnarray}
Considering now the sound horizon dependence within the early-time physics given by $\Lambda$CDM with fixed number of neutrino species and fixed baryon density we can write the above expressions as, 
\begin{eqnarray}
   \label{dvrs_first} \frac{D_V(z=0.38)}{r_{\rm d}}&\simeq&\mid_{\Lambda {\rm CDM}}10.04
    \left(\frac{\Omega_m}{0.316}\right)^{0.12}\left(\frac{h}{0.674}\right)^{-0.54};\\
   \frac{D_V(z=0.70)}{r_{\rm d}}&\simeq&\mid_{\Lambda {\rm CDM}}16.15
    \left(\frac{\Omega_m}{0.316}\right)^{0.047}\left(\frac{h}{0.674}\right)^{-0.54};\\
\frac{D_V(z=1.48)}{r_{\rm d}}&\simeq&\mid_{\Lambda {\rm CDM}}25.90
    \left(\frac{\Omega_m}{0.316}\right)^{-0.058}\left(\frac{h}{0.674}\right)^{-0.54};\\
 \label{dvrs_last}  \frac{D_V(z=2.4)}{r_{\rm d}}&\simeq&\mid_{\Lambda{\rm CDM}}31.30
    \left(\frac{\Omega_m}{0.316}\right)^{-0.11}\left(\frac{h}{0.674}\right)^{-0.54}.
\end{eqnarray}
These scalings are illustrated and discussed  below, in figure~\ref{fig:fig1}.
Note that measurements of the unnormalized and uncalibrated $D_V/r_{\rm d}$ at different redshifts, $\mathfrak{D}_V(z)$, yield a constraint on $\Omega_m$, whereas the normalized (but uncalibrated) $D_V(z)/r_{\rm d}$ measurements additionally yield a constraint on $hr_{\rm d}$, both solely based on late-time physics and late-time observables.\footnote{While the measurements are based on late-time quantities and physics, $r_{\rm d}$ is an early-time quantity sensitive to early-time physics.}

\subsection{Shape information -  \texorpdfstring{$h$}{} from the equality epoch } \label{sec:SF-shape}
The main new ingredient of \textit{ShapeFit}, the compressed variable capturing the shape, $m(z)$, is obtained from the data in a model-independent way and its measurement can be compared to any cosmological model of choice in a later step.

There is a variety of physical effects that modify the power spectrum shape in a trivial or non-trivial way. These include: the scale-independent scalar tilt $n_s$ of the primordial power spectrum, scale-dependent shape variations due to the scale of equality between matter and radiation, the baryon suppression or the free-streaming scale of massive neutrinos. While all of these effects have very different signatures on the matter power spectrum when considering the full wave-vector range in principle  accessible by observations, their behaviour is qualitatively similar in the range of interest for galaxy clustering, $0.02 < k\,\hoverMpc < 0.20$, (see for  example figures 1 and 4 of \cite{ShapeFit}). This justifies the interpretation of the measurement of the effective parameter  $m(z)$ within any model that does not involve severe scale-dependent changes in slope. Still, it is important to note that the shape is \textit{not} a pure late-time or early-time parameter, but rather captures several physical effects ranging from the epoch of inflation until today \cite{BriedenPRL21}.

Nevertheless, under certain assumptions, the shape can be used to obtain a (model-dependent) constraint of $h$. First of all we need to assume that the measured  large-scale clustering (at $k<0.1$ $h$Mpc$^{-1}$) reflects faithfully the shape of the matter transfer function (e.g., the galaxy bias is scale-independent at these large, linear scales).  
Second, a prior on the scalar tilt $n_s$, which under the inflation-motivated assumption of a power-law primordial power spectrum is very well measured by CMB data \cite{Aghanim:2018eyx,SPT-3G:2021eoc,ACT:2020gnv} due to the wide range of scales covered, needs to be imposed. This prior is needed given that, for the current data sensitivity, the degeneracy between the scale-independent ($n$) and scale dependent ($m$) slopes introduced in \cite{ShapeFit} cannot be effectively broken within \SF, although this may change with future surveys probing larger volumes. 
Third, a prior on the baryon density, $\Omega_\mathrm{b}h^2$, is needed, because this parameter controls the baryon suppression in the matter transfer function, which is degenerate with $m$. This can be obtained either from CMB observations or adopting a BBN-motivated prior as discussed in section~\ref{sec:SF-geom}.
Finally, the presence of massive neutrinos leads to a signature in the linear power spectrum which is degenerate with $m$; we study this in more detail in section \ref{sec:results-geoshapegrowth}.

To summarize, by assuming fixed fiducial values for, or imposing priors on, $n_s$, $\Omega_{\rm b}h^2$ and the sum of neutrino masses $\Sigma m_\nu$ (or giving it some upper limit), we effectively calibrate the shape variable, $m(z)$. This is the model-dependent, but necessary step, before making $m$ a variable that captures the signature of early-time physical processes. With this, this variable can be related directly to how fast modes re-enter the horizon at the end of the radiation dominated era, where their logarithmic growth is gradually suppressed along with the increasing importance of the matter density relative to radiation.
This is the physical signature that the large-scale shape of the matter transfer function captures: the evolution of the growth suppression of the last few modes that enter the horizon just before matter domination. 
In this sense the calibrated $m$ is a {\it speedometer} of how fast the horizon was expanding at that epoch, and this is the physical signature that 
will be used to measure the Hubble expansion rate.  
In particular, if the Universe is effectively $\Lambda$CDM at times around matter-radiation equality, $k_{\rm eq}$ is set by the parameter combination $\Omega_\mathrm{m}^{\rm}h^2\mid_{\rm eq}$. Here we have explicitly indicated that this early-time quantity is set by  matter-radiation equality physics. 
With the above caveats and assumptions, indeed we find that the relation between $m$ and $k_{\rm eq}$ is well approximated by,
\begin{align}
    \frac{k_\mathrm{eq}}{k_\mathrm{eq}^\mathrm{fid}} = 0.77 m^4 + 0.80 m^3 + 0.96 m^2 + 1.16 m + 1\,.
    \label{eq:mscaling}
\end{align}
Note that the equality scale can be used as a standard ruler \citep{cunnington22}, with its own scaling parameters  $F_{\rm AP}$ and $D_V(z)k_{\rm eq}$, yielding a self-contained constraint on $\Omega_m$ (for uncalibrated and unnormalized geometric measurements) and on $h$ for the calibrated and normalized quantities, in the exact way as being done for the sound horizon scale. Such a measurement has not been done yet on real data, although this would be close to the parameterisation chosen by \cite{Cuceu:2021hlk} for Lyman-$\alpha$ analyses beyond the BAO. We investigate the prospect of such an equality based standard ruler approach in section \ref{sec:keq}. Instead, the $m$-based analysis does not employ this scale as a standard ruler, but gives an indirect determination of this early-time quantity through its feature on the matter power spectrum (the shape once calibrated on $n_s$, $\Omega_{\rm b}h^2$ and $\sum m_\nu$) rather than its position in scale. 

As previously mentioned, the shape is directly coupled to the scale of equality in coordinates $m \propto \keq \, [\mathrm{Mpc}^{-1}] \propto \Om h^2$.
While under a change of coordinate system towards $h\mathrm{Mpc}^{-1}$ the equality scale obviously transforms as $\keq \, [h\mathrm{Mpc}^{-1}] \propto \Om h$, the scaling of the unit-less {\it shape} variable $m$ with $\Om h^2$ is invariant under such a coordinate transformation: for a given power spectrum, any multiplicative coordinate transformation only changes the horizontal and vertical position in $\log(k)-\log(P(k))$ space while leaving the {\it shape} intact.  

This is the reason why our `indirect' $m$-based equality measurement is sensitive to the parameter $\Om h^2$.\footnote{Note that in the discussion in our previous work \cite{BriedenPRL21} $m$ was presented to be proportional to $\Omega_\mathrm{m}h$. This is incorrect. While the large-scale structure measurement of $k_{\rm eq}$ as a standard ruler does scale $\propto \Omega_\mathrm{m}h$, $m$ itself scales $\propto \Omega_\mathrm{m}h^2$.} On the other hand, a `direct' equality measurement based on the standard ruler method (converting angles and redshifts into distances and using the equality scale as a unit) would depend on $\Om h$. This is further explained and explored in section \ref{sec:keq}.

To summarize, in a $\Lambda$CDM model the measurement of $m$ can be interpreted as a constraint on $\Omega_mh^2$, which, in conjunction with the $\Omega_\mathrm{m}$ constraint from the geometrical, ``unnormalized and uncalibrated BAO'' information, results in a constraint on $h$. This constraint is hence independent of the absolute value of the sound horizon, but still (because of the adopted $\Omega_m$ constraint) makes use of the fact that the sound horizon is a standard ruler\footnote{In principle an $\Omega_m$ constraint could be obtained from other standards such as the standard ruler explored in section~\ref{sec:keq} or non-standard-ruler probes such as supernovae. Here we prefer to derive all the constraints as much as possible internally to large-scale structure data currently available.}. The same reasoning applies to recent full-modeling approaches  for sound horizon-free determinations of the Hubble constant \cite{Philcox_Sherwin_Farren_Baxter_21,Farren:2021grl,Philcox:2022sgj,Smith:2022iax}.

\subsection{Geometry vs. shape  - a tale of two \texorpdfstring{$h$}{}'s}

Given the observables related to geometry and growth that \SF provides, in this section we would like to explore their cosmological implications and their role in providing an internal consistency check of the underlying model.

\begin{figure}[t]
    \centering
    \includegraphics[width=0.49\textwidth]{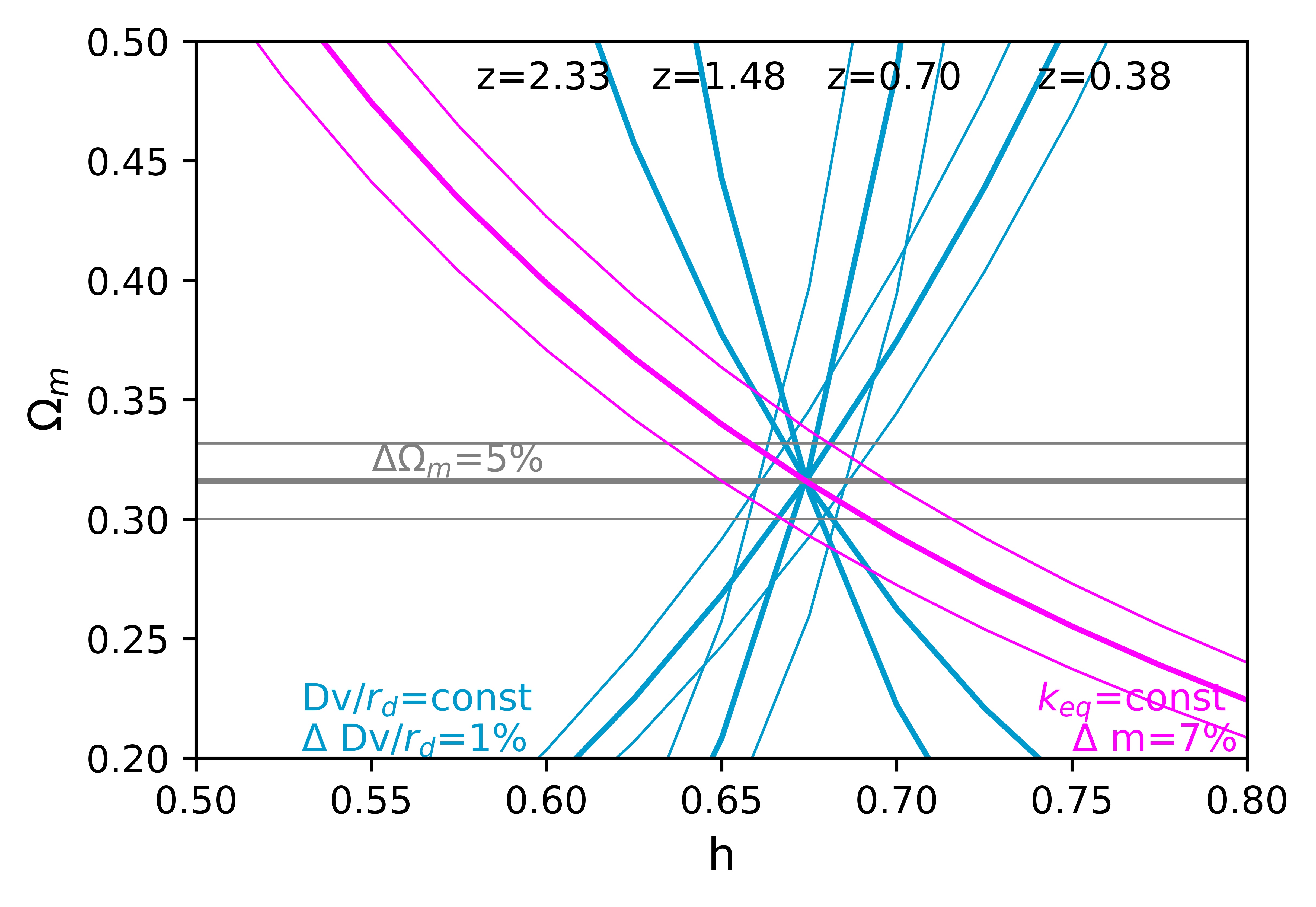}
    \includegraphics[width=0.49\textwidth]{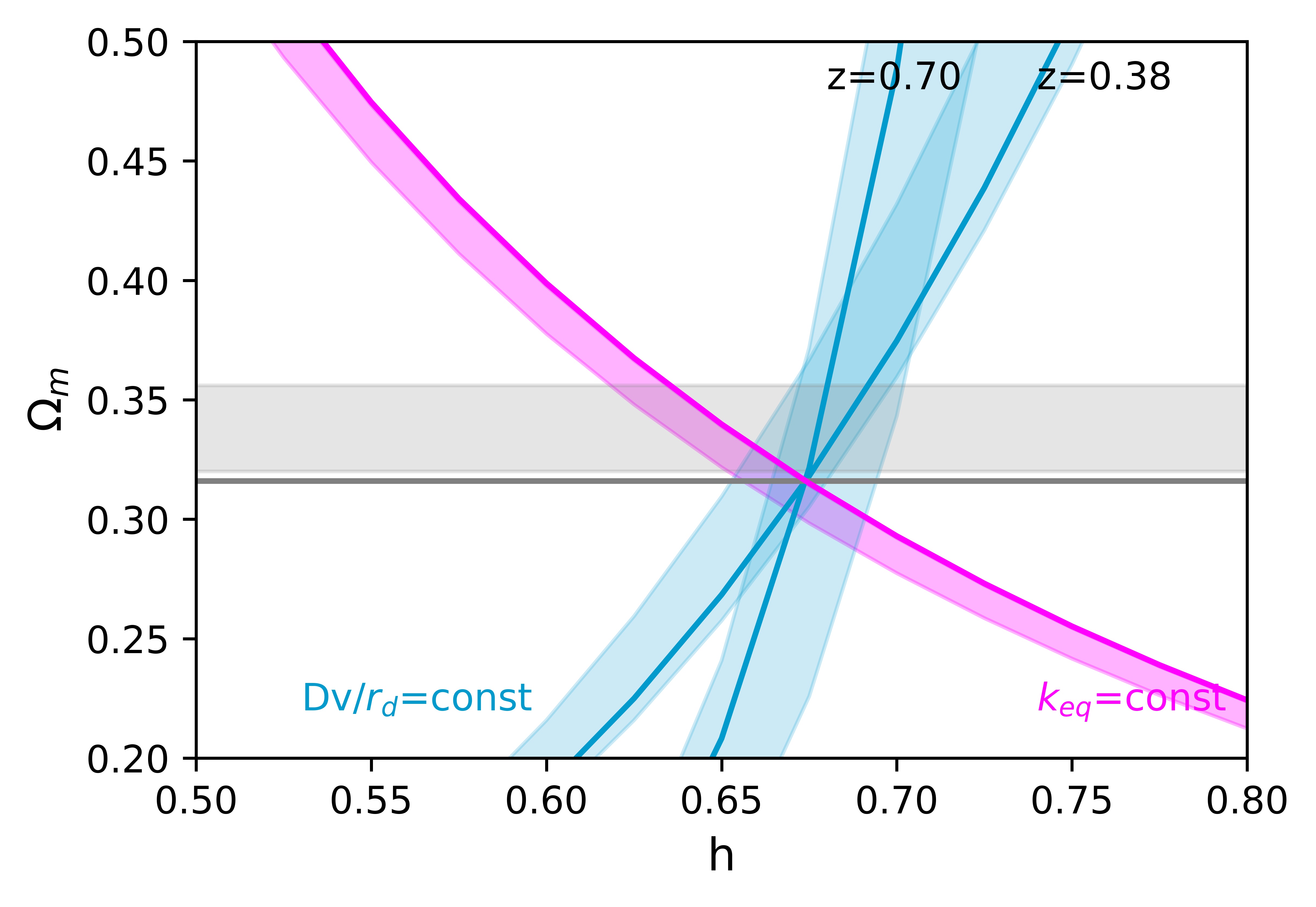}
\caption{Left panel:} Expected $\Omega_\mathrm{m}-h$ plane relations  within a  $\Lambda$CDM model: thick cyan lines show the relation imposed by $D_V/r_{\rm d}$ predicted by the Planck-calibrated $\Lambda$CDM model at representative  redshifts (see eqs.~\eqref{dvrs_first}-\eqref{dvrs_last}),  the  thick  magenta line is $m$ (or $k_{\rm eq}$) constant  line which corresponds to a constant $\Omega_mh^2$, also predicted by  the  Planck-calibrated $\Lambda$CDM model; the thick grey line indicates the $\Omega_m$ best-fit Planck value. The thinner lines represent shifts from that model as annotated. In particular a 7\% shift in $m+1$ corresponds to a $\sim 9$\% shift in $k_{\rm eq}$. Right panel: same, but with semi-transparent colour bands representing selected  measurements (1$\sigma$ confidence levels) from the literature as detailed in the text. The fact that the measured values agree with the expectations offer a powerful consistency check of the model.
    \label{fig:fig1}
\end{figure}

In a Planck-calibrated $\Lambda$CDM model,  the expected relation between $\Omega_m$ and $h$  given by a fixed $D_V/r_{\rm d}$ (calibrated) is shown in the left panel of
figure~\ref{fig:fig1} for few  representative redshifts, $z=0.38,\,0.70,\,1.48,\,2.33$ (in solid cyan lines), following the expressions given by eqs.~\eqref{dvrs_first}-\eqref{dvrs_last}. The lines intersect at the Planck-calibrated `true' value for $\Omega_m$ and $h$. Similarly, the expected Planck-calibrated relation for $m$ is shown as magenta solid line, which also intersects the $D_V/r_{\rm d}$ lines at the same point.  If the measured value of $D_V/r_{\rm d}$ were to differ from the Planck $\Lambda$CDM prediction by 1\% then the cyan lines would shift as indicated by the respective fainter and thinner cyan lines. If the measured value of $m$ were to be different by 0.07 (or a change of $7\%$ in $m+1$) from the expected value, the  magenta line would shift as indicated by the  fainter and thinner magenta lines. The $\Omega_m$ value for the Planck-calibrated model is shown by the grey horizontal line, with a $\pm 5\%$ change also shown. The shaded regions in the right panel of figure ~\ref{fig:fig1} represent the constraints from BOSS+eBOSS samples (combination of all the redshift bins)~\cite{Brieden:2022lsd} for $m$ (in magenta) and for $D_V/r_{\rm d}$ (in cyan)  for  $z=0.38$ and $z=0.70$ at 1$\sigma$ (other redshifts are not shown for visibility purposes), the horizontal grey region represents the $\Omega_m$ constraint from the Pantheon+ sample \cite{pantheonplus,brout22}. Also note that when considering $D_V/r_{\rm d}$ (cyan lines/bands), if we were marginalizing over the absolute size of the sound horizon scale, $r_{\rm d}$ ($\mathfrak{D}_V$ case), the cyan lines/bands would  move, freely but coherently, horizontally in figure~\ref{fig:fig1}; $h$ would be totally unconstrained as the system would be uncalibrated). Still, the  constraints for the different redshift bins  would cross at the same $\Om$ value, constraining thus $\Om$ as anticipated.

\label{sec:2h}

From eq.~\eqref{eq:mscaling} it follows that, if $k_{\rm eq}$ were to be 1, 2, or 5\% different from the fiducial value assumed, $m$ would change by $\pm (8.5\times 10^{-3}, 1.7\times 10^{-2}, 4.3\times 10^{-2})$, respectively.  A change of 7\% in $m+1$ as reported in figure~\ref{fig:fig1} would correspond to a change of about 9\% in $k_{\rm eq}$ and  thus, a  9\% change in $\Omega_m h^2$. 
In fact, within a $\Lambda$CDM model $k_{\rm eq}$ is related to $\Omega_mh^2$ (and hence to $m$ through eq.~\eqref{eq:mscaling}) as,
\begin{equation}
k_{\rm eq}\,{ {\rm [Mpc^{-1}]}}\simeq\mid_{\Lambda{\rm CDM}} 0.010398 \frac{\Omega_mh^2}{0.1432}\left(\frac{2.379\times 10^4}{\Omega_{\rm rad}h^2}\right)^{1/2}\,,
\end{equation}
where $\Omega_{\rm rad}$ is the radiation density. The fact that the measured $h$ and $\Omega_m$ parameter combinations
agree with the $\Lambda$CDM prediction is already a powerful consistency check for the model, as figure~\ref{fig:fig1} illustrates.
On the other hand, given the measured values for $D_V/r_{\rm d},\, F_\mathrm{AP}$ and $m$, we can ask ourselves what can be inferred about the model.\footnote{As explained in section \ref{sec:SF-geom}, $F_\mathrm{AP}$ also provides a measurement of $\Om$, but we leave it out of our discussion in this section for simplicity.}

In order to tie together all these quantities, specific aspects of a model have to be assumed at different interpretation stages; a late-time  $\Lambda$CDM-type expansion history, standard expansion rate around matter-radiation equality, or standard early-type components, such as the sound horizon scale predicted by an early-time $\Lambda$CDM model (just) before recombination.  But what if one or more of the adopted aspects of the model were incorrect?  In this case, when fitting the data, the various parameters, in particular $\Omega_m$ and $h$, would have to be seen as effective parameters: they would be biased, because of the specific shortcomings of the adopted model. 

\begin{figure}[t]
    \centering
    \includegraphics[width=0.45\textwidth]{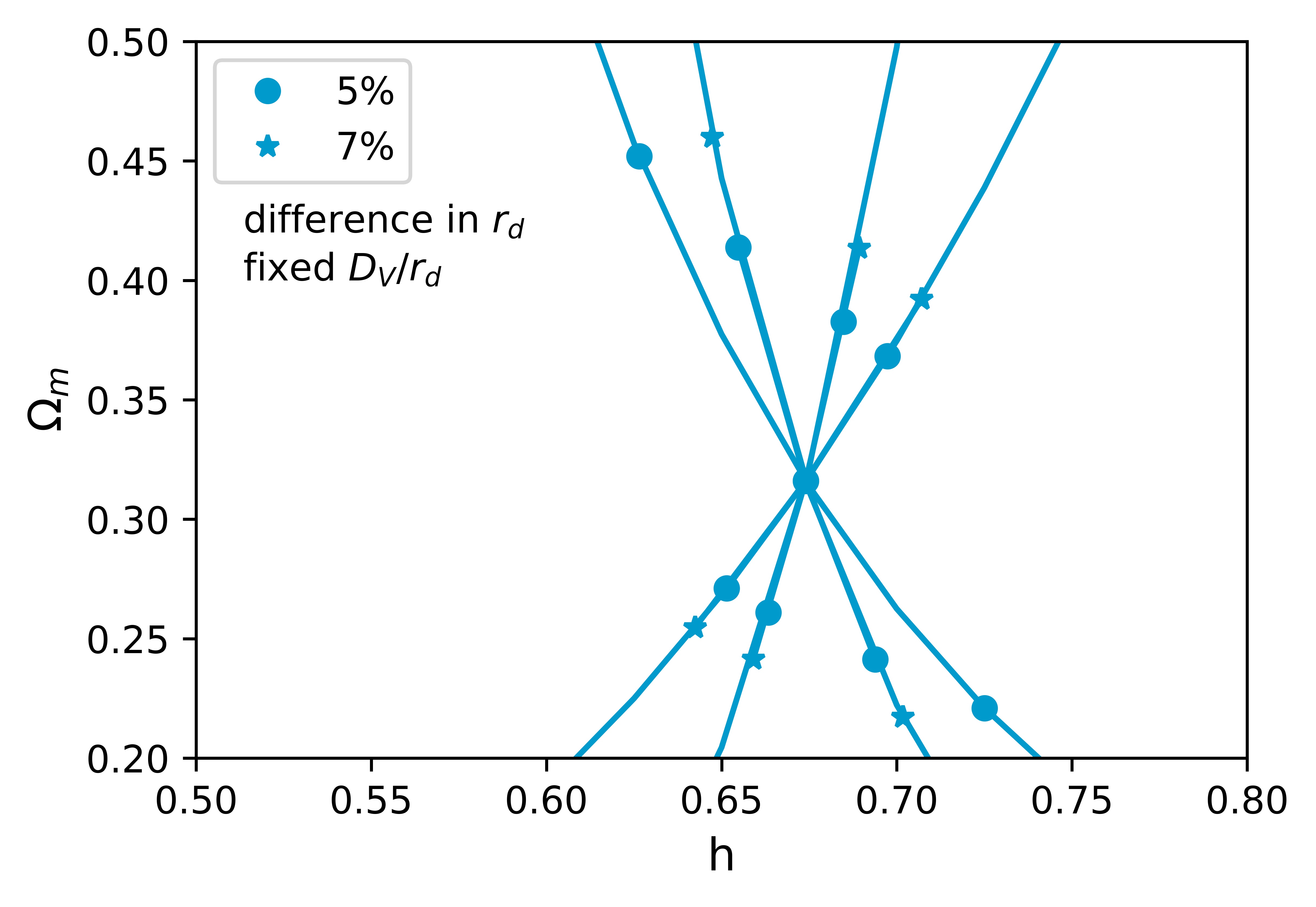}
    \includegraphics[width=0.45\textwidth]{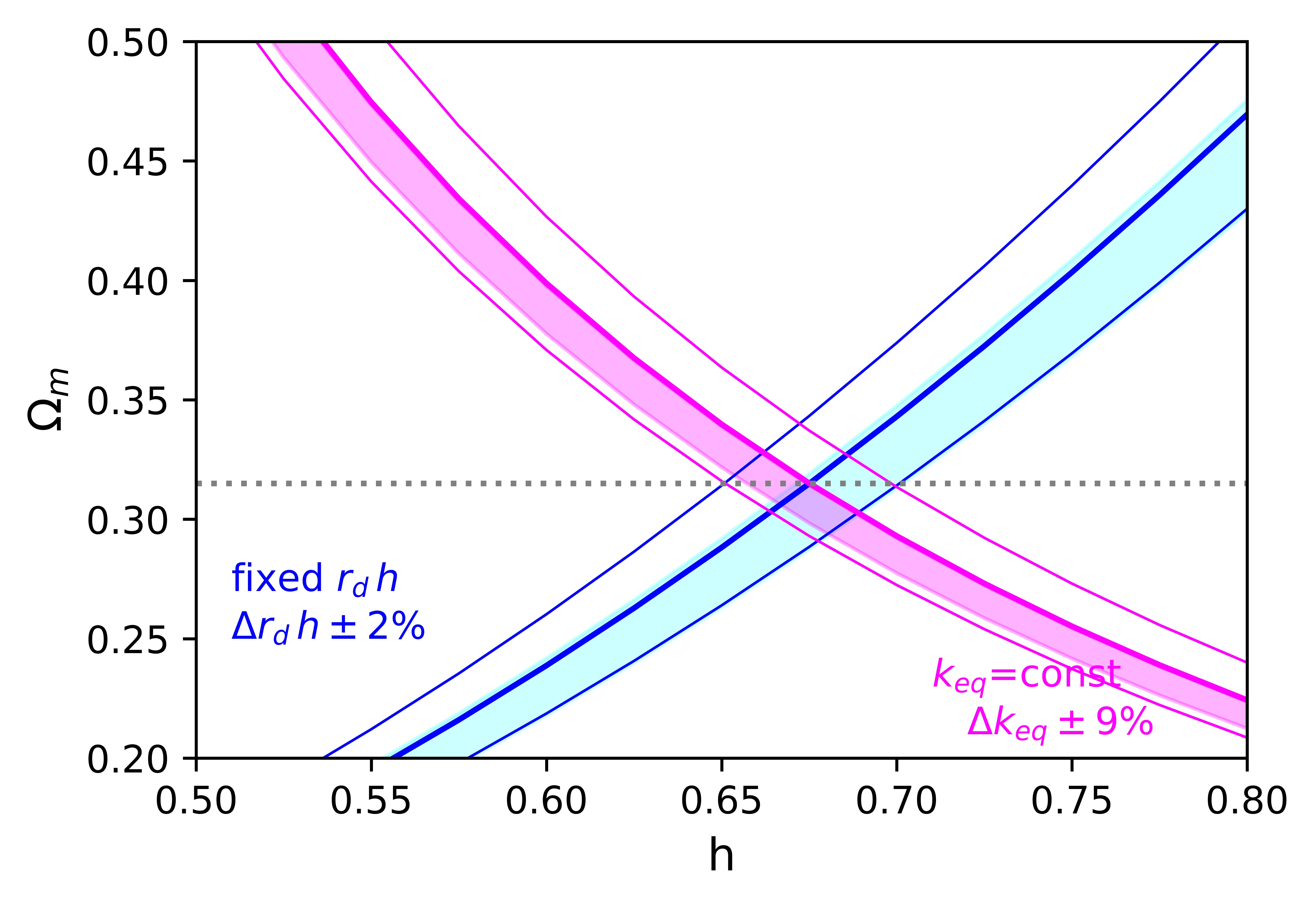}
\caption{Expected $\Omega_\mathrm{m}-h$ plane relations  under the assumption of a $\Lambda$CDM model. On the left panel: the cyan lines are the same as in the left panel of} figure~\ref{fig:fig1}, now indicating  how a change in the $r_{\rm d}$ value (because of new physics in the early universe) would reflect on the $\Omega_m$ and $h$ values to keep the BAO-measured quantity $D_{\rm V}/r_{\rm d}$ fixed when both $r_{\rm d}$ and $D_V$ are interpreted in a standard $\Lambda$CDM model.
This exemplifies how the combination of $D_V/r_{\rm d}$ measurements at different redshift yield the correct late-time $\Omega_m$ even in the case that unknown (or incorrect) early-time physics bias $r_{\rm d}$.
On the right panel: expected $\Omega_\mathrm{m}-h$ plane relations  for constant $hr_{\rm d}$ (cyan) and  constant $k_{\rm eq}$ (hence $\Omega_m h^2$, magenta). Despite being constrained by late-time observables, these actually depend on purely early-time quantities. Thinner lines show 2 and 9\% (respective) departures from the $\Lambda$CDM expected values. The semi-transparent cyan band represents the measured constraints on $hr_{\rm d}$ obtained by \cite{BernalTriangles} assuming a model-agnostic 4 knots (6 parameters)  reconstruction  of the late-time expansion history and marginalizing over curvature. The magenta semi-transparent band corresponds to the $m$ constraint from BOSS+eBOSS also shown in the right panel of fig.~\ref{fig:fig1}.
The horizontal dotted grey line at the fiducial $\Omega_m$ value serves to guide the eye; it shows that in a $\Lambda$CDM model  a 2\% change in $hr_{\rm d}$ should be compensated by a 9\% change in $k_{\rm eq}$ to keep $\Omega_m$ unchanged.
    \label{fig:fig2}
\end{figure}

For example, let us assume some new physics at early time such that the true underlying $r_{\rm d}$ were to be significantly different (5 or 7\%) from the Planck-calibrated $\Lambda$CDM value, but the matter density $\Omega_m$ was unchanged. If one were to interpret the  $D_V/r_{\rm d}$ constraints  within a $\Lambda$CDM model, the recovered values of  $\Omega_m$ and $h$ would move along the $D_V/r_{\rm d}=\,{\rm constant}$ degeneracy line to compensate: they would be biased. This bias would be different for different redshifts as illustrated in the left panel of figure~\ref{fig:fig2}. These shifts may not be observable: the combination of  BAO measurements at different redshifts will  still yield the (correct) late-time $\Omega_{\rm m}$ value and a biased $h$ value. Hence a direct (local) measurement of $h$  at late-time would signal an issue \cite{riess20}. The new physics at early times postulated in this example, however, would have to leave $k_{\rm eq}$ (almost) unchanged, given the $m$ constraints. For example, additional dark radiation would change $r_{\rm d}$ and would bring the early-Universe inferred $h$ value in closer agreement with the local determinations \cite{Riess:2021jrx}. However enough additional dark radiation  to yield a $\sim 7$\% change in $h$, would also likely change $k_{\rm eq}$ and BBN: \cite{NilsBBN2} shows that for dark radiation models constraints broaden, but large systematic shifts are disfavoured. Early dark energy models that solve the Hubble tension leave BBN unchanged, modify the expansion rate at matter-radiation equality, but fade away quickly after recombination: this affects therefore $r_{\rm d}$, but also $k_{\rm eq}$. 
 
This argument is parallel to that of \cite{Knox_Millea20}, but now can be made only referring to late-time observations and  independently of the CMB constraints (see also \cite{Smith:2022iax}). Of course the $m$ constraints can be `evaded' by  allowing the primordial power spectrum spectral slope $n_s$ to deviate from the canonical $n_s\sim 0.96$ values and/or allowing the $\Omega_{\rm b}h^2$ constraint to deviate from BBN.
On the other hand recall that, without assuming an early-time 
value on $r_{\rm d}$, the late-time observations constrain the late-time $\Omega_m$ and the combination $hr_{\rm d}$, which is an early-time quantity, sensitive to early-time physics. 

The right panel of figure~\ref{fig:fig2} illustrates this last point. The cyan lines correspond to constant $hr_{\rm d}$ values and the magenta lines to constant $k_{\rm eq}$ both under the assumption of a model consistent with $\Lambda$CDM at early times (with fixed baryon density), but not necessarily with the same values of cosmological parameters as the late time. In other words the $h$ and $\Omega_m$ represented in the right panel of figure~\ref{fig:fig2} are early-time quantities. 

To summarize,  measurements of the compressed variables of eq.~\eqref{eq:theta} can be used to provide a constraint on the late-time matter density parameter, $\Omega_m^{\rm late}$, assuming only isotropy (i.e., condition {\it i)}, through relative measurements of $D_M(z_i)/D_H(z_i)$ at the same redshift) and that the late-time expansion history is that of a flat $\Lambda$CDM model. A more stringent constraint can be obtained on $\Omega_m^{\rm late}$ if  also assuming that the ruler does not evolve with time (i.e., condition {\it ii)}, through relative measurements of $D_V(z_i)/D_V(z_j)$ at the different redshifts), what we refer as the uncalibrated-and-unnormalized' BAO. Additionally, the normalized-but-uncalibrated BAO yields a measurement of the combination $hr_{\rm d}$, sensitive to recombination-time physics. From this a $h$ measurement can be obtained given the knowledge of $r_{\rm d}$ (condition {\it iii)}, through an early-time physics assumptions, for e.g., eq.~\eqref{eq:rw_lcdm}). For standard early-time physics (during the decade of expansion just before recombination and a baryon density prior \citep{Knox_Millea20}), the $hr_{\rm d}$ constraint yields a degeneracy in the $h^{\rm early,r}-\Omega_m^{\rm early,r}$ plane.\footnote{The superscript $r$ explicitly denotes the quantity's sensitivity to the physics of the recombination epoch.} If $\Omega_m^{\rm early}=\Omega_m^{\rm late}$, then $r_{\rm d}$ can be constrained from late-time observations only. This procedure yields the sound-horizon based Hubble parameter, $h_{r_{\rm d}}$.

Under the assumption of a $\Lambda$CDM-like model at early times, an anchored shape parameter (anchored by a prior on $n_s$ and $\Omega_{\rm b}h^2$) estimated from galaxy clustering yields constraints on $\Omega_m^{\rm early,eq} (h^{\rm early,eq})^2$, where we have specified that these are quantities sensitive to physics at equality times (and BBN) rather than at around recombination times.  These constraints do not rely on any assumptions about the late-time expansion history, but assume standard early-time physics at the time of the matter-radiation transition, and that late-time processes do not alter the large-linear scales.  This degeneracy in the $h-\Omega_m$ plane is different from the one given by $D_V/r_{\rm d}$. Thus, another Hubble parameter arises, $h_m$, when combining this measurement with the $\Omega_m^{\rm late}$ determination from geometry, again assuming it is an estimate for $\Omega_m^{\rm early}$ ($\Omega_m^{\rm early,eq}$ in this case).

Moreover, the two degeneracies in the $h-\Omega_m$ plane yield a joint constraint i.e., $h_{(m,r_{\rm d})}$ and  $\Omega_m$ (assuming $\Om^{\rm{early,r}}=\Om^{\rm{early,eq}}$), the latter can be compared with the geometric determination of $\Omega_m^{\rm late}$. 
These three $h$ are not statistically independent, as to obtain $h_{r_{\rm d}}$ and $h_m$ we have used the same $\Omega_m^{\rm late}$ geometric determination.
In section \ref{sec:results-geoshape} we extract all these quantities from state of the art data.   

\subsection{Adding the growth information. Yet another \texorpdfstring{$h$}{}?} \label{sec:SF-growth}

The redshift space distortion signal is efficiently captured by the compressed variable $f\sigma_{\rm s8}$, which is readily interpreted as the growth rate $f\equiv d\ln D /d\ln a$, where $D(z)$ is the linear growth factor of perturbations, combined with the {\it rms} amplitude of matter fluctuations, which is closely related to the $\sigma_8$ cosmological parameter.
The extraction of this  compressed variable relies on relatively weak assumptions (valid  even in models where gravity is significantly away from GR): that clustering is intrinsically isotropic with respect to the LOS, that in the linear regime growth is scale-independent,\footnote{If growth were to be scale-dependent the recovered value would correspond to some effective average.} that the Euler equation holds, that gravity is the only force at play on large scales, and that there is no velocity bias of dark matter tracers (galaxies and quasars in this case).

However, the interpretation of this compressed variable as related to the amplitude of matter perturbations, $\sigma_8$, also requires the Poisson equation to hold. 
As such, this compressed variable does not constrain the matter power spectrum amplitude {\it per se}, but, for observations at different redshifts,  carries information about the growth of structure throughout redshift. Alternatively, within a cosmological model and  in combination with the primordial amplitude of perturbations, a primordial power spectrum spectral slope, a theory of gravity and a matter transfer function, it can  yield information about the Universe composition and in particular about the matter density parameter. For example in a  $\Lambda$CDM model, $f(z)=\mid_{\Lambda {\rm CDM}}\Omega_m(z)^\gamma$,  with $\gamma\simeq 0.56$ and $\sigma_{s8}(z)\propto D(z)/D(z_i)\sigma_{s8}(z_i)$ where $D(z)/D(z_0)$, depends only on $\Omega_m$.
Measurements of $f\sigma_{s8}$ at different redshifts hence constrain $\Omega_m$.

If the true underlying model were instead to be richer than $\Lambda$CDM, by interpreting the growth information  while adopting the $\Lambda$CDM model for $D(z)$, the extracted $\Omega_m$ would correspond to a parameter driving the growth of perturbations, which in general for generic gravity models, does not necessarily need to coincide with the background-late-time $\Omega_m$, or the early-time $\Omega_m$, as described in section \ref{sec:2h}. This so-called $\Omega_m^{\rm growth}$ is a pure perturbation-late-time quantity.
On the other hand, within a $\Lambda$CDM model, a constraint on $f\sigma_{s8}$, combined with a constraint on $h$ or $k_{\rm eq}$ (from the BAO or from the shape), leads to a prediction for the amplitude of primordial perturbations. Conversely, a Planck-calibrated $\Lambda$CDM model also predicts a value for $f\sigma_{s8}$ as a derived parameter.

It is important to note that the \SF compressed variable,  $f\sigma_{s8}$, slightly deviates from the traditional approach one, $f\sigma_8$, as explained in detail in \cite{ShapeFit}. In a nutshell, the deviation is twofold: {\it i)} \SF measures the power spectrum amplitude at a reference scale in units of the sound horizon; and {\it ii)} upon changing the shape parameter $m$, only the power spectrum amplitude at the pivot scale $k_p=\pi/\rd$ remains fixed. Therefore, in reality, \SF constrains the variable $f A_{sp}^{1/2}$, where $A_{sp}$ is the power spectrum amplitude evaluated at $k_p$. This quantity can be transformed into our desired compressed variable $f\sigma_{s8}$, considering how deviations from the fiducial template shape choice (i.e., $m \neq 0$) affects the amplitude on scales of $R = 8 \, \mathrm{Mpc}/h$,
\begin{align} \label{eq:fAsp_to_fss8}
    f\sigma_{s8} \simeq f\sigma_{s8}^\mathrm{fid} \frac{f A_{sp}^{1/2}}{\left(f A_{sp}^{1/2}\right)^\mathrm{fid}} \times {\exp \left(\frac{m}{2a_{m}} \tanh \left(a_m \ln \left(\frac{\rd^\mathrm{fid} \, \left[\mathrm{Mpc}/h \right]}{8 \, \mathrm{Mpc}/h} \right) \right)\right)},
\end{align}
where for the \SF standard parametrization we have chosen $a_m$ to be 0.6 \cite{ShapeFit}.\footnote{Note that eq. \ref{eq:fAsp_to_fss8} is not exact, since $\sigma{s8}$ is an integrated quantity of the power spectrum, while $A_{sp}$ is not. For that reason it is recommended to always use $fA_\mathrm{sp}$ for cosmological inference rather than $f\sigma_{s8}$. }

\begin{figure}[t]
    \raggedright
    \includegraphics[width=0.49\textwidth]{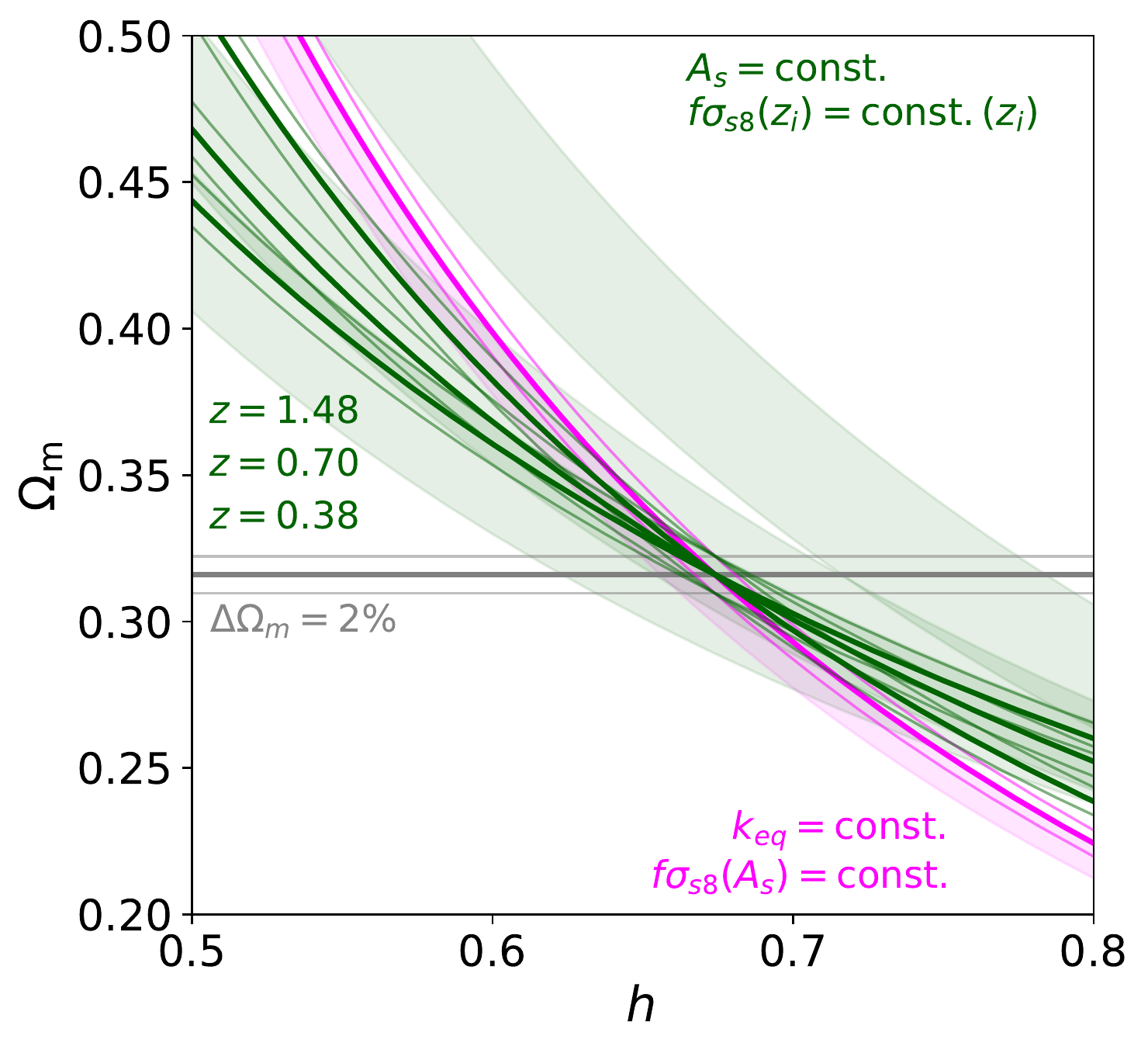}
    \includegraphics[width=0.49\textwidth]{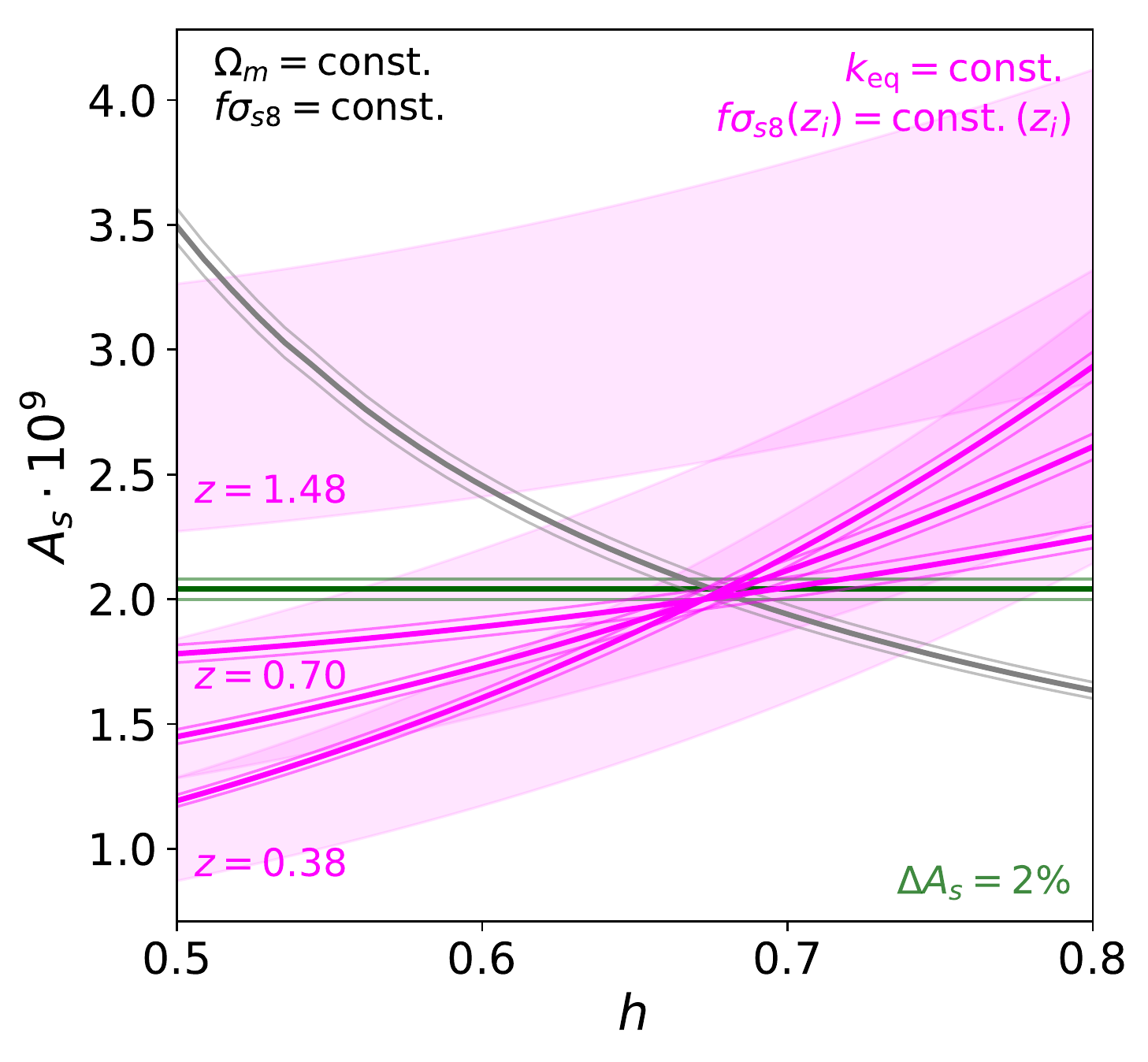}
    \caption{We show the cosmological information encoded in $f\sigma_{s8}$ measurements at three redshift bins. The left and right panels show the $\Om-h$ and $A_s-h$ planes respectively, where thick lines represent those values in parameter space that keep fixed corresponding annotated quantities. In particular, grey lines display fixed $\Om$, magenta lines fixed $\keq \propto \Om h^2$, and green lines fixed $A_s$. All lines represent $f\sigma_{s8}$-isocurves upon considering the full three-dimensional parameter space $(\Om, A_s, h)$. Thin lines indicate 1$\sigma$ deviations from a Planck-calibrated $\Lambda$CDM model and semi-transparent regions show the parameter space constrained by independent late-time probes as specified in the text.
    }
    \label{fig:fixed_fss8}
\end{figure}

We illustrate the information that \SF $f\sigma_{s8}(z_i)$ measurements add to the geometry and shape in figure~\ref{fig:fixed_fss8}. The left panel displays the $\Om-h$ plane, similar to both panels of figure~\ref{fig:fig1} combined. Since the additional growth data allow us to constrain the primordial amplitude $A_s$, we also display the $A_s-h$ plane in the right panel. The colour scheme is consistent between both panels, which helps to appreciate the three-dimensionality of the underlying system, $\{\Om, A_s, h\}$. Similarly to previous figures, grey, magenta and green curves keep $\Om$, $\Om h^2$ and $A_s$ fixed, respectively to Planck-$\Lambda$CDM reference values, as indicated in the plot. All lines show those parameter combinations that keep $f\sigma_{s8}$ fixed, where in some cases (green lines, $A_s=\rm{constant}$) the degeneracy directions are redshift-dependent. All these  $f\sigma_{s8}$=constant curves are derived from the following scalings for a Planck-like $\Lambda$CDM universe,

\begin{equation} \label{eq:fss8_scalings}
\begin{aligned}
    f\sigma_{s8}\, (z=0.38) =\mid_{\Lambda {\rm CDM}} 0.475  \left( \frac{\Om}{0.316} \right)^{1.76} \left( \frac{h}{0.674} \right)^2 \left( \frac{A_s}{2.04 \times 10^{-9}} \right)^{1/2}, \\
    f\sigma_{s8}\, (z=0.70) =\mid_{\Lambda {\rm CDM}} 0.461  \left( \frac{\Om}{0.316} \right)^{1.52} \left( \frac{h}{0.674} \right)^2 \left( \frac{A_s}{2.04 \times 10^{-9}} \right)^{1/2},  \\
    f\sigma_{s8}\, (z=1.48) =\mid_{\Lambda {\rm CDM}} 0.375  \left( \frac{\Om}{0.316} \right)^{1.22} \left( \frac{h}{0.674} \right)^2 \left( \frac{A_s}{2.04 \times 10^{-9}} \right)^{1/2}.
\end{aligned}
\end{equation}

The scalings with $h$ and $A_s$ are redshift-independent and arise from the fact that the galaxy power spectrum amplitude scales like $P_g \propto \sigma_8^2 \propto A_s h^4$; the redshift dependence enters through the scaling of $\Om$, which governs the growth history of cosmic structures.

In the left panel of figure~\ref{fig:fixed_fss8}, the thick grey line refers to constant $\Om$ (which can be constrained from a late-time geometry) and the thick magenta line to constant $\keq$ (which is inferred from the {\it shape} and corresponds to early-time quantities), equivalent to figure~\ref{fig:fig1}. The thin lines (for all cases) represent changes in $\Om$ by $2\%$, which correspond to the 1$\sigma$ sensitivity of Planck within the $\Lambda$CDM model.
Green thick lines display the combinations of ($\Om, h$) values that for a given redshift (see the individual relations in \eqref{eq:fss8_scalings}) keep $f\sigma_{s8}(z)$ fixed, given a constant value of the primordial amplitude $A_s$. Again, thin green lines represent deviations induced by a variation of $\Om$ by $2\%$ around the fiducial value. The green, semitransparent regions show the parameter space constrained at the 1$\sigma$ level by our \SF measurements at redshifts $z=(0.38, 0.7, 1.48)$ (we do not show the $z=0.51$ bin for better visibility). The $z=1.48$ bands display a $\sim2\sigma$ offset value with respect to Planck predictions. This is not unexpected and is caused by the high value of $f\sigma_{s8}$ for the QSO sample in \cite{Brieden:2022lsd} (see also original eBOSS works on this sample \cite{houetal2021,neveuxetal2020}).

From the similarity between the theoretical expectation at different redshifts (green curves) alone, we can see that inferring $\Om$ from the combination of $f\sigma_{s8}(z)$ measurements is challenging. 
Also, the green curves are quite similar to the magenta $\keq \propto \Om h^2$ lines. In fact, for fixed $A_s$, which fixes the power spectrum amplitude at $k_\mathrm{piv} = 0.05 \mathrm{Mpc}^{-1}$, most of the behaviour of $f\sigma_{s8}$ is driven by the exponential term in eq.~\eqref{eq:fAsp_to_fss8} because $k_\mathrm{piv}$ is rather close to the `\SF pivot scale', $k_p \approx 0.03 \, h\mathrm{Mpc}^{-1}$:  most of the amplitude variation at larger wavenumber $\approx  1/(8\mathrm{Mpc}/h)$ comes from the variation in shape when $A_s$ is fixed. This similarity between $\keq$ and $f\sigma_{s8}$ constraints appears to be stronger for higher-redshifts (where the sensitivity of the growth rate to $\Omega_{\rm m}$ is weaker).

In the right panel of figure~\ref{fig:fixed_fss8} we show the behavior of the $f\sigma_{s8} = \mathrm{constant}$ curves in the ($A_s, h$) plane considering three different cases. 
\begin{itemize}
    \item We keep $\Om$ fixed (grey lines), which fixes the growth history, such that the redshift-dependence vanishes. Hence, these lines (where thick lines refer to the fiducial cosmology and thin lines to deviations in $A_s$ by 2\%) are identical for each redshift bin.
    \item We keep $\keq \propto \Om h^2$ fixed (magenta lines), meaning that $\Om$ and hence the growth history changes for varying $h$. In order to keep $f\sigma_{s8}$ fixed, the change in $\Om$ is compensated by varying $A_s$. Since the impact of $\Om$ varies with redshift, this compensation via $A_s$ is redshift-dependent. This is why we show a different magenta isocurve for each redshift. The thin magenta lines represent a variation of $A_s$ by $2\%$.  
    \item For reference, we also show the case of fixed $A_s$ in green, where thin lines represent, again, a variation of $A_s$ by $2\%$. It is possible to fix $f\sigma_{s8}$ in this case by adjusting $\Om$ as explained before and displayed in the left panel.
\end{itemize}
In this ($A_s, h$) plane the redshift dependence of $f\sigma_{s8}$ reveals itself within the $\keq = \mathrm{const.}$ (magenta) case. Again, we can see the trend that for higher redshifts the green and magenta theory predictions come closer together while for later times the degeneracy is broken: at high-$z$ the compressed variable $f\sigma_{s8}$ is purely determined by the primordial amplitude $A_s$ and the shape $\Omega_m h^2$.

So how does the inclusion of the growth measurements impact our sound-horizon independent measurement of $h$? In principle, one could measure $h$ by combining $f\sigma_{s8}$ measurements at different redshifts and derive from figure~\ref{fig:fixed_fss8} at which value of $h$ they `cross'.\footnote{Here, it is important to note that by definition the $f\sigma_{s8}$ measurements are independent of the absolute value of the sound horizon. They represent the velocity fluctuation amplitude evaluated at a certain scale in units of the sound horizon, without making any assumption of the absolute value of the sound horizon.}

Considering the ($\Om, h$) plane in the left panel, we can see that, given the BOSS and eBOSS sensitivity (i.e., the size of the semitransparent bands), the combination of growth information from different redshifts cannot efficiently resolve the degeneracy and only provide a very loose constraint on the matter density of $\Om^\mathrm{growth} = 0.19\pm 0.06$. This corresponds to the point in parameter space where all green bands would overlap.
As already mentioned, it is not possible to derive $h$ from the growth of structure measurements alone. But, as evident from eq. \eqref{eq:fss8_scalings}, such a measurement can in principle be obtained by using a prior on $A_s$ (and also on $n_s$, which we do not take into account in this discussion for simplicity). In this way, $A_s$ serves as our new early-time anchor, we hence call the attributed $h$-measurement $h_{A_s}$. Such a constraint on $h_{A_s}$ is however very loose: we find an uncertainty of order $\sigma_{h_{A_s}} \approx 0.6$. For that reason we do not consider it in what follows.

On the other hand, the ($A_s, h$) plane in the right panel reveals an additional $h$ measurement.
While $h$ cannot be inferred from $f\sigma_{s8}$ alone, the combination with the measurements of $m$ indeed provides a measurement of $h$. In particular, we find $h_{m,f\sigma_{s8}} = 0.92_{-0.18}^{+0.13}$, where this high value is driven by the low $\Om^\mathrm{growth}$ measurements given above. Once we combine this with the uncalibrated BAO information we obtain $h_m = 0.701_{-0.021}^{+0.019}$ consistent our $h_m$ measurement from geometry and shape only from section \ref{sec:2h}.
In summary, we can see that the addition of growth data amplifies the possibilities for further internal consistency checks of the model:
\begin{itemize}
    \item In the ($\Om, h$) plane, do the green $f\sigma_{s8}$= constant constraints cross at the same position where the magenta $m$-derived $\keq$=constant  constraint crosses the grey $\Om$ constraint derived from geometry?
    \item In the ($A_s, h$) plane, is the crossing point of the magenta $f\sigma_{s8}$= constant constraint consistent with the $A_s$ value that returns the same $f\sigma_{s8}$ given $\Om$ is constrained via geometry (grey line)?
    \item Also in the ($A_s, h$) plane, is the $A_s$ value inferred from the crossing of the magenta $f\sigma_{s8}$=constant constraints and/or their crossing with the $\Om$ constraint  consistent with the Planck-derived value?
\end{itemize}
We can see that, once we start including other independent datasets and external priors, this list may become even longer.

One important question in this context is, of course: how could  the $\Lambda$CDM model be altered or extended to make the model consistent with the higher SH0ES value of the Hubble parameter while preserving agreement with the BOSS+eBOSS sound-horizon-free dataset, in particular our $f\sigma_{s8}$ measurements? 

From the left panel of figure~\ref{fig:fixed_fss8} we can draw the same conclusion as from figure~\ref{fig:fig1}. Without touching the $\Lambda$CDM model at or prior to equality, to increase $h$ we need to lower $\Om$, such that the model is still consistent with our $\keq$ and $f\sigma_{s8}$ measurements. But this automatically induces a tension with the (late-time) $\Om$ value preferred by geometry. From the right panel we can see that the problem is even more severe: increasing $h$ while still fitting the geometry and growth data forces $A_s$ to decrease. On the other hand, fitting shape and growth data forces $A_s$ to increase. Apparently, the $\Lambda$CDM model does not allow us to increase $h$ and at the same time fit the full geometry+shape+growth dataset.  
A related `curiosity' can be seen from our current BOSS+eBOSS dataset. The high $f\sigma_{s8}$ value reported by the QSO sample pushes $\Om$ towards smaller values (and $h$ and $A_s$ towards larger values, where the semitransparent bands overlap), but this is in contradiction with the expected behaviour of a $\Lambda$CDM model fed with the best-fitting geometrical constraints of the same QSO sample: a calibrated $D_V$ QSO-only analysis actually yields higher values of $\Omega_m$ and lower values of $h$ (see figure~9 of \cite{Brieden:2022lsd}). Since this `anomaly' is not (yet) statistically significant, only future data may allow us to flag this as an inconsistency of the $\Lambda$CDM model.

\section{Results from BOSS and eBOSS data} \label{sec:results}

\subsection{The data} \label{sec:results-data}
We use the same dataset as in \cite{ShapeFit:data} (see table 1 and figure~1 therein). This includes the LRG samples (three redshift bins) covering $0.2<z<1.0$ in a wavevector range of $0.02 < k \hoverMpc < 0.15  $ pre- and post- reconstruction, as well as the QSO sample (one redshift bin) covering $0.8<z<2.2$ in a wavevector range of $0.02 < k\hoverMpc < 0.30$ pre-reconstruction only. We directly use the compressed variables $\mathbf{\Theta}$ of eq.~\eqref{eq:theta} provided by \cite{ShapeFit:data}, and summarized in  appendix E therein. 
We conveniently transform the compressed datavector (and its covariance) using eqs.~\eqref{eq:DVrd} and \eqref{eq:FAP} to obtain the following modified compressed datavector, 

\begin{align} \label{eq:thetaprime}
    \mathbf{\Theta}^\prime(z) = \left\lbrace D_V(z)/\rd, F_\mathrm{AP}, f(z)\sigma_{s8}(z), m(z) \right\rbrace .
\end{align}
We extend the redshift range of our dataset by using the eBOSS DR16 Lyman-$\alpha$ BAO data from \cite{dumasdesBorboux2020}. As our baseline choice (that is except in sec.~\ref{sec:results-nu} where we  explicitely vary the number of effective neutrino species), we include the BBN prior adopted from \cite{Cuceu20,Cooke:2018,Adelberger:2010qa},
\begin{align} \label{eq:bbn-prior}
    \Omega_\mathrm{b}h^2 = 0.02235 \pm 0.00037 ~,
\end{align}
and fix the spectral index $n_s$ to our baseline choice of,
\begin{align} \label{eq:ns-prior}
n_s = 0.97.
\end{align}
In Appendix \ref{app:ob-ns} we explore the effect of varying these priors.
Our cosmological likelihood is based on a modified version of the Boltzmann-code \textsc{CLASS} \cite{2011JCAP...07..034B} within the \textsc{MontePython} sampler \cite{Brinckmann:2018cvx}.

Unless otherwise stated, we report inference of cosmological parameters obtained from the combination of the constraints on the physical variables at different redshifts. In particular, $m$ is not predicted to change with redshift in any of the $\Lambda$CDM extensions considered in this paper; similarly the $m$ measurements reported in \cite{ShapeFit:data} do not show any significant redshift-dependence. 

\subsection{\texorpdfstring{Piecewise-$\Lambda$}{}CDM constraints from geometry and shape} \label{sec:results-geoshape}

\begin{figure}[t]
    \centering
    \includegraphics[width=\textwidth]{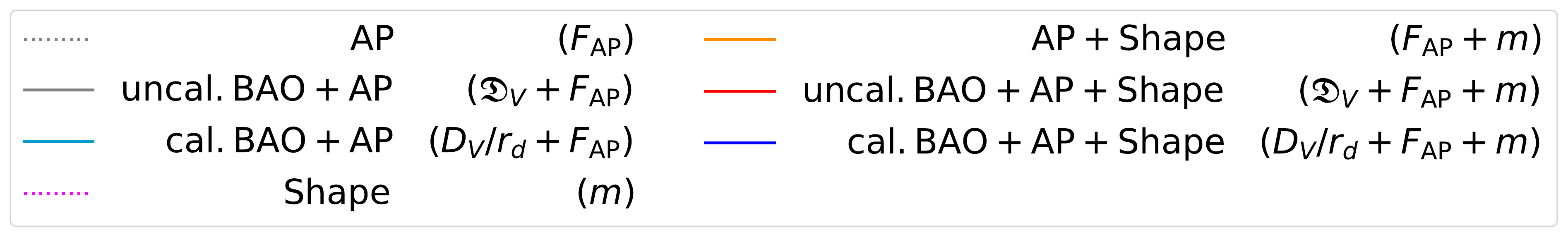} \\
    \includegraphics[width=0.49\textwidth]{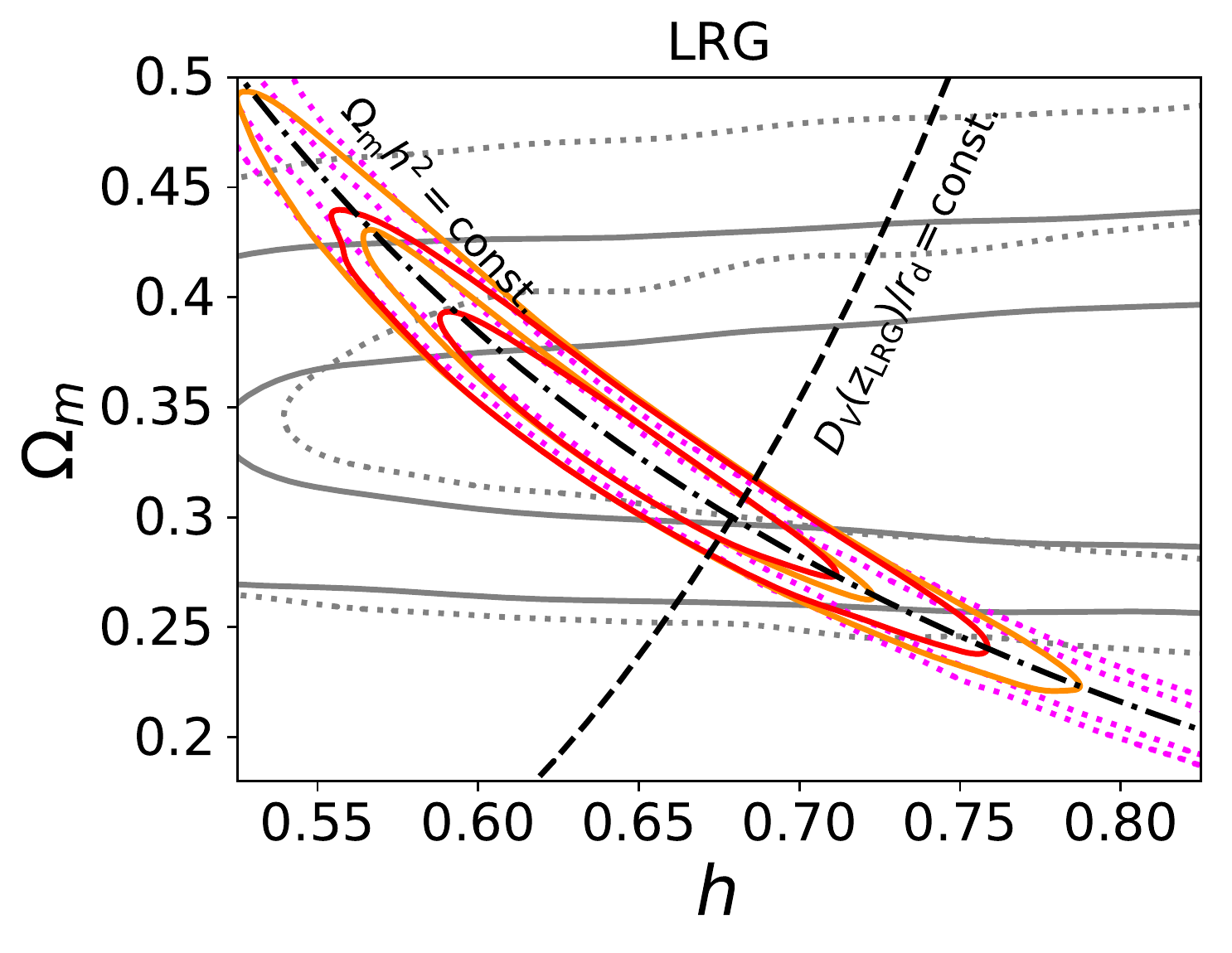}
    \includegraphics[width=0.49\textwidth]{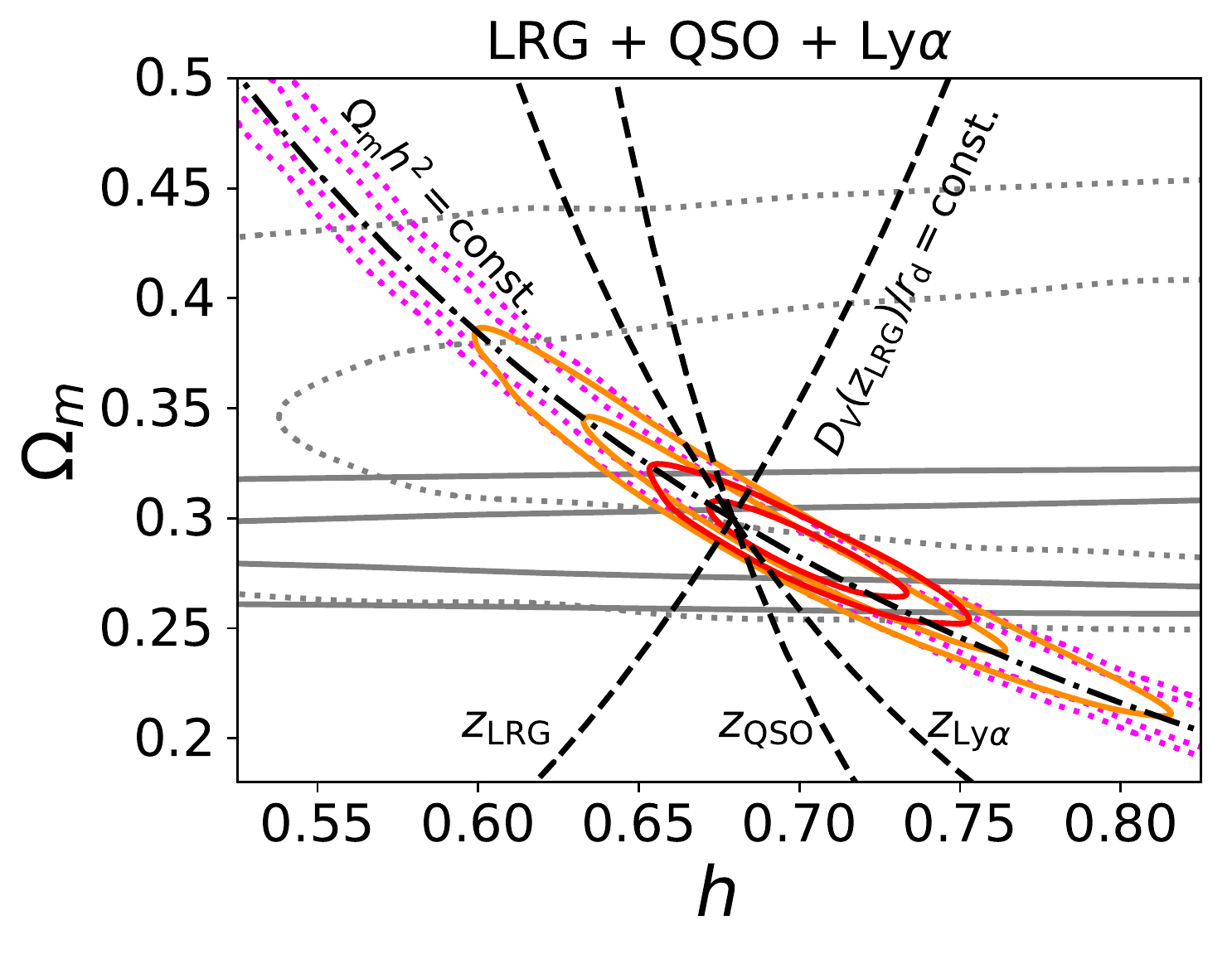} \\
    \includegraphics[width=0.49\textwidth]{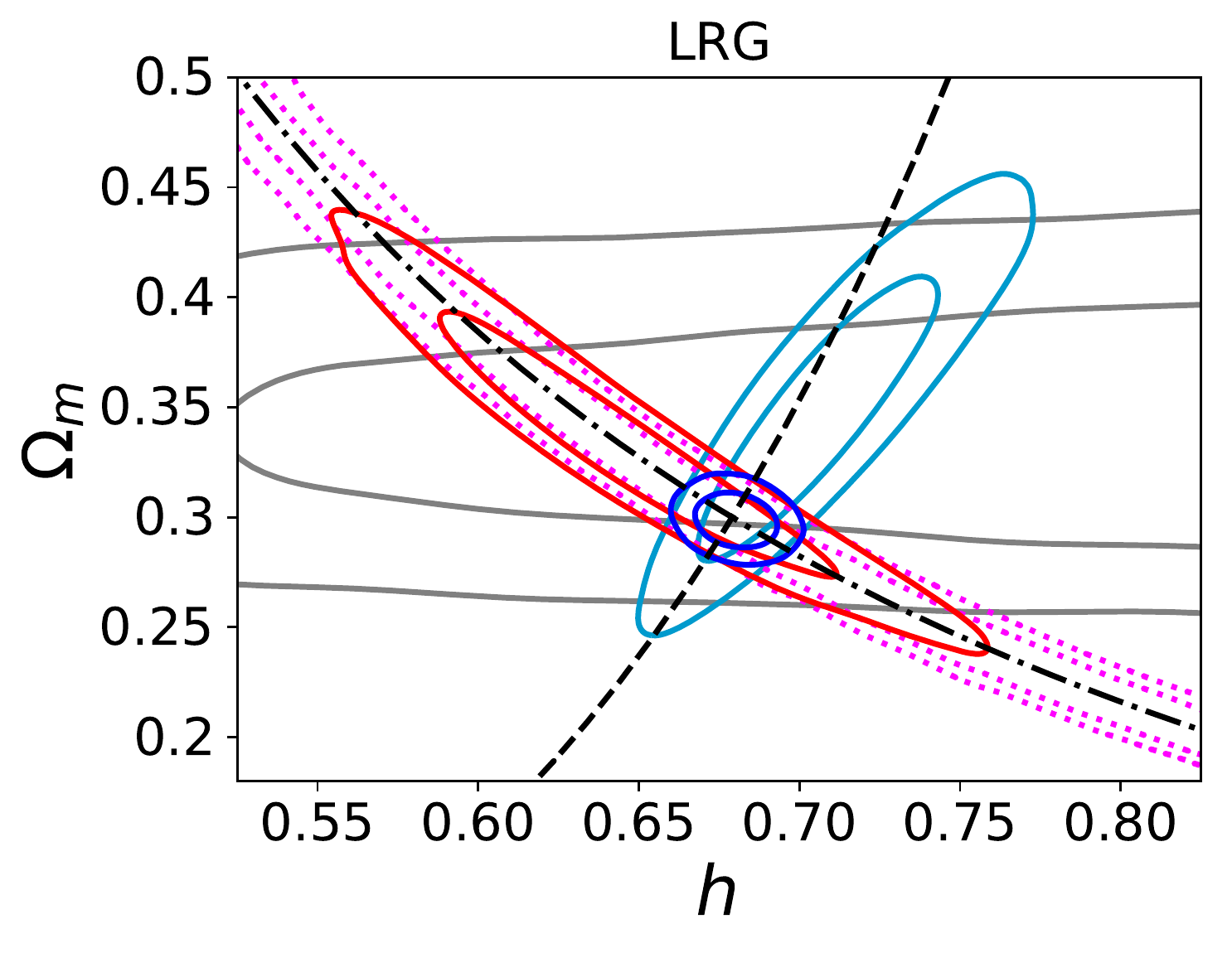}
    \includegraphics[width=0.49\textwidth]{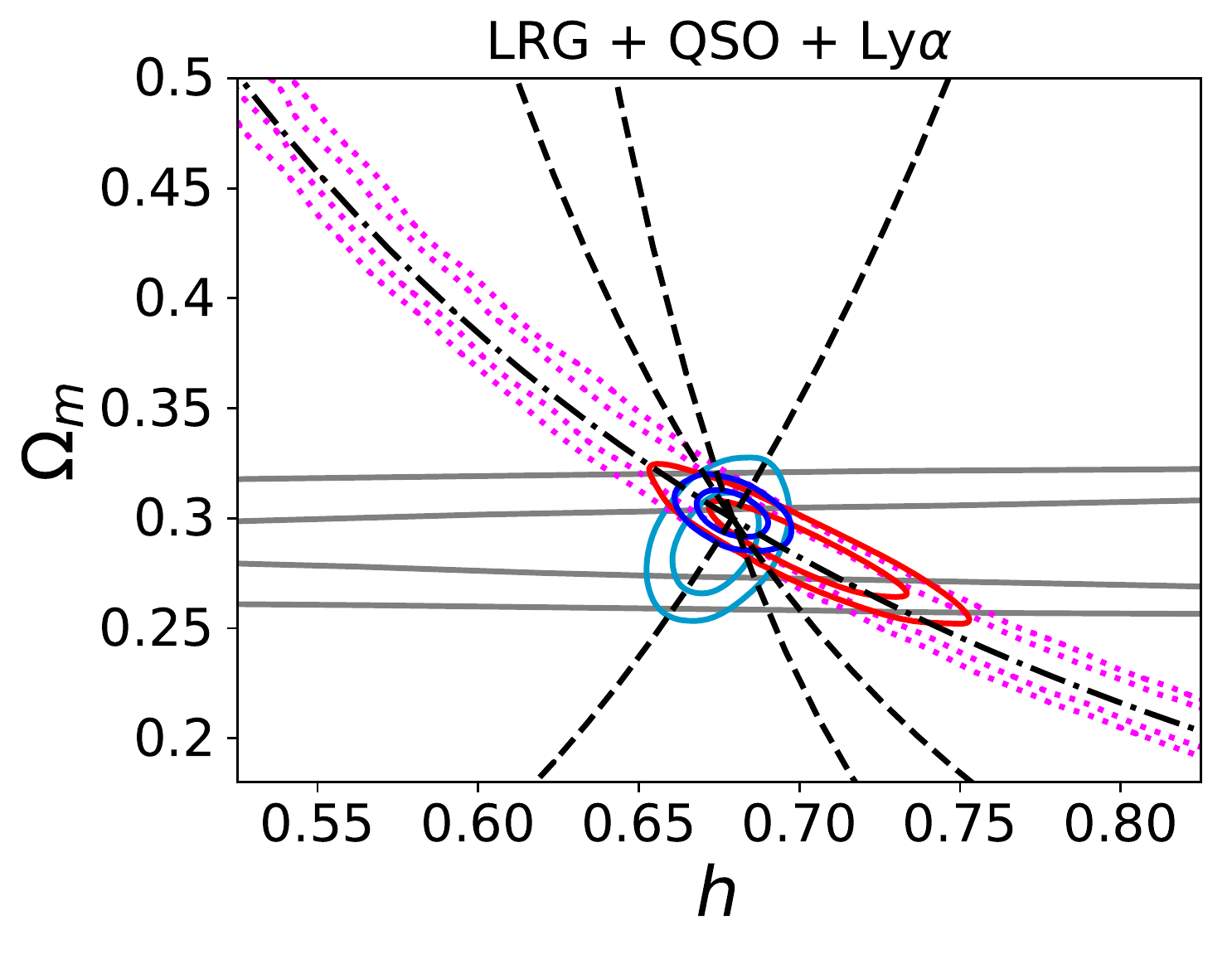} \\
    \caption{$\Om-h$ plane derived from piecewise $\Lambda$CDM fits to the BOSS+eBOSS LRGs (left panels) and all samples (right panels) considering different combinations of compressed variables. The upper panels show all constraints except the ones relying on BAO calibration, which are shown in the lower panels only. Grey contours refer to purely late-time constraints, either only using the AP parameter, $F_\mathrm{AP}$ (grey dashed contours, upper panels), that does not assume any information on the sound horizon scale physics (only its existence as an isotropic feature), or combining $F_\mathrm{AP}$ with the uncalibrated BAO information, $\mathfrak{D}_V$, standing for $D_V/\rd$ with $\rd$ marginalised out (grey solid contours), which additionally assumes that $\rd$ does not change with redshift. These results (grey curves), by construction, are independent of the employment of any BBN prior. Note, however, that all colored lines assume BBN (not included in the legend for conciseness). By adding information on the absolute size of $\rd$ we obtain our first measurement of $h$ (light-blue contours), commonly known as BAO+BBN. On the other hand, the shape $m$ alone (magenta dash-dotted line) constrains the combination of $\Omega_\mathrm{m}h^2$ (upon priors on $n_s$ and $\Omega_{\rm b}h^2$ are set). By adding either the $F_\mathrm{AP}$ (orange contours) or $F_\mathrm{AP} + \mathfrak{D}_V$ (red contours) information on $\Om$, we obtain sound-horizon-independent constraints on $h$.  All combined (dark-blue contours) give the full BAO+BBN+shape result for $h$. All these $h$-measurements are provided in table \ref{tab:hmeasurements}.}
    \label{fig:contibutions}
\end{figure}

As first diagnostic test we present constraints on the matter density $\Omega_\mathrm{m}$ and the Hubble parameter $h$  under a piecewise-flat $\Lambda$CDM  model from the different pieces of information introduced in section \ref{sec:SF}
in figure~\ref{fig:contibutions}. The left panels show results from the three LRG samples alone ($0.2<z<1.0)$, while the right panels contain our full LRG+QSO datasets and Lyman-$\alpha$ BAO ($0.2<z<3.5$).
In this section we explore, how adding different ingredients of the datavector $\Theta$ (representing different physical processes) results in different constraints of the model's parameters.  
We start with the purely geometrical information without BAO calibration (achieved via the sound horizon marginalisation)\footnote{This can be achieved by introducing the sound horizon as a nuisance parameter within the cosmological likelihood.} represented by `$\mathfrak{D}_V$' and `$F_\mathrm{AP}$'. The latter alone (grey dotted contours) and its combination with `$\mathfrak{D}_V$' (grey solid contours) constrain the matter density $\Omega_\mathrm{m}$. Considering only the limited redshift range of LRG only (left panels), the constraints are rather weak ($\sigma_{ \Omega_\mathrm{m}} \approx 0.05$). Conversely, when considering the full redshift range from all samples (right panels, grey solid contours), the geometrical constraint tightens to $\Omega_\mathrm{m} = 0.289\pm0.016$.
Not unexpectedly, the `$\mathfrak{D}_V$'-derived constraints benefit significantly from the extended redshift range 
compared to `$F_\mathrm{AP}$'.
This is directly related to the fact that `$\mathfrak{D}_V$' builds on property \textit{ii)} of the standard ruler definition (see section \ref{sec:SF-geom}), while `$F_\mathrm{AP}$' does not.
These $\Lambda$CDM constraints on $\Omega_\mathrm{m}$ are independent of any early universe assumptions, as the sound horizon information is either cancelled out (grey dotted contours) or marginalized out (grey solid contours). The only assumption entering here is that the late-time background expansion is dictated by standard flat-$\Lambda$CDM, hence by $\Om$. Therefore, these represent geometric constraints on the late-time $\Om^{\rm late}$ quantity.

Next, we add the
sound horizon scale to make use of the full 
late+early-time information `$D_V/\rd + F_\mathrm{AP}$' (light-blue solid contours, lower panels). If the physical baryon density $\Ob h^2$ is varied freely, the system would be under-constrained and the light-blue solid contour would become indistinguishable from the grey solid contour.
But, by fixing the baryon density via
the BBN prior, the system becomes constrained and $\rd$ is `calibrated'. Now, $D_V(z)/\rd$ constrains the parameter combinations of eqs. \eqref{dvrs_first}-\eqref{dvrs_last} displayed via black dashed lines. Together with the uncalibrated BAO constraint on $\Omega_\mathrm{m}$, this delivers our first measurement of the Hubble parameter, $h_{\rm r_d}=0.6742_{-0.0094}^{+0.0088}$, known as the traditional BAO+BBN technique, equivalent to the $h$-measurement delivered by the eBOSS collaboration team \cite{eboss_collaboration_dr16}, $h=0.6735\pm0.0097$. This  measurement relies on the assumptions of standard late-time background expansion, standard pre-recombination physics and that the $\Om$ value at pre-recombination times is effectively the same as the late-time one, $\Om^{\rm late}=\Om^{\rm {early},r}$. This constraint can also be compared with \cite{NilsBBN2} $h=0.676^{+0.0094}_{-0.0103}$ where the central values are in excellent agreement and the small difference in error-bars arises from a slightly different treatment of the BBN prior and with \cite{Brieden:2022lsd} $h=0.6742 ^{+0.0084}_{-0.0091}$. 

Finally, we carry out the same exercise (still using the BBN prior), but including the shape measurement $m(z)$ in our cosmological analysis. The shape alone (magenta bands) constrains the parameter combination $\Omega_\mathrm{m}h^2$ indicated by the black dash-dotted lines. Combined with the uncalibrated BAO (purely late-time) measurements of $\Omega_\mathrm{m}$ and using the full LRG+QSO+Lyman-$\alpha$ samples, we find $h_m = 0.695_{-0.051}^{+0.042}$ in the  `$F_\mathrm{AP}+m$' case (orange contours, upper panels) and $h_m = 0.702_{-0.021}^{+0.019}$ in the `$\mathfrak{D}_V+F_\mathrm{AP}+m$' case (red contours).\footnote{Note that the sound horizon marginalisation applied to obtain the uncalibrated BAO also changes the pivot scale $k_{sp}$ at which the shape $m$ is defined. We find that this does not impact our $h_m$ measurement, since $m$ represents an almost scale-independent slope (see also \cite{Farren:2021grl} for a more detailed discussion).} 
The latter represents our second $h$-measurement, which is comparable in information content to the sound-horizon-independent constraint of \cite{Philcox_Sherwin_Farren_Baxter_21,Farren:2021grl,Philcox:2022sgj}, although using a larger redshift range and being independent of the Pantheon+ dataset; this can also be compared with the complementary approach and results of \cite{Smith:2022iax}. These measurements rely on the assumptions of a standard late-time background expansion, standard matter-radiation equality physics, the fact that $r_{\rm d}$ is a standard ruler and that the values of $\Om$ at equality and late-time are the same, $\Om^{\rm early,eq}=\Om^{\rm late}$.

Consistently combining these two measurements we obtain the full BAO+BBN+shape constraint $h_{(m,r_{\rm d})} = 0.6790_{-0.0075}^{+0.0076}$ (`$D_V/\rd + F_\mathrm{AP}+m$' case), which relies on standard late-time background expansion, standard pre-recombination and equality epoch physics, and that $\Om$ is the same in all those epochs, $\Om^{\rm late}=\Om^{\rm early,r}=\Om^{\rm early,eq}$. 
As appendix~\ref{app:ob-ns} shows,  while the  adopted baryon abundance prior is important for $h$ (but not for $\Omega_m$), a reasonably wide prior on $n_s$ has a small effect on both parameters.

The individual $h$-measurements are summarized in table~\ref{tab:hmeasurements} (the full results are reported in appendix~\ref{app:s8_values}, table~\ref{tab:full}). We see that under the assumption of the flat $\Lambda$CDM model and the employed BBN prior, all these measurements are consistent with each other. In fact, the two degeneracy directions (that enable our two $h$-measurements) displayed by the black dashed and dash-dotted lines in figure~\ref{fig:contibutions}, exactly meet where they are in agreement with the $\Omega_\mathrm{m}$ band obtained from the expansion history. On one hand this indicates the strong self-consistency of the $\Lambda$CDM model. On the other hand this exemplifies how difficult it is to reconcile the high-$h$ value obtained by SH0ES, $h=0.7304\pm0.0104$ \cite{Riess:2021jrx}, with indirect determinations of $h$ by postulating new physics in the early Universe. Any extension to the standard model must consistently change the sound horizon and the matter-radiation equality physics, in such a say that both `anchors' deliver a SH0ES-consistent value for $h$, as well as a consistent $\Omega_m$ measurement with the uncalibrated late-time universe measurements. Indeed this is  what \cite{Smith:2022iax} finds for a range of popular beyond-LCDM models.
Looking forward, as the error-bars continue to  shrink, this implies that a new value of $h$ cannot be reached by moving along one of the two degeneracy directions displayed in figure~\ref{fig:contibutions}; nor it can be achieved by recalibrating or changing  either the length of the sound horizon scale, or the interpretation of the shape. Instead, to accommodate a significantly higher value of the Hubble parameter, both degeneracy directions must be coherently shifted towards higher values of $h$. This implies that the physics governing the sound horizon at radiation drag (as seen in the BAO) and the physics governing the shape, hence to first approximation physics around  matter-radiation equality, should be changed in a coherent manner as not to disturb the (currently evident) agreement. Alternatively, two different mechanisms would need to be invoked, one (early-time)  that shifts $r_{\rm d}$ and another one completely unrelated that shifts $m$ `just so' to preserve the agreement.
This agreement also includes obtaining  a consistent $\Om$ value with respect to that delivered by the uncalibrated BAO, based entirely on geometric late-time physics and the early-time one.

These considerations can be seen as a straitjacket for extended models that could potentially resolve the Hubble tension.   

\begin{table}[t]
    \centering
    \large 
    \onehalfspacing    
    \begin{tabular}{|L|C|C|C|}
 \multicolumn{4}{c}{$\boldsymbol{H_0}$  {\bf[km s$^{-1}$} $\boldsymbol{{\rm Mpc}^{-1}]}$ $\boldsymbol{\pm}$ \bf{68\%~CL}    }\\

 \hline
\mathrm{Probe/feature}& \mathrm{LRG} & 
\mathrm{ALL} & \mathrm{Assumptions} \\ 
        \hline \hline
        & & &  \Omega_{\rm m}^{\rm eq}=\Omega_{\rm m}^{\rm late} \,\, \& \\
        F_\mathrm{AP}+m \,\,\,\, (h_m)& 63.9_{-6.4}^{+4.7} & 69.5_{-5.1}^{+4.2} & {\it i)};\, \mathrm{priors ~ on ~} \omega_\mathrm{b}, n_s\\
        \mathfrak{D}_V + F_\mathrm{AP}+m \,\,\,\, (h_m) & 64.5_{-4.5}^{+3.5} & 70.1_{-2.1}^{+2.1} & {\it i)},\, {\it ii)};\, \mathrm{priors ~ on ~} \omega_\mathrm{b}, n_s \\
        \hline
        &&& \Omega_{\rm m}^{\rm r}=\Omega_{\rm m}^{\rm late}\, \& \\
        D_V/\rd + F_\mathrm{AP} \,\,\,\,\,\, (h_{r_{\rm d}})  & 70.8_{-2.8}^{+2.2} & 67.42_{-0.94}^{+0.88} & {\it i)},\, {\it ii)},\, {\it iii)};\, \mathrm{prior ~ on ~} \omega_\mathrm{b} \\
        D_V/\rd + F_\mathrm{AP} + m \,\,(h_{m,r_{\rm d}}) & 67.99_{-0.85}^{+0.83} & 67.90_{-0.75}^{+0.76} & {\it i)},\, {\it ii)},\, {\it iii)};\, \mathrm{priors ~ on ~} \omega_\mathrm{b}, n_s \\
        \hline
        \hline
        &&&  \Omega_{\rm m}^{\rm eq}=\Omega_{\rm m}^{\rm late}\,\,\&\\
        F_{\rm AP}+m+f\sigma_{\rm s8} & 63.8_{-5.8}^{+4.5} & 72.8_{-4.8}^{+4.0} & {\it i)},\,{\rm GR};\, \mathrm{priors ~ on ~} \omega_\mathrm{b}, n_s\\
        \mathfrak{D}_V+F_{\rm AP}+m+f\sigma_{\rm s8} & 64.5_{-4.5}^{+3.5} & 70.2_{-2.1}^{+1.9} & {\it i)},\, {\it ii)},{\rm GR};\, \mathrm{priors ~ on ~} \omega_\mathrm{b}, n_s \\
        \hline
        &&& \Omega_{\rm m}^{\rm r}=\Omega_{\rm m}^{\rm late} \,\,\& \\
        D_V/r_{\rm d}+F_{\rm AP}+f\sigma_{\rm s8} & 70.9_{-2.8}^{+2.2} & 67.37^{+0.86}_{-0.95} &  {\it i)},\, {\it ii)},\, {\it iii)},\mathrm{GR};\, \mathrm{prior ~ on ~} \omega_\mathrm{b} \\
        D_V/r_{\rm d}+F_{\rm AP}+m+f\sigma_{\rm s8} & 68.00^{+0.73}_{-0.73} & 67.9^{+0.76}_{-0.75} & {\it i)},\, {\it ii)},\, {\it iii),{\rm GR}};\, \mathrm{priors ~ on ~} \omega_\mathrm{b}, n_s \\
        \hline
    \end{tabular}
    \caption{Constraints for the Hubble expansion rate, $H_0=100h\,[{\rm km}\, {\rm s}^{-1}\,{\rm Mpc}^{-1}]$, considering different combinations of compressed variables from either LRGs only ($0.2<z<1.0$) or all samples combined ($0.2<z<3.5$) and different anchors. The top two rows show constraints on $H_0$ independent of the sound horizon, whereas the two rows below include the calibrated BAO. The four bottom rows report those constrains including the $f\sigma_{s8}$ variable (first two independent of sound horizon, second two with the calibrated sound horizon scale), which additionally assume GR at the perturbation level. Adding the growth  information does not chance the anchor, of course.  All results are based on the flat $\Lambda$CDM model assumption (at the background level) and include the BBN prior from eq.~\eqref{eq:bbn-prior}. Those results including the shape parameter $m$ also include the $n_s$ prior from eq.~\eqref{eq:ns-prior}. The assumptions {\it i}), {\it ii}), {\it iii}) have been introduced in section~\ref{sec:standard_ruler}. All results are graphically displayed in figure~\ref{fig:contibutions} (four top rows) and figure~\ref{fig:growth} (three bottom rows, for `ALL' only).}
    \label{tab:hmeasurements}
\end{table}

\subsection{Piecewise \texorpdfstring{$\Lambda$}{}CDM constraints from geometry, shape and growth} \label{sec:results-geoshapegrowth}

We investigate how adding the growth signal affects measurements of $h$. Adding the (relative) growth signal does not change the $h$-anchor, so this route provides effectively another determination of  the same quantities as above, $h_m$, $h_{\rm r_d}$ or $h_{(m,r_d)}$.
As explained in section \ref{sec:SF-growth}, the \SF $f\sigma_{s8}(z)$ measurements carry two types of information: 1) the global amplitude of perturbations, which is given by the scalar amplitude $A_s$ of fluctuations generated by inflation, and by $h$ through the units of $\sigma_{\rm s8}$ in ${\rm Mpc}h^{-1}$; and 2) the {\it relative} redshift evolution of $f\sigma_{\rm s8}(z)$, which only depends on the matter density $\Om$ within a flat-$\Lambda$CDM model (see equations \ref{eq:fss8_scalings}). Note that since in our baseline setup we vary $A_s$ freely, we can only constrain $\Om$ through  the {\it relative} redshift dependence of $f \sigma_{\rm s8}$ across redshift bins, which is, by construction, independent of the overall scalar amplitude $A_s$. On the other hand, the variation of $A_s$ does not allow us to obtain a third independent $h$ measurement, as the system still needs to be calibrated, either through $m$ or the sound horizon scale, as we have described in section~\ref{sec:results-geoshape}. Note that in this case, the intrinsic $\Lambda$CDM assumption of $\Omega_m^{\rm growth}=\Omega_m^{\rm  late}$ or $\Omega_m^{\rm growth}=\Omega_m^{\rm  early}$ will be exploited. 

\begin{figure}[t]
    \centering
    \includegraphics[width=0.49\textwidth]{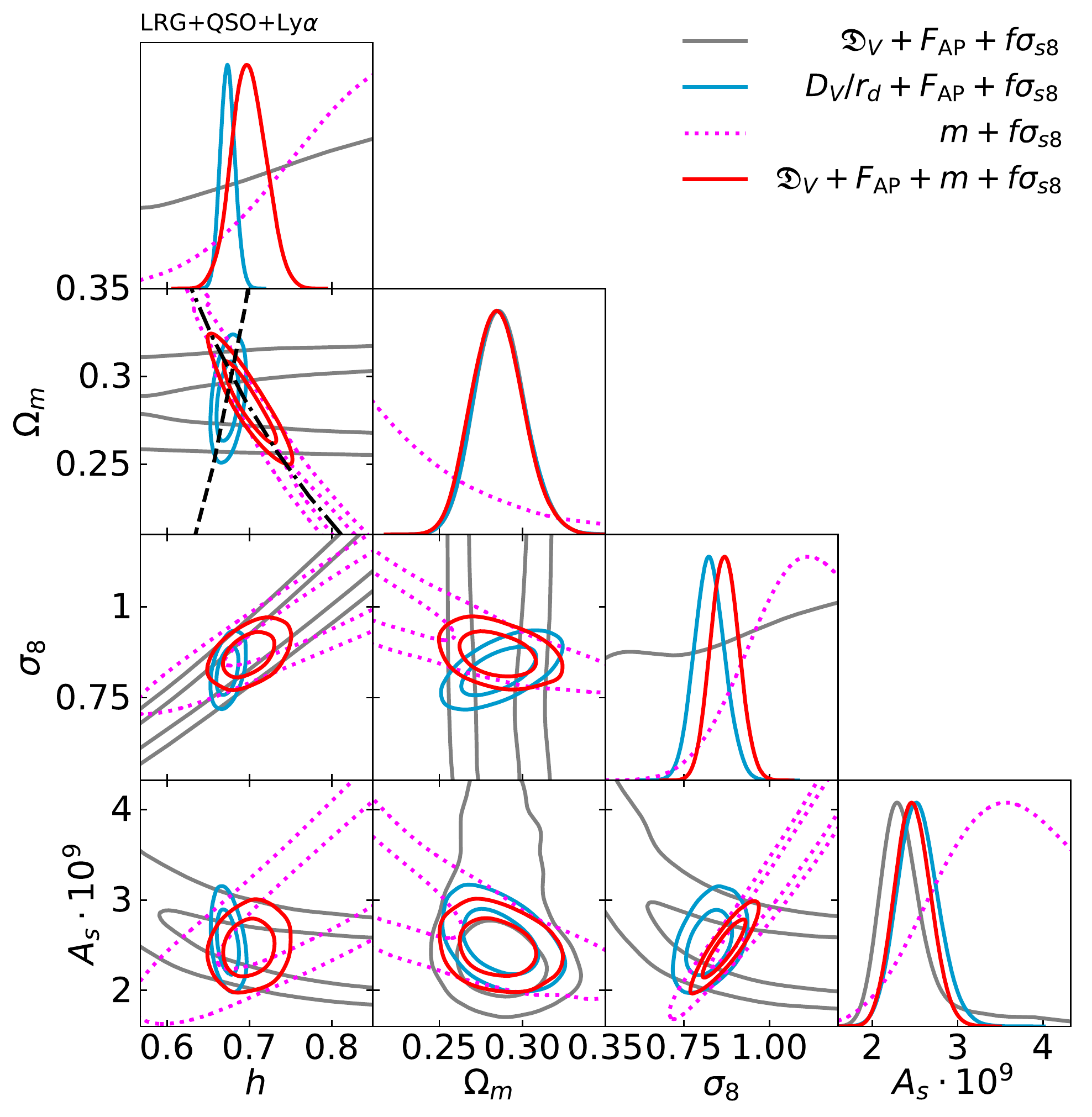}
    \includegraphics[width=0.49\textwidth]{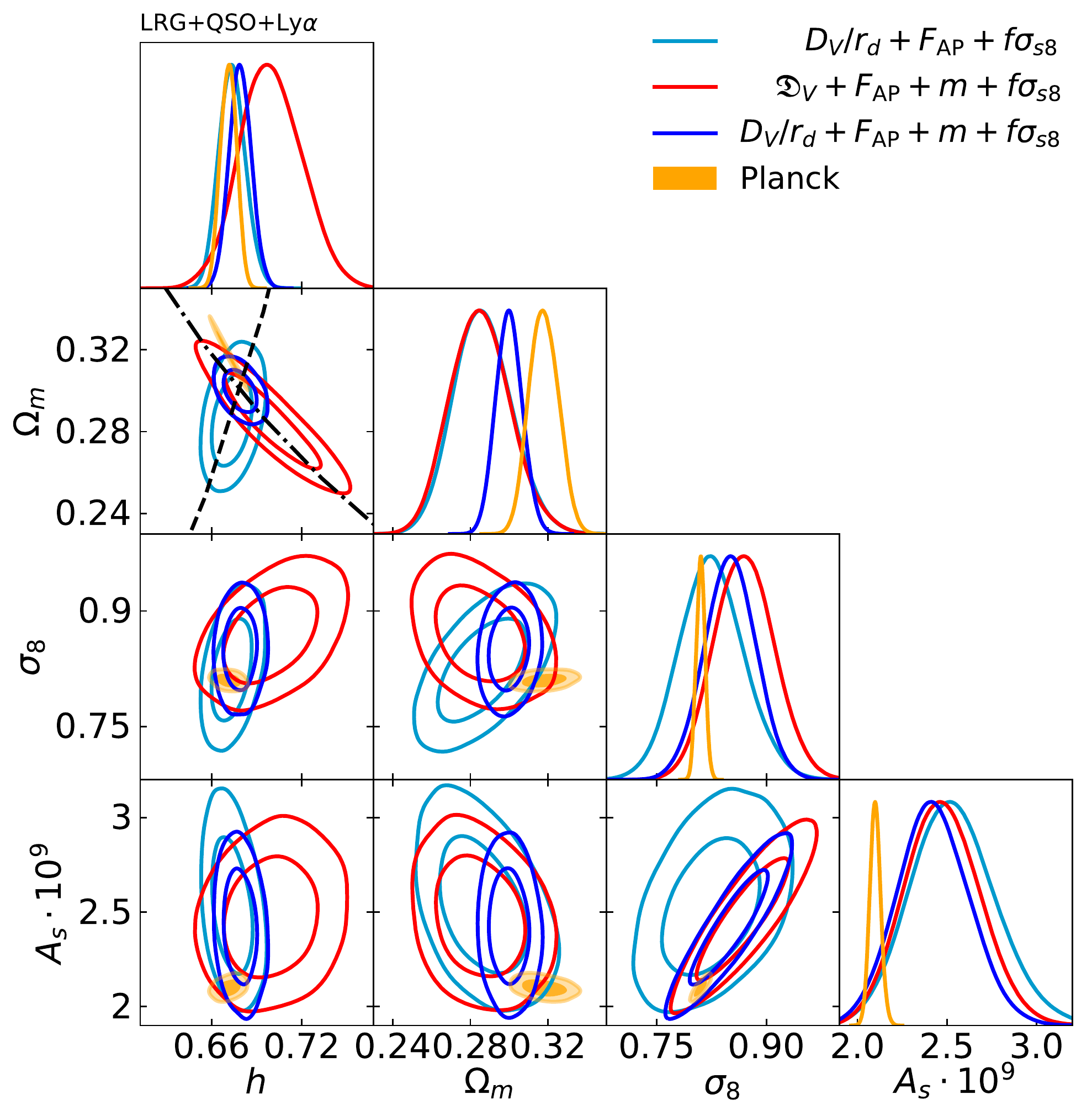}
 \caption{Derived constraints from piecewise $\Lambda$CDM fits to the BOSS+eBOSS samples (LRG+QSO+Lyman-$\alpha$), analogous to the right plots} of figure~\ref{fig:contibutions}, using the same colour 
 scheme. The right panel is a zoom into the left panel, where we also show the full ShapeFit (empty dark-blue contours) and Planck \cite{Aghanim:2018eyx} $\Lambda$CDM (filled orange contours) constraints. All cases include the perturbation-like variable $f\sigma_{\rm s8}$, which allow us to set constraints on the amplitude of perturbations ($\sigma_8$ or $A_s$) when the GR assumption is being made to connect the matter at background and perturbation levels: $\Omega_m(z)=f(z)^{0.56}$. The black dashed lines in the $\Om-h$ panels are the same as those in figure~\ref{fig:contibutions}. Priors on $\Omega_{\rm b}h^2$ and $n_s$ are set when they are relevant, according to eqs.~\eqref{eq:bbn-prior} and \eqref{eq:ns-prior}, respectively.
 The Hubble parameter can be constrained efficiently only if either the system is calibrated using the sound horizon scale as a ruler, or the shape parameter, $m$, is employed to determine how fast the horizon grows at matter-radiation equality epoch. Thus, adding the perturbation information through $f\sigma_8$ only helps to improve the constraints on $h$ via an improvement on the $\Om$ determination. In principle a determination of $h$ based on $m+f\sigma_{\rm s8}$ would also be possible, but in practice $f\sigma_8$ alone constraints $\Om$ poorly, in part due to the lack of precise measurements at very distinct redshifts. As in figure~\ref{fig:contibutions} all  $h$ constraints are fully consistent. These constraints are reported in table \ref{tab:hmeasurements} and table \ref{tab:full}.
    \label{fig:growth}
\end{figure}

The resulting constraints on the flat-$\Lambda$CDM parameters are displayed in figure~\ref{fig:growth} for the whole set of BOSS+eBOSS tracers (LRG+QSO+Lyman-$\alpha$; $0.2<z<3.5$), for different combinations of parameters 
(analogously to the right panels of figure~\ref{fig:contibutions}, but all including $f\sigma_{\rm s8}$)
. As before, some of the parameter combinations require either a prior on the baryon density, or the spectral index, exactly in the same way as previously described in section~\ref{sec:results-geoshape}, and as explicitly displayed in the right column of Table~\ref{tab:hmeasurements}. The information is split in two panels, where the right panel is a zoom into the left panel, which allows for a comparison of our ShapeFit constrains with Planck.

The addition of the  $f\sigma_{\rm s8}$ constraint does not change significantly the constraints on $\Omega_{\rm m}$ or $h$. Nevertheless it is interesting to note that, in principle, if $f\sigma_{\rm s8}$ could be measured over a wide redshift range and with enough accuracy,  the combination $m+f\sigma_{\rm s8}$ could yield a qualitatively different $h$ determination  whereby the $\Omega_{\rm m}h^2=$constant degeneracy given by $m$ is broken by a determination of $\Omega_{\rm m}^{\rm growth}$ that only relies on the growth of  perturbations. For the redshift range available, with baseline from $z=0.38$  to  $z=1.48$, and at the current precision level, this is not yet possible. However, forthcoming data and the addition of higher-order statistics may be able to break the degeneracy between $f$ and $\sigma_{s8}$ \cite{gil-marin_bispectrum_dr12}  making this new $h$ determination possible. 

To summarize, when the $f\sigma_{\rm s8}$ information is combined with the calibrated BAO, or the uncalibrated BAO+shape, $h$ is fully determined, and $f\sigma_{\rm s8}$ moderately helps to increase the precision of the $h$ determination. These measurements allow for an extra determination  of $h$ which relies on the $\Lambda$CDM assumptions: 1) the growth of structure is connected to the matter density according to GR predictions; 2) this perturbation-derived matter density is the same as the late-time-background density, or also additionally to the early-time matter density. However, note that these `new' values of $h$  so obtained are fully correlated with those described in section~\ref{sec:results-geoshape}. 

The bottom two rows of both panels of figure ~\ref{fig:growth} are useful to shed light on another mild tension involving $\sigma_8$, the late-time amplitude of perturbations (see \cite{Amon_Efstathiou22} and references therein).
In particular, weak-lensing observations (for e.g., \cite{des_collaboration_dark_2017}) tend to report a lower value of $\sigma_8$ compared to that estimated from CMB data \cite{Aghanim:2018eyx}.  The constraints reported in figure~\ref{fig:growth}, which include the redshift space distortion information through $f\sigma_{\rm s8}$, enable another diagnostic test or internal consistency check. Moreover, any proposed solution to fix the Hubble tension should also help to improve, or at least not exacerbate, the $\sigma_8$ tension. 

For example, the $m+f\sigma_8$ constraint (combining equality physics with late-time growth rate) would need a higher $A_s$ for higher $h$ values, but also  higher $\sigma_8$ values. On the other hand, the purely late-time, uncalibrated BAO signal, combined with the growth rate constraints, would need a lower $A_s$ for higher values of $h$, but also a lower $\sigma_8$. This type of `scissor' behaviour of the constraints is particularly useful for diagnostic tests of the model.

By combining the un-calibrated BAO information with the shape and relative growth, we obtain a $\sigma_8$ value of $\sigma_8=0.872_{-0.042}^{+0.042}$ (red lines). This result relies upon the velocity at which the modes re-enter the horizon at matter-radiation scales, and other late-time quantities (growth and geometry), but is  independent of sound-horizon scale physics.
On the other hand, the sound horizon scale calibrated results (cyan lines) in combination with $f\sigma_{\rm s8}$ yield a highly consistent value  $\sigma_8=0.825_{-0.047}^{+0.043}$. This determination is also highly consistent with the official eBOSS reported value, $\sigma_8=0.850_{-0.033}^{+0.033}$ \cite{eboss_collaboration_dr16}. The full combination of sound-horizon calibrated BAOofLSS and the shape $m$, relying on processes at equality, and the late-time growth rate, provides $\sigma_8=0.850_{-0.038}^{+0.035}$, again very similar to the eBOSS official results. The values of $\sigma_8$, as well as other parameters of interest are reported in Appendix~\ref{app:s8_values}, table~\ref{tab:full}.

\subsection{Beyond \texorpdfstring{$\Lambda$}{}CDM: massive neutrinos and dark-radiation.}
\label{sec:results-nu}

Two popular $\Lambda$CDM extensions involve considering a free sum of neutrino masses parameter, $\Sigma m_\nu$, or varying the effective number of neutrino species, $N_\mathrm{eff}$, incorporating dark radiation. We explore the constraints \SF impose on $h$ within these models. 

\begin{figure}[t]
    \centering
    \includegraphics[width=0.49\textwidth]{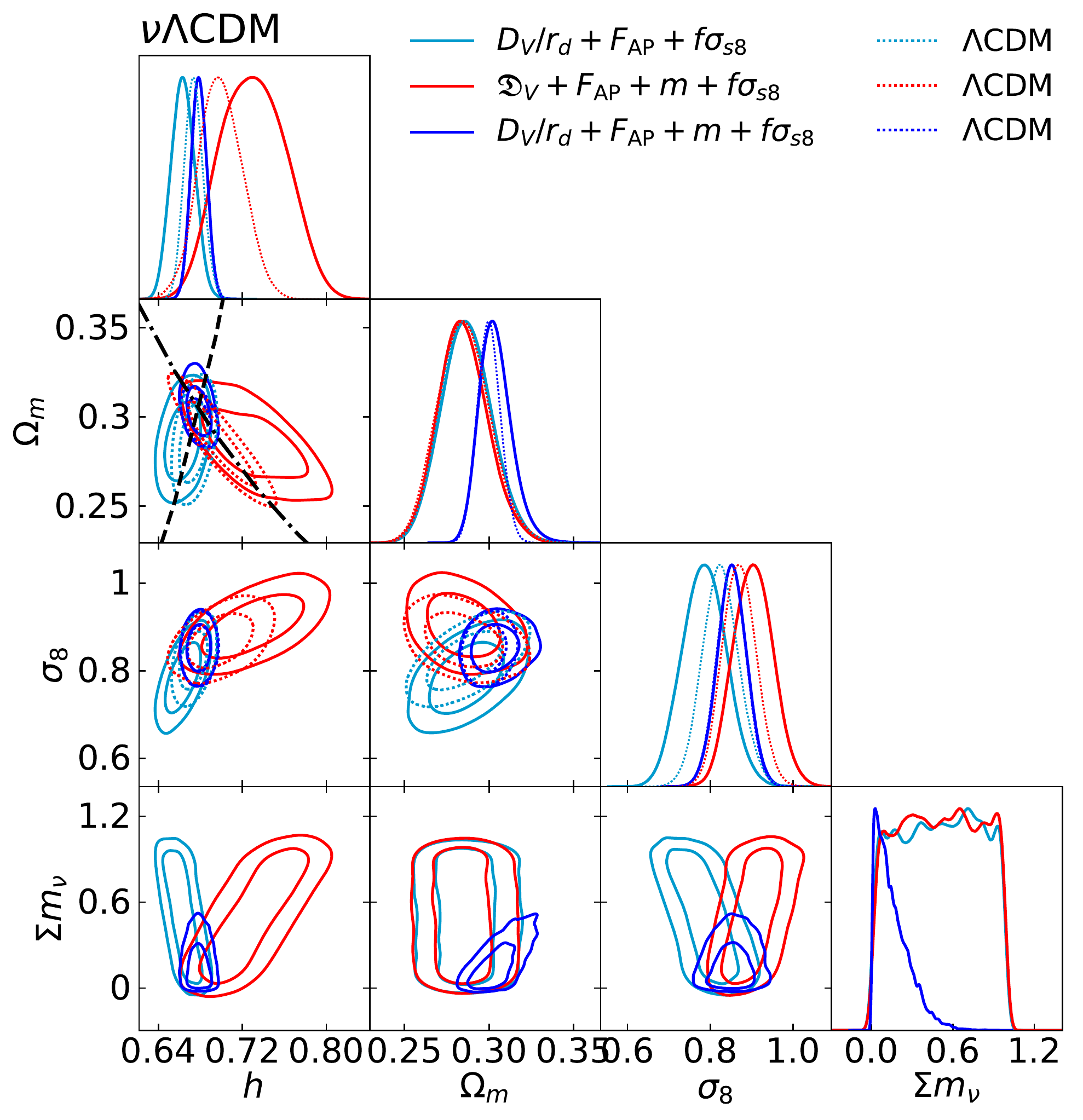}
    \includegraphics[width=0.49\textwidth]{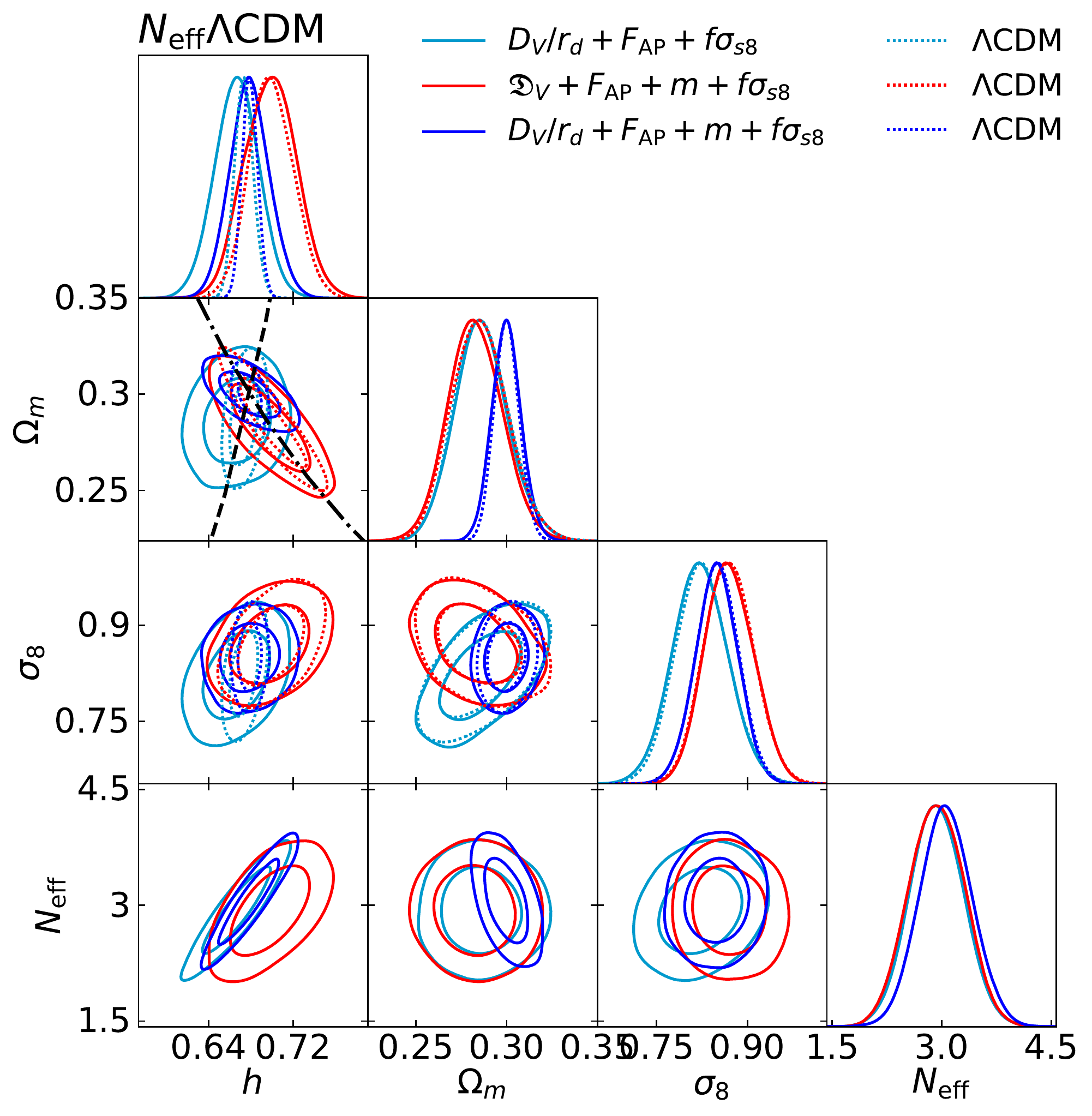}
 \caption{Left panel: parameters of the $\nu\Lambda$CDM model (with a BBN prior) for the most relevant cases when the shape, geometry and growth parameters have been added on the BOSS+eBOSS samples ($0.2<z<3.5$), following the same colour scheme as in the previous figures. When the $\sum m_\nu$ parameter space is varied, a degeneracy with $h$ appears. However this $\sum m_\nu-h$ degeneracy is different when using the  sound horizon scale anchor (cyan contours) and when using the shape parameter $m$: higher neutrino masses prefer a lower(higher) value of $h$ for the sound horizon(shape) informed analysis. This `scissor' behaviour is precisely which allows to break the degeneracy and put moderate constraints on $\sum m_\nu$ without needing any other external dataset (such as CMB data). A uniform prior of $0<\sum m_\nu<1$ has been applied in all cases.
 Right panel: parameters of the $N_\mathrm{eff}\Lambda$CDM model (with a BBN prior) for the same cases and same colour scheme as in the left panel. We adopt priors on $\Omega_\mathrm{b}h^2$, $N_\mathrm{eff}$  provided in eq. \eqref{eq:bbn-neff-prior} motivated by the full BBN treatment of \cite{NilsBBN2}.}
    \label{fig:varying-Mnu}
\end{figure}

The effect of varying the sum of the neutrino masses is displayed in the left panel of figure~\ref{fig:varying-Mnu} for the geometry, shape and growth parameters, as in figure~\ref{fig:growth}, using the same colour scheme. For comparison, the dotted lines are the constraints for a standard $\Lambda$CDM model. Interestingly, varying the sum of neutrino masses has a very distinct effect on the  constraints provided by the different features used by \SF. Features that depend on the sound horizon yield cosmological parameters degeneracy directions (cyan contrours) very different from those obtained considering features that depend on the shape (red). 
A large $\sum m_\nu$ forces $h$ to decrease(increase) for the sound horizon(shape) physical parameters combinations, implying that a large value of $\sum m_\nu$ cannot solve the $h$ tension. Both cases consistently overlap at $h=0.6786^{+0.0078}_{-0.0074}$ (and $\sum m_\nu$ consistent with zero), consistent with Planck low-$h$ values. A similar effect appears for $\sigma_8$, where the constraints based on the sound horizon ruler(shape) tend to have lower(higher) values of $\sigma_8$, although both are consistent with the combined preferred value $\sigma_8=0.853^{+0.035}_{-0.036}$. This `scissor' behaviour offers a  useful  diagnostic test of the model.

\begin{figure}[t]
    \centering
    \includegraphics[width=0.69\textwidth]{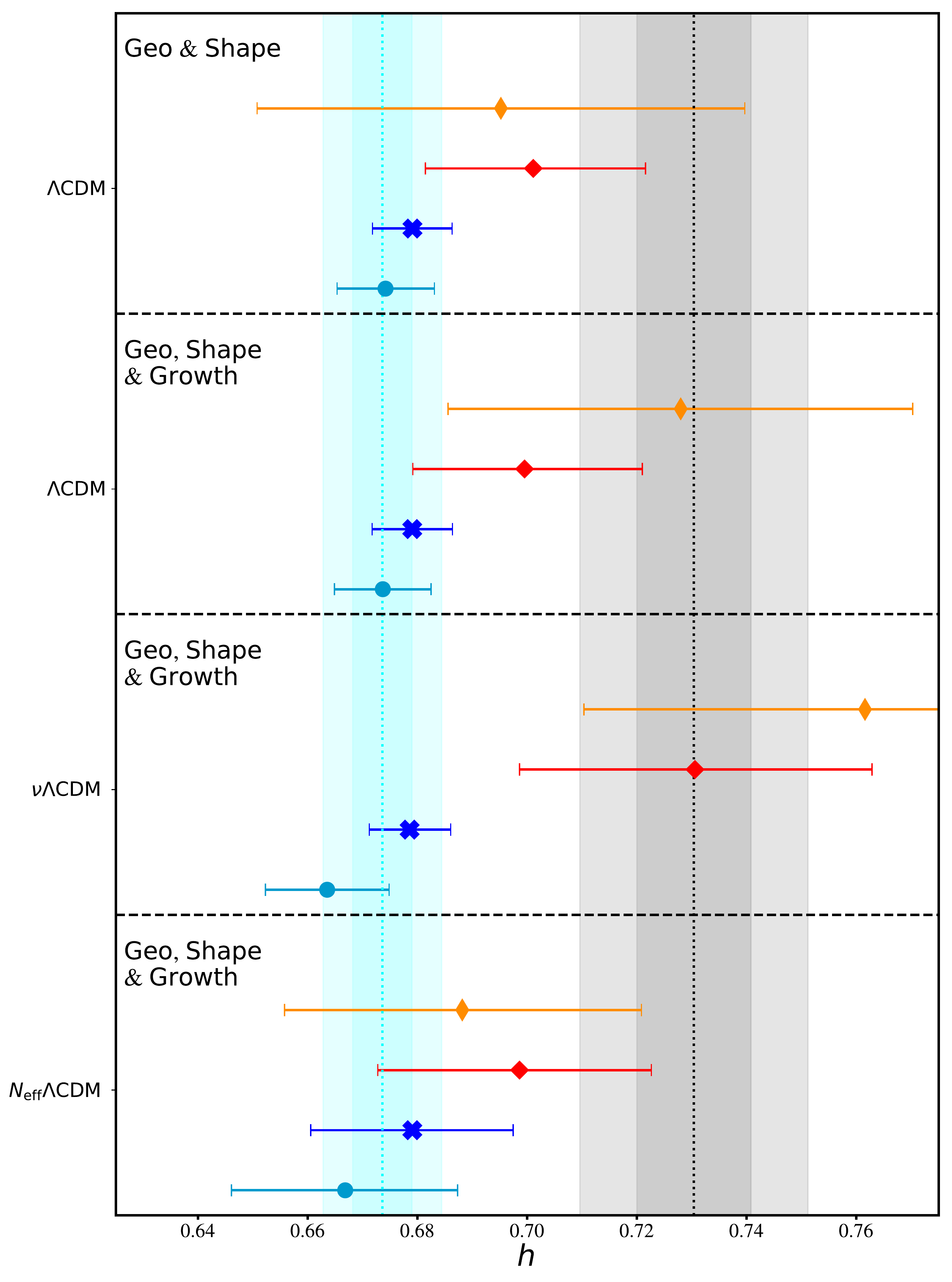}
     \includegraphics[width=0.29\textwidth]{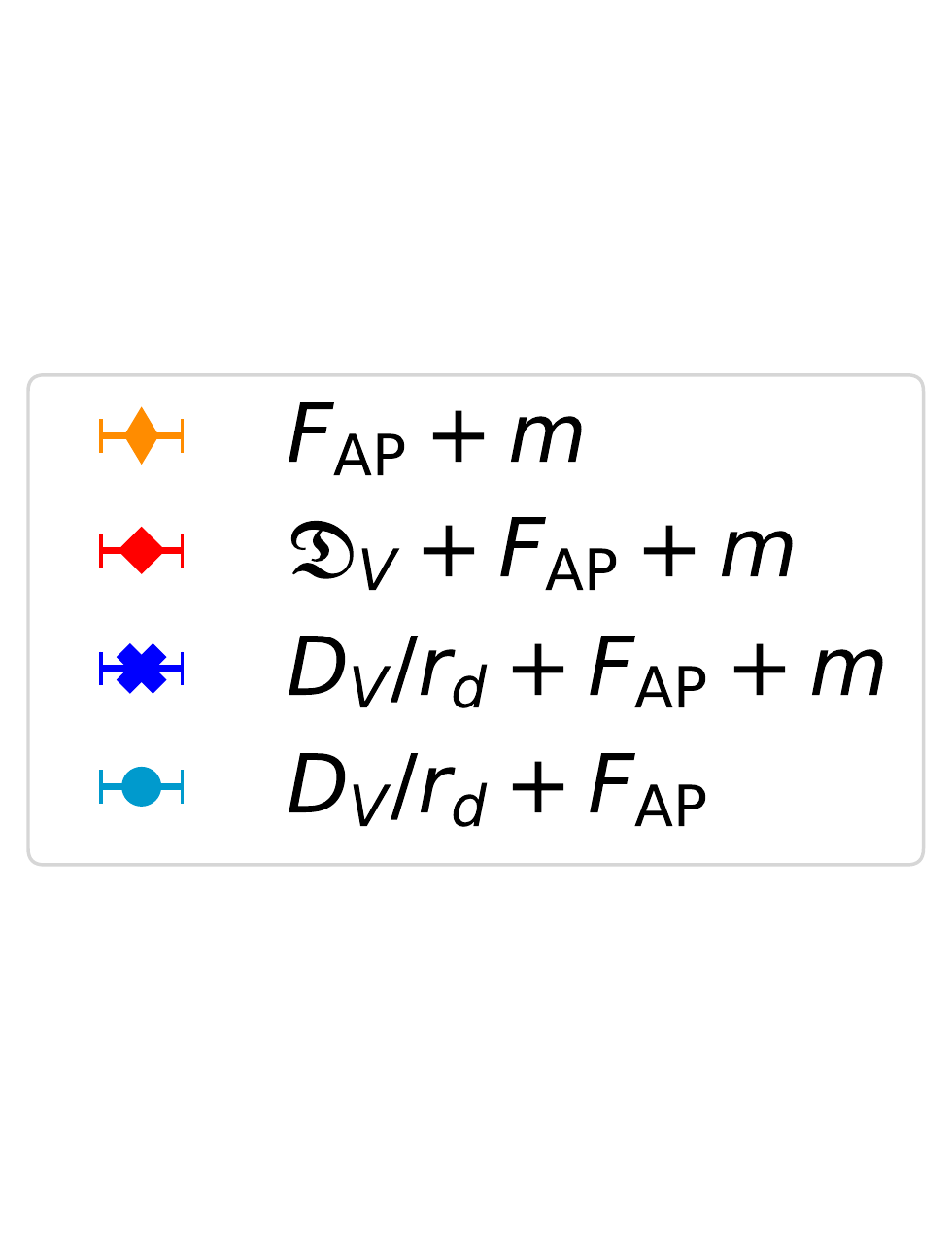}
 \caption{Summary plot of the most relevant results on $h$ of section~\ref{sec:results} obtained from BOSS and eBOSS data in combination with a BBN prior on $\Omega_{\rm b}h^2$ and with $n_s$ set to $0.97$. The top and middle panels display the results for a $\Lambda$CDM model when geometry and shape parameters are used (background variables only $D_V/r_{\rm d},\,F_{\rm AP},\,m)$ and when the growth is also added (perturbation variable, $f\sigma_{\rm s8}$), respectively. The two bottom panels display the same results for two extra cosmologies, a $\nu\Lambda$CDM and $N_{\rm eff}\Lambda$CDM, when geometry, shape and perturbation variables are considered. The symbols, whose errors represent the $1\sigma$ confidence level, follow the same colour scheme described in figures~\ref{fig:contibutions}-\ref{fig:varying-Mnu}. Additionally, the constraints from Planck \cite{Aghanim:2018eyx} and SH0ES \cite{Riess:2021jrx} have been added as light-blue and grey bands, for the $1\sigma$ and $2\sigma$ confidence levels. The relatively large central value for  the orange constraints in the two central panels is driven by the high $f\sigma_{s8}$ for the  quasars sample (see figure~\ref{fig:fixed_fss8}). 
 This is not reflected in the red constraints, in which most of the impact on $h$ coming from the $\Om$ determination from $f\sigma_{s8}$ is heavily superseded by the one from $\mathfrak{D}_V + F_\mathrm{AP}$.   } 
    \label{fig:whiska}
\end{figure}

Next, in the right panel of \ref{fig:varying-Mnu} we show the effect of additional dark radiation, parameterised by the effective number of neutrino species $N_\mathrm{eff}$. We include the BBN prior, but we adjust our baseline choice of  eq.~\eqref{eq:bbn-prior} taking into account the BBN constraints on $\Ob h^2$ and $N_\mathrm{eff}$ when considering Deuterium \cite{Cooke:2018} and Helium \cite{Aver:2015iza} data (see figure 1 in \cite{NilsBBN}). For simplicity, instead of undertaking the full BBN likelihood calculation as in \cite{NilsBBN2}, we choose the following Gaussian priors 
\begin{align} \label{eq:bbn-neff-prior}
    \Omega_\mathrm{b}h^2 = 0.0222 \pm 0.0005 \qquad N_\mathrm{eff} =  3.0 \pm 0.3, 
\end{align}
where we also incorporate a correlation coefficient between both parameters of $\rho = 0.6$ to correctly reproduce the findings of \cite{NilsBBN}.

Again, we show the constraints obtained from geometry and growth (cyan contours), shape and growth (red contours), and their combination (blue contours). The $\Lambda$CDM degeneracies discussed extensively in section \ref{sec:2h} are also indicated here via the black lines in the $(\Om,h)$ plane. We can see that, despite the inclusion of dark radiation, the posteriors are still restricted to these degeneracies. The posterior widths inflate, subject to the  fact that the BBN prior now allows for a certain correlation between $\Ob h^2$ and $N_\mathrm{eff}$ \cite{NilsBBN,NilsBBN2}. In particular, we measure $h_{\rd}=0.667 \pm 0.021$ in the geometry and growth case; $h_m=0.699 \pm 0.025$ in the shape and growth case, and $h_{m,\rd}=0.679 \pm 0.019$ in the combined case (for full results see appendix~\ref{app:s8_values}, table~\ref{tab:full}).

Figure~\ref{fig:whiska} summarizes the different values of $h$'s obtained in this section, where the colour code follows that of figures~\ref{fig:contibutions}-\ref{fig:varying-Mnu}. The first upper panel shows the results summarized in section~\ref{sec:results-geoshape} based exclusively on background geometry and shape; the second panel displays the results described in section~\ref{sec:results-geoshapegrowth}, which consists of adding the growth information. Both upper panels display the $h$ results for a flat-$\Lambda$CDM model. The two bottom panels display the results based as well on background (geometry and shape) and perturbation (growth),  for the $\nu\Lambda$CDM and $N_\mathrm{eff}$ models as described in this section.
We also include the light-blue and grey bands, corresponding to the measurements of Planck \cite{Aghanim:2018eyx} and SH0ES \cite{Riess:2021jrx}, respectively. In general, measurements calibrated with the horizon scale (cyan symbols) are in good agreement with the low-$h$ value measured by Planck, whereas the shape-informed measurements (red) is, given the error-bars, consistent with both.

\section{Beyond \texorpdfstring{$m$}{} - a geometrical interpretation of the shape.}  \label{sec:keq}

As indicated already in section \ref{sec:SF-shape}, our $h_m$ measurement based on the equality scale is indeed independent of the absolute value of the sound horizon $\rd$, but still makes use of the fact that $\rd$ is a standard ruler (in order to provide a constraint on $\Omega_m$). In this section we would like to explore the option of using instead the direct measurement of the equality scale $k_\mathrm{eq}^{-1}$ itself (from the turnover of the power spectrum) as a standard ruler. Then we investigate how this characteristic scale relates to the shape parameter $m$, and a potential additional measurement of $h$.

\subsection{Mock analysis setup}
\begin{figure}[t]
    \centering
    \includegraphics[width = 0.9\textwidth]{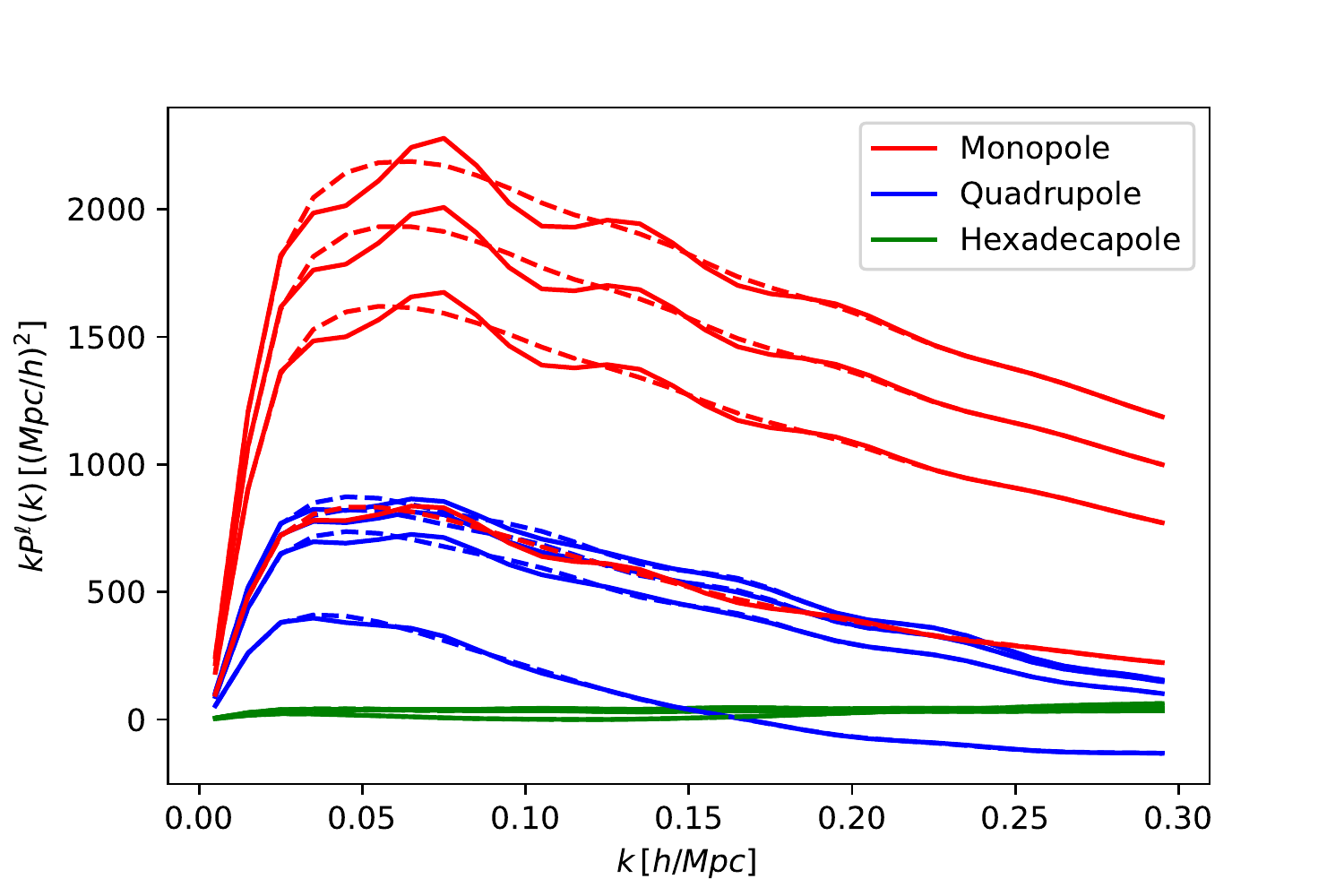}
    \caption{Linear theory prediction for the galaxy power spectrum multipoles at the four considered eBOSS redshift bins (from lowest redshift bin at top to highest redshift bin in the bottom) used to generate our mock dataset. Solid lines include the BAO wiggles (`withBAO'), while they are removed for the dashed lines (`noBAO').
    }
    \label{fig:mockdata}
\end{figure}

For this purpose, we create a mock dataset, `withBAO', coinciding with the linear theory prediction of the galaxy power spectrum multipoles $P^\ell (k)$ for the fiducial eBOSS cosmology and galaxy bias (motivated by table F3 of \cite{Gil-Marin:2020bct}) at each redshift bin centered at $z = \left\lbrace 0.38, 0.51, 0.70, 1.48 \right\rbrace$, which we refer as LRG+QSO sample. Another mock data set, `noBAO', is created using the same parameters, but with  BAO wiggles removed  using the numerical smoothing method provided in appendix D of \cite{ShapeFit:data}. The theory prediction used to produce these idealized set of mocks are displayed in figure~\ref{fig:mockdata}. When  analysing this set of mocks we  employ the same covariances as for the original BOSS and eBOSS data.

For the fixed-template fits we assume the same model that has been used to generate the mock power spectra: we use linear perturbation theory, the second order bias expansion parameterized by $b_1, b_2$ in the same way as in \cite{Gil-Marin:2020bct,ShapeFit:data} and assume the non-local bias parameters $\bs, \bnl$ to follow the local Lagrangian predictions \cite{McDonald_2009,Saito:2014qha}. We incorporate the redshift-space distortion via the Taruya-Nishimichi-Saito (TNS) model \cite{Taruya:2010mx} and a Lorentzian Fingers-of-God (FoG) damping term parameterized by the dispersion scale $\sigma_{\rm FoG}$. Finally, we model deviations from Poissonian shot noise via the parameter $A_\mathrm{noise}$ as defined in \cite{ShapeFit:data}. For all samples we fit the wavevector range $0<k\hoverMpc<0.15$.\footnote{Note that for the actual BOSS and eBOSS data we always set $k_{\rm min}=0.02\,h{\rm Mpc}^{-1}$.} We analyse each set of mocks as follows.

\begin{itemize}
    \item `withBAO'. Using the fiducial linear power spectrum as template, we apply both {\it ShapeFit} and the classic fit (where $m$ is kept to be fixed to 0) to each redshift bin varying the physical parameters $\{\alpha_\parallel, \alpha_\perp, f, (m)\}$ and the nuisance parameters $\{b_1, b_2, \sigma_{P}, A_\mathrm{noise}\}$.
    \item `noBAO'. Using the dewiggled linear power spectrum as template, we apply both {\it ShapeFit} and the classic fit to each redshift bin consisting of the physical parameters $\{\alpha_\parallel, \alpha_\perp, f , (m)\}$ and the same nuisance parameters as above.
\end{itemize}

In the `withBAO' case the physical parameter constraints can be interpreted in the usual way (see section 3.3 of \cite{ShapeFitPT} for a concise overview). In particular, the scaling parameters $\{\alpha_\parallel, \alpha_\perp\}$ can be transformed easily to the parameter base $\mathbf{\Theta}$ introduced in eq.~\eqref{eq:theta} via
\begin{align} \label{eq:trafo-withBAO}
    \frac{D_H(z)}{\rd} = \alpha_\parallel^\mathrm{withBAO} \frac{D_H^\mathrm{fid}(z)}{\rd^\mathrm{fid}}~, \qquad \qquad
    \frac{D_M(z)}{\rd} = \alpha_\perp^\mathrm{withBAO} \frac{D_M^\mathrm{fid}(z)}{\rd^\mathrm{fid}}~.
\end{align}
However, in the `noBAO' case the interpretation of the scaling parameters $\{\alpha_\parallel, \alpha_\perp\}$ needs to be modified, since the sound horizon is not a measurable quantity anymore. Instead, the equality scale earns the role of a standard ruler, such that the physical interpretation of the generic scaling parameters changes to
\begin{align} \label{eq:trafo-noBAO}
    \frac{D_H(z)}{\keq^{-1}} = \alpha_\parallel^\mathrm{noBAO}
    \frac{D_H^\mathrm{fid}(z)}{(\keq^{-1})^\mathrm{fid}}~, \qquad \qquad
    \frac{D_M(z)}{\keq^{-1}} = \alpha_\perp^\mathrm{noBAO} \frac{D_M^\mathrm{fid}(z)}{(\keq^{-1})^\mathrm{fid}}~.
\end{align}
Any cosmological BAO likelihood can be adjusted to the equality-based formulation of eq.~\eqref{eq:trafo-noBAO}, simply by replacing the sound horizon by $\keq^{-1}$.

\subsection{Compressed-variables results}

\begin{figure}[t]
    \centering
    \includegraphics[width=0.49\textwidth]{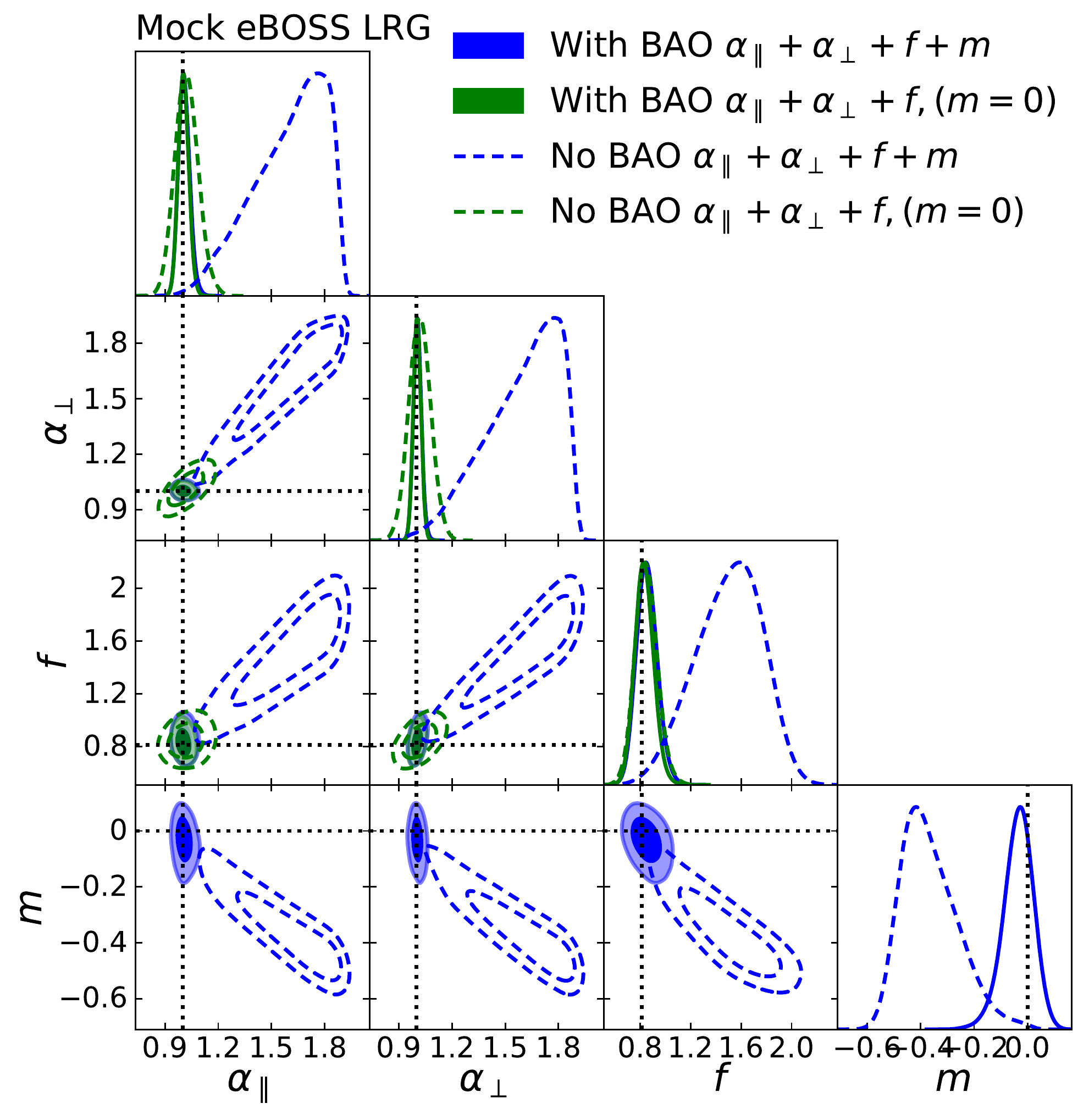}
    \includegraphics[width=0.49\textwidth]{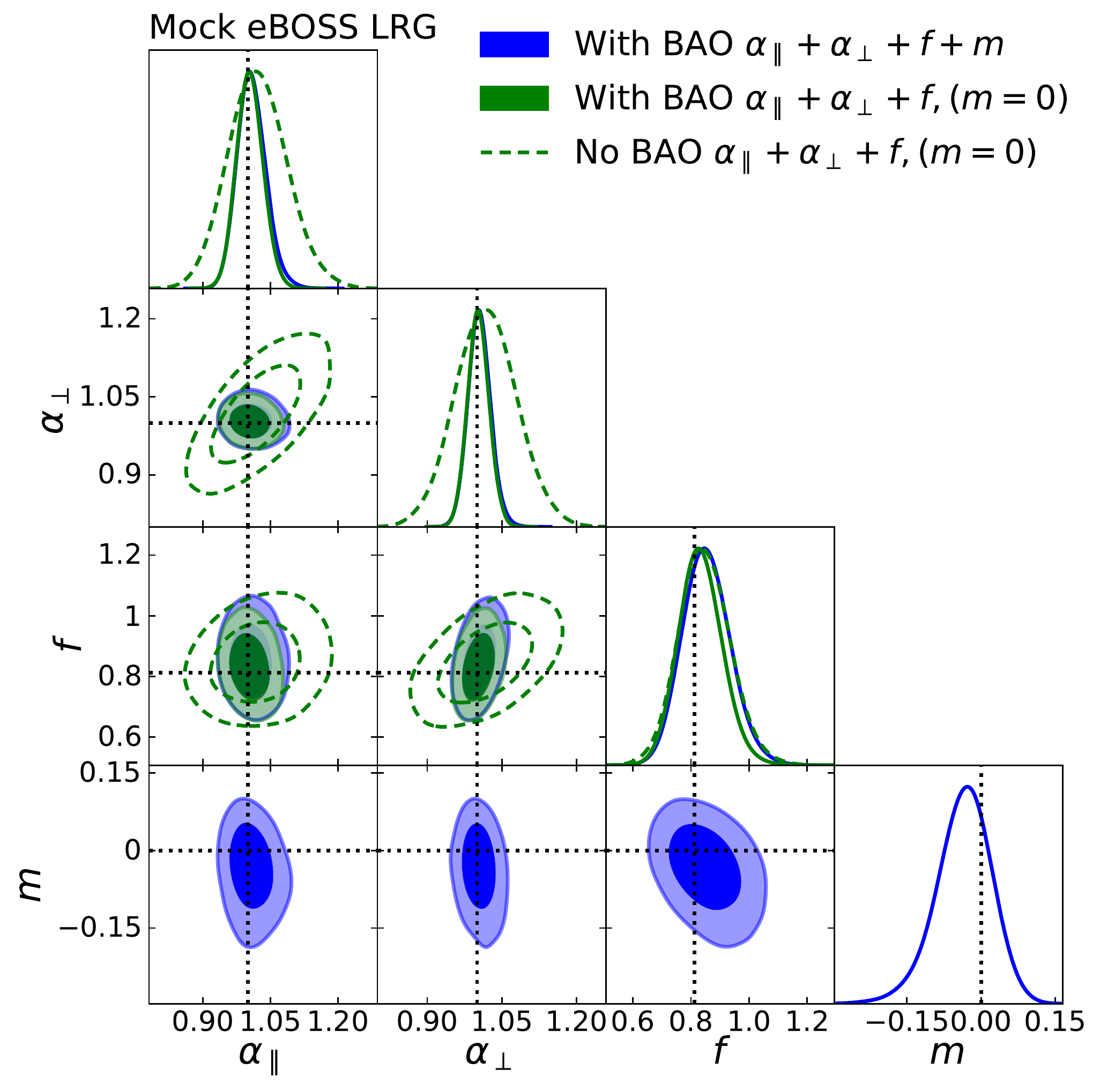}
    \caption{Tests on mocks: Fixed-template bestfit parameters for the mock power spectra of figure~\ref{fig:mockdata}, corresponding to the eBOSS LRG sample at effective redshift $z=0.70$. The physical parameters $\left\lbrace \alpha_\parallel, \alpha_\perp, f, m\right\rbrace$ constraints for the fixed template fits to the `withBAO' and `noBAO' are represented by filled and empty contours, respectively. The fit including $m$ is shown in blue, the other two fits, where $m$ has been fixed to its fiducial value, are shown in green. The right panel is a zoom-in of the left panel, with the `noBAO \SF' case excluded. The black dotted lines indicate the `true' parameter values, from which the mocks were generated.}
    \label{fig:mockresults_compressed}
\end{figure}

The compressed results of the four fits are shown in the left panel of figure~\ref{fig:mockresults_compressed}: `withBAO {\it ShapeFit}' (filled blue contours), `withBAO classic' (filled green contours), `noBAO {\it ShapeFit}' (empty blue contours), and `noBAO classic' (empty green contours), for the eBOSS LRG sample at effective redshift $z=0.70$ (recall that the error-bars have been derived by assuming the actual eBOSS LRG covariance on these mocks). The right panel is a zoom-in of the left panel, but without the `noBAO {\it ShapeFit}' case. Note that the physical interpretation of the scaling parameters is different between the `withBAO' and `noBAO' cases, according to eqs.~\eqref{eq:trafo-withBAO} and \eqref{eq:trafo-noBAO}. 
The fact that the filled contours are tighter than the empty contours, reveals that --within the standard ruler analysis-- the equality scale is significantly less constraining than the sound horizon. This is expected, as the BOSS and eBOSS galaxy power spectra are not very sensitive to the turnaround scale $\keq \sim 0.01\hoverMpc$, even in our optimistic case without a scale cut on the minimum $k$ (and fully systematics-free).  Instead, they are sensitive to the broadband shape for scales $k>\keq$. 
Note that the green dashed lines have $m$ fixed at the fiducial value, in the blue dashed lines $m$ is a free parameter. The  skewed and asymmetric constraints shown by the blue dashed lines are due to a combination of a non-linear response of the likelihood to changes in $m$ around the fiducial value and to prior volume effects.    
 
This is the reason why we observe such a strong correlation between the scaling parameters $\{\alpha_\parallel, \alpha_\perp\}$ and $m$ in the `noBAO \SF' case, in which these parameters remain unconstrained, meaning that the cosmological interpretation of $m$ and $D_V/\keq^{-1}$ are indeed closely related.

\subsection{Cosmological interpretation of the mock data}
\begin{figure}[t]
    \centering
    \includegraphics[width=\textwidth]{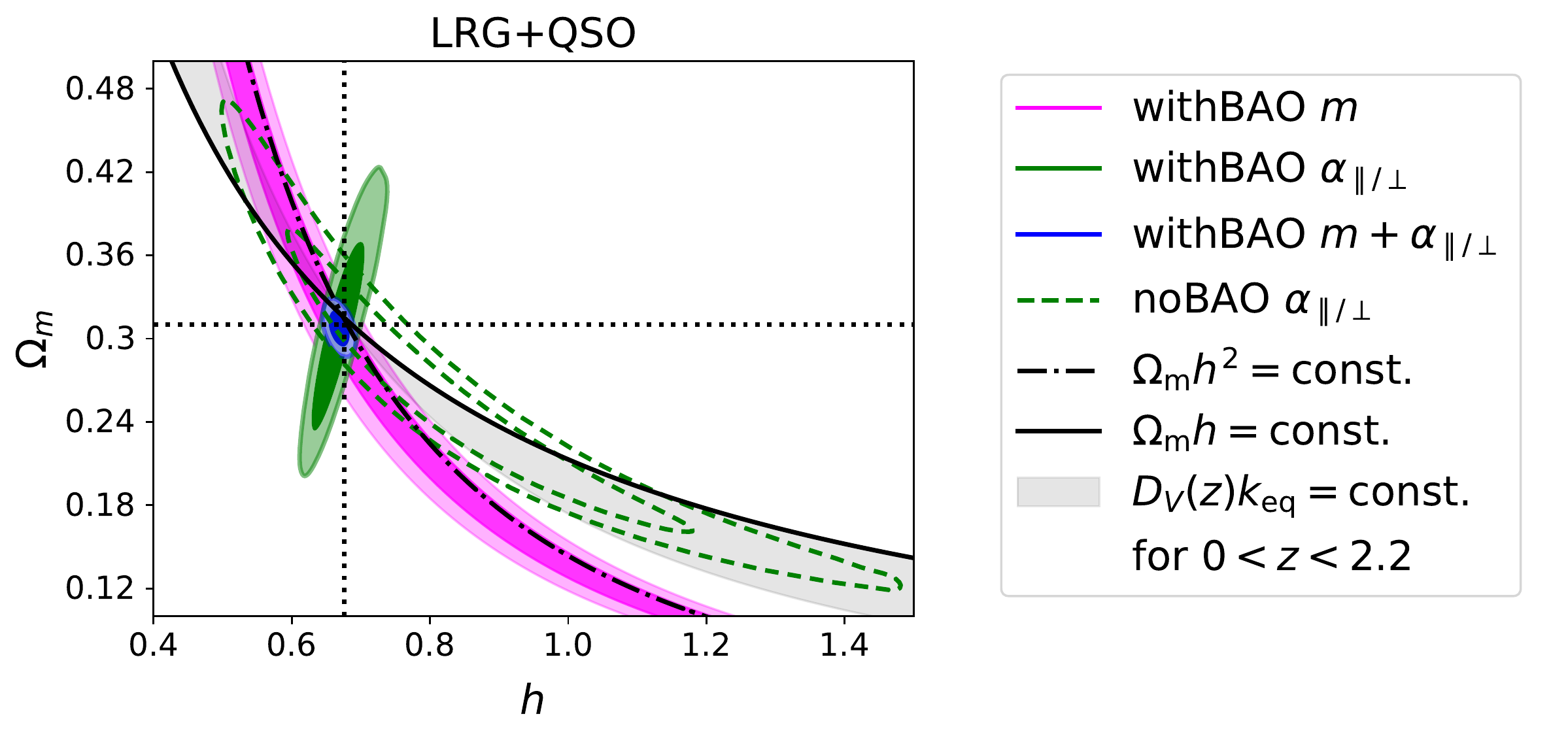}
    \caption{
    Interpretation of the results of the compressed variables in figure~\ref{fig:mockresults_compressed} for a flat-$\Lambda$CDM model assumption. We employ the same colour code, but this time also adding in magenta the `withBAO' results from $m$-only. These follow the $\Omega_\mathrm{m}h^2={\rm const.}$ line indicated by the black dash-dotted line, while the $\keq^{-1}$-derived constraints (empty green dashed contours) correspond to $D_V(z) k_\mathrm{eq} = {\rm const.}$ (grey region), where the $z\rightarrow 0$ limit, $\Omega_\mathrm{m}h={\rm const.},$ is indicated by the black solid line. The underlying values used to generate the power spectrum mock signal are shown by the horizontal and vertical black dashed lines.}
    \label{fig:mockresults-cosmo}
\end{figure}

We explore how the compressed mock data results compare in light of a flat $\Lambda$CDM model. We fix $n_s$ to the underlying value and use a prior on $\Omega_\mathrm{b}h^2$ with the same width as in eq.~\eqref{eq:bbn-prior}, but centered around the underlying value. Using eqs.~\eqref{eq:trafo-withBAO} and \eqref{eq:trafo-noBAO} for the different sets of scaling parameters, we obtain the constraints on $\Omega_\mathrm{m}$ and $h$ shown in figure~\ref{fig:mockresults-cosmo}.

First, we show the results from interpreting $m$ only in the `withBAO' case (magenta contour). This constraint exactly follows the relation $\Omega_\mathrm{m}h^2={\rm const.}$ indicated (for the true cosmology) by the black dash-dotted line: it is the same behaviour already observed in figure~\ref{fig:contibutions}. The filled green(blue) contours show the `withBAO' results from the scaling parameters via eq.~\eqref{eq:trafo-withBAO}, excluding(including) the shape. The empty green dashed contours arise from interpreting the scaling parameters in the `noBAO' case via eq.~\eqref{eq:trafo-noBAO}. Interestingly, they do not overlap with the `withBAO' $m$-only constraints, but approximately follow the combination $\Omega_\mathrm{m}h={\rm const.}$, indicated by the black dashed line that represents the limit $z \rightarrow 0$. The grey region indicates how this line evolves across redshifts.

This behavior can be explained as follows. The wavevector of equality $\keq$ in units of $1/\mathrm{Mpc}$ is proportional to
\begin{align}
    \keq \propto \Omega_\mathrm{m}h^2~,
\end{align}
which is what $m$ measures through the \textit{feature} associated to the wavevector of equality: the broadband shape. In the `noBAO' case, however, the scaling parameters are not sensitive to the absolute scale of equality $\keq^{-1}$, but to the cosmological distance in units of the equality scale (see eq.~\eqref{eq:trafo-noBAO}). Assuming the low redshift limit $z \rightarrow 0$, where all cosmological distance are proportional to the inverse Hubble parameter, $D \propto 1/h$, we find the following proportionality in the `noBAO' case,
\begin{align}
    \frac{D}{\keq^{-1}} = D\keq \propto \frac{\Omega_\mathrm{m}h^2}{h} = \Omega_\mathrm{m}h ~.
\end{align}

In reality, cosmological distances at non-zero redshifts show the additional $\Om$-dependence of eqs.~\eqref{dv_first}-\eqref{dv_last}, represented by the grey semitransparent region in figure~\ref{fig:mockresults-cosmo}. We see that this region aligns well within $1\sigma$ with the green dotted contour.

There is another interesting difference between the `withBAO $m$' and the `noBAO $\alpha_{\parallel/\perp}$' cases. The former only constrains the combination $\Omega_\mathrm{m}h^2$ and is not able to disentangle the two parameters within the prior range. The latter, on the other hand, is able to distinguish between them and provides measurements on $\Omega_\mathrm{m}$ and $h$ of order $\sigma_{\Omega_\mathrm{m}} \approx 0.08$ and $\sigma_h \approx 0.19$. 

These constraints demonstrate that, in principle, the equality scale can be used as an uncalibrated standard ruler (see \cite{cunnington22} and references therein) to measure the late-time expansion history, which is determined by $\Omega_\mathrm{m}$ in the case of a flat-$\Lambda$CDM model. Still, the traditional BAO constraints (filled green contours) are tighter by factors 2 and 10 for $\Omega_\mathrm{m}$ and $h$, respectively, which is a remarkable result and demonstrates the utility and robustness of the sound horizon as a standard ruler.

\begin{figure}[t]
    \centering
    \includegraphics[width=0.49\textwidth]{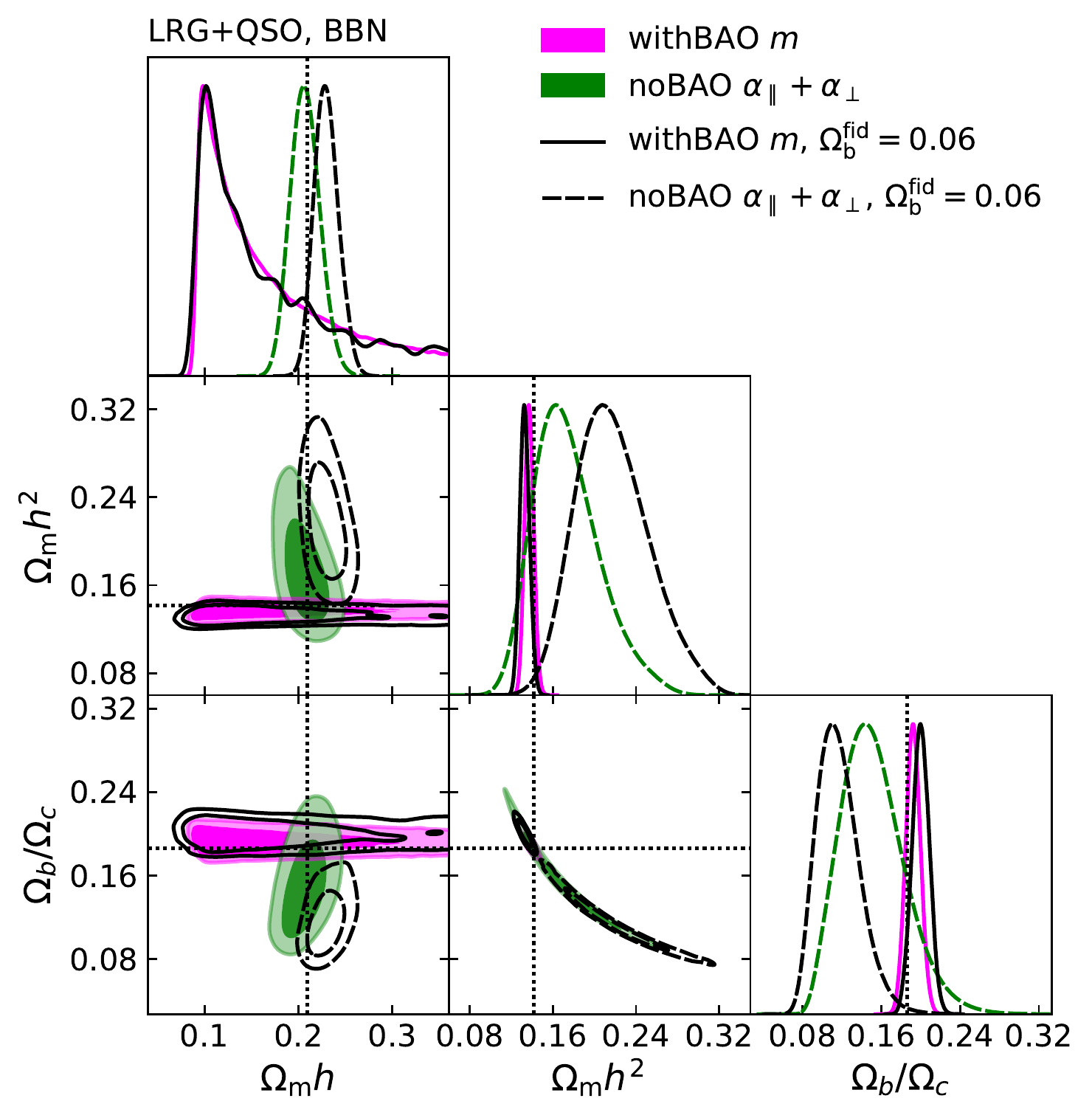}
     \includegraphics[width=0.49\textwidth]{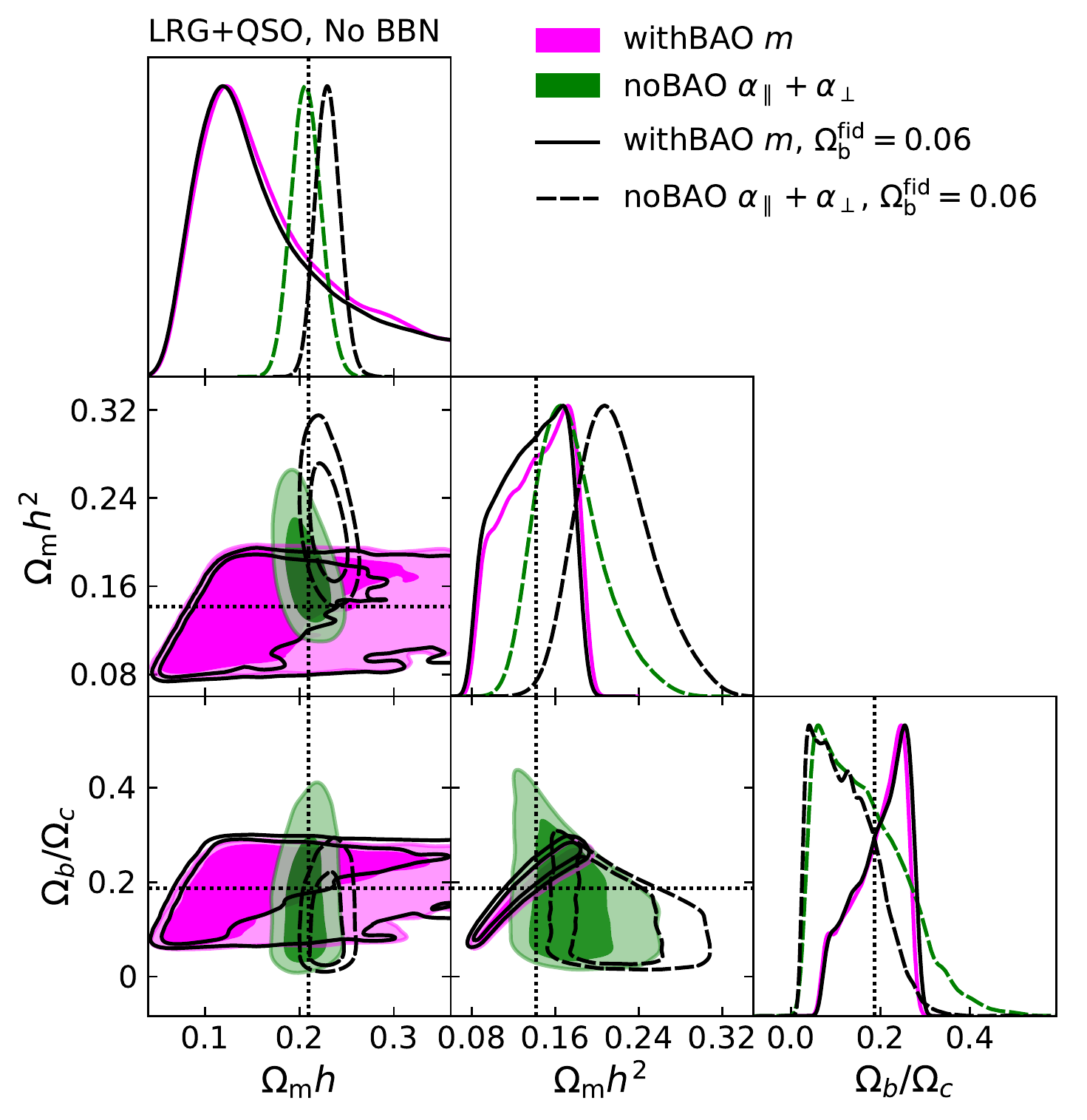}
    \caption{Impact of the choice of the fiducial value of $\Omega_\mathrm{b}$ within the fixed-template fits, when employing $\keq$ as a ruler (green and dashed) or $m$ as an anchor (magenta, solid black),  with (left panel) or without (right panel) BBN prior during the cosmology inference. The results derived from a template-fit with a {\it correct} fiducial value for the baryon density, $\Omega_{\rm b}=0.048$, are shown in magenta contours (for the mocks with BAO, where only the shape parameter $m$ is being used), and in green filled, dashed contours (for the mocks without BAO, using the scaling variables $\alpha_{\parallel,\perp}$ through eqs.~\eqref{eq:trafo-noBAO}). Conversely, the cases where a {\it displaced} fiducial value for the baryon density is assumed for the shape of the template, $\Omega_{\rm b}=0.06$, are shown in black lines: solid for mocks with BAO when only considering the shape parameter $m$, and dashed for mocks without BAO when considering the scaling $\alpha_{\parallel,\perp}$ variables through eqs.~\eqref{eq:trafo-noBAO}. The true cosmology of the mocks is represented by the vertical and horizontal dotted lines. On one hand, the cosmology-derived contours from which $\keq$ is indirectly inferred from $m$ (magenta bands and solid black lines for mocks with BAO) do not present any systematic shift with respect to the expected true values. On the other hand, the results derived from a direct inference of $\keq$ from the scaling $\alpha_{\parallel,\perp}$ variables (dashed lines) show a systematic offset when the value of $\Omega_{\rm b}$ assumed on the fixed-template deviates from its true underlying value.
    }
    \label{fig:mockresults-cosmo-nobbn}
\end{figure}

Because the power spectrum turnaround  is at very large scales where cosmic variance is large, it is unlikely that this approach will be superior to the tried and tested BAO one. But there is a more severe problem with the $\keq^{-1}$-based standard ruler approach. Cosmological constraints obtained from the calibrated equality scale measurement suffer a strong template-dependence. We visualize this issue in figure~\ref{fig:mockresults-cosmo-nobbn}, where we explore the behavior of the relevant `withBAO' and `noBAO' cases when including (left panel) or not (right panel) the BBN-inspired prior on $\Omega_\mathrm{b}h^2$.   

The consequences of such prior for the `withBAO $m$-only' case (magenta contours) are as follows. The power spectrum slope is now not only determined by $\Omega_\mathrm{m}h^2$, but also by the baryon-to-cold-dark matter ratio, $\Ob/\Oc$. Hence, the cosmological fit shows a complete degeneracy between these two parameter combinations.

This evident degeneracy, however, is not at all captured in the `noBAO' case (filled green contours), where $\Omega_\mathrm{m}h$ seems overly well constrained in the case that the BBN prior is not included. Of course this is not surprising, since $\keq\propto \Omega_\mathrm{m}h$ does not depend on the baryon density $\Omega_\mathrm{b}h^2$ (or ratio $\Ob/\Oc$) in any way. But since the fixed template fit at the previous compression step is carried out with a fixed $\Omega_\mathrm{b}h^2$, the cosmological constraints from the so-obtained $\keq$ are not able to capture the correct degeneracies in cosmological parameter space.
The \textit{ShapeFit} parameterisation of the power spectrum shape through $m$, on the other hand, is flexible enough to capture a variety of models in a template-independent way. 

We visualize this issue by carrying out additional fits using a template with an intentionally chosen displaced fiducial value for $\Omega_\mathrm{b}^\mathrm{fid} = 0.06$, deviating by $25\%$ from the `true' value $\Omega_\mathrm{b}^\mathrm{true} = 0.048$.  Figure~\ref{fig:mockresults-cosmo-nobbn} demonstrates that, in the cases for which the template corresponds to  $\Omega_\mathrm{b}^\mathrm{fid} = 0.06$, the $m$-derived cosmological constraints (black solid contours) do not deviate from the baseline results (magenta contours) by more than $1\sigma$, whereas the $\keq^{-1}$-derived constraints (black dotted contours) exhibit a clear shift (with respect to the green dashed lines) of $1.6\sigma$ in $\Om h$ and $2.0\sigma$ in $\Om h^2$ due to the incorrect value of $\Omega_\mathrm{b}^\mathrm{fid}$ in the template.  This clearly demonstrates that calibrated $\keq^{-1}$-based standard ruler methods would appear severely biased in case a wrong template is assumed. This could be remedied by either varying $\Ob$ at the stage of the template fit (and hence worsening the constraints) or adjusting the interpretation of $(\alpha_\parallel \alpha_\perp^2)^{1/3}$ towards $D_V k_\mathrm{eq}$ by taking into account the baryon suppression consistently. We leave such an investigation for future work. 

To summarize, we have shown that in theory it is possible to measure $h$ by way of direct inferences of $\keq^{-1}$ as standard ruler, even for cosmologies with no sound horizon information at all (see the empty green dashed contour of figure~\ref{fig:mockresults-cosmo}). However there are a few reasons for scepticism. First, this $h$-measurements would not be competitive with respect to other methodologies and datasets (BAO, GW, SN, etc). Second, it is very hard
to obtain such a measurement in a template-independent way using the standard methodology, as the one used for BAOofLSS. For that reason, previous works that have already tried to infer $\keq^{-1}$-derived constraints on $h$, have employed model-dependent approaches, where the template shape varies at the same time as the cosmology inference (see for e.g., \cite{Philcox_Sherwin_Farren_Baxter_21,Farren:2021grl,Philcox:2022sgj}). These measurements should not be interpreted as direct $\keq^{-1}$-based standard ruler measurements though, but rather as measurements of $h$ without the absolute value of the sound horizon (while still making use of the fact that the sound horizon is a standard ruler). Instead, their measurements are equivalent to our $h_m$ measurement in section \ref{sec:results}.

However, the (uncalibrated) $\keq^{-1}$-based standard ruler analyses come with the prospect of delivering a sound horizon-independent $\Omega_\mathrm{m}$ measurement. Obtained from the isotropic component `$D_V$' -and hence relying on standard ruler property \textit{ii)}-, this is also subject to some real-world subtleties as elaborated before. But if only the anisotropic `$F_\mathrm{AP}$' is used, which relies on assumption \textit{i)}, these issues can be remedied. In fact, in the work of \cite{Cuceu:2021hlk} which includes a scaling of the broadband-only part of the power spectrum very similar to our treatment of the `noBAO' mocks, they only use the anisotropic `$F_\mathrm{AP}$' part for cosmological interpretation for exactly the same reasons mentioned here. A promising research direction would hence be to combine their approach with our \textit{ShapeFit} method, and as such enable a template-independent interpretation of the isotropic component as well. Since this is beyond the scope of this paper, we leave this for future research.

\section{Discussion and conclusions}\label{sec:conclusions}

Direct, cosmic-distance ladder-based determinations of the Hubble parameter $H_0$, anchored to $z\sim0$ calibrators,  are in tension with indirect determination of $H_0$ seen as a global parameter of the (extremely successful) standard $\Lambda$CDM model. Indirect determinations are usually anchored to early-time physics. This mismatch  has motivated several proposed extensions to the $\Lambda$CDM model. A comprehensive analysis \cite{H0olympics} clearly indicates that early-time solutions, models that involve new physics beyond $\Lambda$CDM model before recombination, are favored over late-time solutions.

The distance ladder-based $H_0$ determinations have several different anchors (Cepheids, TRGB, Masers, etc.), yet they mostly cluster around the high-$H_0$ `camp'. The early-time physics determinations have one well-established and exquisitely determined anchor,  the sound horizon at radiation drag. Within the standard $\Lambda$CDM model this quantity is determined by CMB observations with a $\sim$0.2\% error, it is, however, strongly model-dependent. It can also be determined independently of CMB observations e.g., \cite{NilsBBN,NilsBBN2} by resorting to BBN, but it must still assume standard pre-recombination physics and a $\Lambda$CDM model or small  parametric extensions to it.
It is therefore of value to provide an independent early-time  anchor for global indirect $H_0$ determinations. 

This other anchor is provided by the physical processes at and around the matter-radiation equality era.  We have shown that, in principle, the equality scale could be used as a standard(-izable) ruler, its signature being the turn-around of the matter power spectrum on large scales. This approach, however, at the moment is not  really competitive. On the other hand the broadband shape of the matter power spectrum at large, linear scales is related to how fast modes re-entered the horizon at the end of the radiation era. We have shown how, under specific assumptions,  the logarithmic slope of the power spectrum at those scales is related to how fast the horizon was expanding and can thus be seen as a ``speedometer'' at that epoch.  This is a promising route  to an equality-anchored $H_0$ determination.   

We have performed such measurement using the \SF approach and the state-of-the-art BOSS and eBOSS galaxy clustering data, finding the equality-anchored Hubble constant to be $H_0=70.1\pm 2.1$ km s$^{-1}$ Mpc$^{-1}$; this result represents the most precise  measurement of $H_0$ to date, being independent of the sound horizon physics when LSS-only data (in combination of a BBN and a $n_s$ priors) are used.

The modular and model-independent approach provided by \SF enables us to perform a set  of diagnostic tests where, for each relevant epoch or physical process, we assume that the $\Lambda$CDM model is a good effective model, but its parameters are not forced to be the same across all epochs and features (we call it piece-wise $\Lambda$CDM model). In this way we obtain two distinct  $H_0$'s: the sound-horizon anchored, the equality anchored (and a third one which is the combination of the other two). We find broad agreement between the equality-anchored and the sound horizon-anchored $H_0$'s and this provides well defined guardrails for new physics beyond $\Lambda$CDM.
To this aim, in table~\ref{tab:conclusion} we summarize our main results, clearly highlighting which physical assumptions are being made and which signatures (and combinations of signals) are being considered.

Early-time exotic models which aim to solve the Hubble tension via early-time modifications  targeted to the sound-horizon scale, must account for matter-radiation equality effects, which in general would modify this other $h$ too. For example, Early Dark Energy (EDE) models promise to reconcile the discrepancy between low- and high-$H_0$ values by invoking a phase of accelerated expansion shortly before recombination \cite{smithetal2020,poulinetal2018,poulinetal2019,Smith:2022iax,Simon:2022adh,Simon:2022lde,herold_ferreira22} and  modifying the size of the sound-horizon ruler. In this way, when $\Lambda$CDM physics is assumed at pre-recombination times, the BAOofLSS measurements of $H_0$ would be incorrectly calibrated. EDE then must dissipate fast enough as to not affect the photon-diffusion scale and the angular scales  that are tightly constrained by measurements of CMB anisotropies  e.g., \cite{Knox_Millea20}. 
Naively, if this  type of new physics leaves the matter-radiation equality physical processes almost unchanged, then (incorrectly) assuming a $\Lambda$CDM model throughout  the data analysis would produce $H_0$ measurements derived from matter-radiation equality  which may be in tension with the BAOofLSS $H_0$ measurements.
We leave a detailed analysis of  this specific case to future work. 

\begin{table}[t]
    \centering
    \begin{tabular}{|p{25mm}|P{52mm}|p{63mm}|}
    \hline
    Method & $\Lambda$CDM baseline model & Main Assumptions \\
    \hline
    \hline
      $D_V/r_{\rm d}+F_{\rm AP}$ \newline+BBN   & \,\newline{\centering $100 \times h_{r_{\rm d}}=67.42^{+0.88}
_{-0.94}$}  & Late-time background expansion; Pre-recombination physics (standard $r_{\rm d}$); $\Om^{\rm early, r}=\Om^{\rm late}$ \\
      \hline
     $m+\mathfrak{D}_V+F_{\rm AP}$\newline$+{\rm BBN}+n_s$\newline+$\sum m_\nu$ &  \,\newline $100\times h_m=70.1\pm 2.1$ & Late-time background expansion; Equality physics; Standard $r_{\rm d}$;  $\Om^{\rm early, eq}=\Om^{\rm late}$ \\
     \hline
     $D_V/r_{\rm d}+F_{\rm AP}$\newline$+m +{\rm BBN}$\newline$+n_s+\sum m_\nu$  &  \,\newline $100 \times h_{(m,r_{\rm d})}=67.90^{+0.76}_{-0.75}$ & Late-time background expansion; Equality physics; Pre-recombination physics (standard $r_{\rm d}$); $\Om^{\rm early, r}=\Om^{\rm early, eq}=\Om^{\rm late}$ \\
      \hline
      \hline
     $\mathfrak{D}_V+F_{\rm AP}$ & $\Om^{\rm late}=0.290^{+0.015}
_{-0.016}$ & Late-time background expansion; \newline Standard $r_{\rm d}$ \\
      \hline
     $m+{\rm BBN}$ \newline $+n_s+\sum m_\nu$ &  $[\Om h^2]^{\rm early,eq}=0.1395\pm0.0036$ & Equality physics \\
     \hline
    \end{tabular}
    \caption{Summary table displaying the most relevant results of this paper on $h$ (3 first rows), and on $\Om$ and $\Om h^2$, in the last two rows, respectively, in all cases under the assumption of a flat $\Lambda$CDM model. The left column displays the set of compressed variables used within the datavector $\Theta$ (see eq.~\eqref{eq:theta}), and also whether some priors are used on $\Ob h^2$ (see eq.~\eqref{eq:bbn-prior} motivated by BBN), $n_s$ (see eq.~\eqref{eq:ns-prior}) and $\sum m_\nu=0.06\,{\rm eV}$. The right column briefly summarize the physical assumptions that each set of variable makes. Finally, in the middle column we stress by the sub- or super-index whether this quantity is sensitive to early- or late-time physics, and whether is based on sound horizon scale's ($r_{\rm d})$, or matter-radiation equality epoch's physics. ($m$ or eq.). }
    \label{tab:conclusion}
\end{table}

Our results can be compared to previous works using an independent methodology to extract a sound horizon-free measurement of $H_0$. In particular \cite{Philcox:2022sgj} reported
$H_0= 69.6^{+4.1}_{-5.4}\, {\rm km\,s}^{-1} {\rm Mpc}^{-1}$ using the BOSS LRGs data at $0.2<z<0.75$\footnote{We choose this value among other values reported in their analysis to be the closest to our analysis setup which relies on minimum assumptions based on external datasets to BOSS. In particular for this value they choose a BBN prior, a neutrino mass sum bound of  $\sum m_\nu<0.26\,{\rm eV}$ and a spectral index prior of  $n_s=0.96\pm0.02$, and assuming a flat-$\nu\Lambda$CDM model.}. This can be compared to our LRG-only measurement (although we also include eBOSS LRGs, thus $0.2<z<1.0$) of $H_0=64.5^{+3.5}_{-4.5} \,{\rm km\,s}^{-1} {\rm Mpc}^{-1}$; and $H_0=70.2^{+1.9}_{-2.1} \,{\rm km\,s}^{-1} {\rm Mpc}^{-1}$ for LRG+QSO+Lyman-$\alpha$ in the full range of $0.2<z<3.5$. Consistently using all BOSS and eBOSS data shrinks the errorbars  by a factor of $\sim 2$ compared to previous results. Even when using only the LRG galaxies, our results is notably tighter, possibly because of the high-$z$ eBOSS galaxies and BAO reconstruction, which significantly help to determine the $\Om$ value through the uncalibrated BAO (Table~\ref{tab:full}). Part of the difference in error bar can also be explained by the slightly different prior choices for cosmological parameters with respect to \cite{Philcox:2022sgj}. If we match their setup, we obtain $H_0=65.5^{+3.9}_{-4.6} \,{\rm km\,s}^{-1} {\rm Mpc}^{-1}$ for the LRG Sample only. 

The $H_0$ tension is truly  a tale of two $h$: the local (direct, late-Universe) and the global (indirect). There are in reality  (many) more than two ways, independent and based on very different physics, to measure $H_0$; yet, they all cluster around the early/low and late/high camps. Importantly, the two  $H_0$ determinations presented here, anchored at early times and based on  different  early-Universe physics ingredients, are consistent with each other. The newer determination, independent of CMB observations and independent of the sound horizon and  anchored at equality,  has now competitive error bars, $\sim$ 3\%; whereas the sound horizon-based one, also independent of CMB observations, has a 1.5\% error. We envision that forthcoming improvements on these large-scale structure-based measurements will act as guardrails on the road to a solution to this  persistent tension.

\section*{Acknowledgments}

We thank  Nils Sch\"oneberg, Jos\'e Luis Bernal, Adam Riess and Tristan Smith  for useful comments on an advanced version of this manuscript. We also thank the anonymous referee for their constructive feedback that helped improve the manuscript. For the purpose of open access, the author has applied a Creative Commons Attribution (CC BY) licence to any Author Accepted Manuscript version arising from this submission.
SB acknowledges support from the European Research Council (ERC) under the European Union’s Horizon 2020 research and innovation program (FutureLSS, grant agreement 853291). HGM and LV acknowledge support of European Union’s Horizon 2020 research and innovation programme ERC (BePreSySe, grant agreement 725327).

Funding for this work was partially provided by the Spanish MINECO under project PGC2018-098866-B-I00MCIN/AEI/10.13039/501100011033 y FEDER ``Una manera de hacer Europa", and the ``Center of Excellence Maria de Maeztu 2020-2023'' award to the ICCUB (CEX2019-000918-M funded by MCIN/AEI/10.13039/501100011033). We acknowledge the IT team at ICCUB for the help with the \textsc{Aganice} cluster where all the calculations presented in this paper where done.

\appendix

\section{The impact of priors on \texorpdfstring{$\Omega_{\rm b}h^2$}{} and \texorpdfstring{$n_s$}{}} \label{app:ob-ns}

We explore the effects of some of the assumptions made in our baseline analysis. First, we investigate the impact of excluding the prior on  $\Ob h^2$ motivated by BBN observations. Second, we evaluate the impact of fixing $n_s = 0.97$ in our baseline analysis versus allowing $n_s$ to vary within a Gaussian prior specified below.

\begin{figure}[t]
    \centering
    \includegraphics[width=0.49\textwidth]{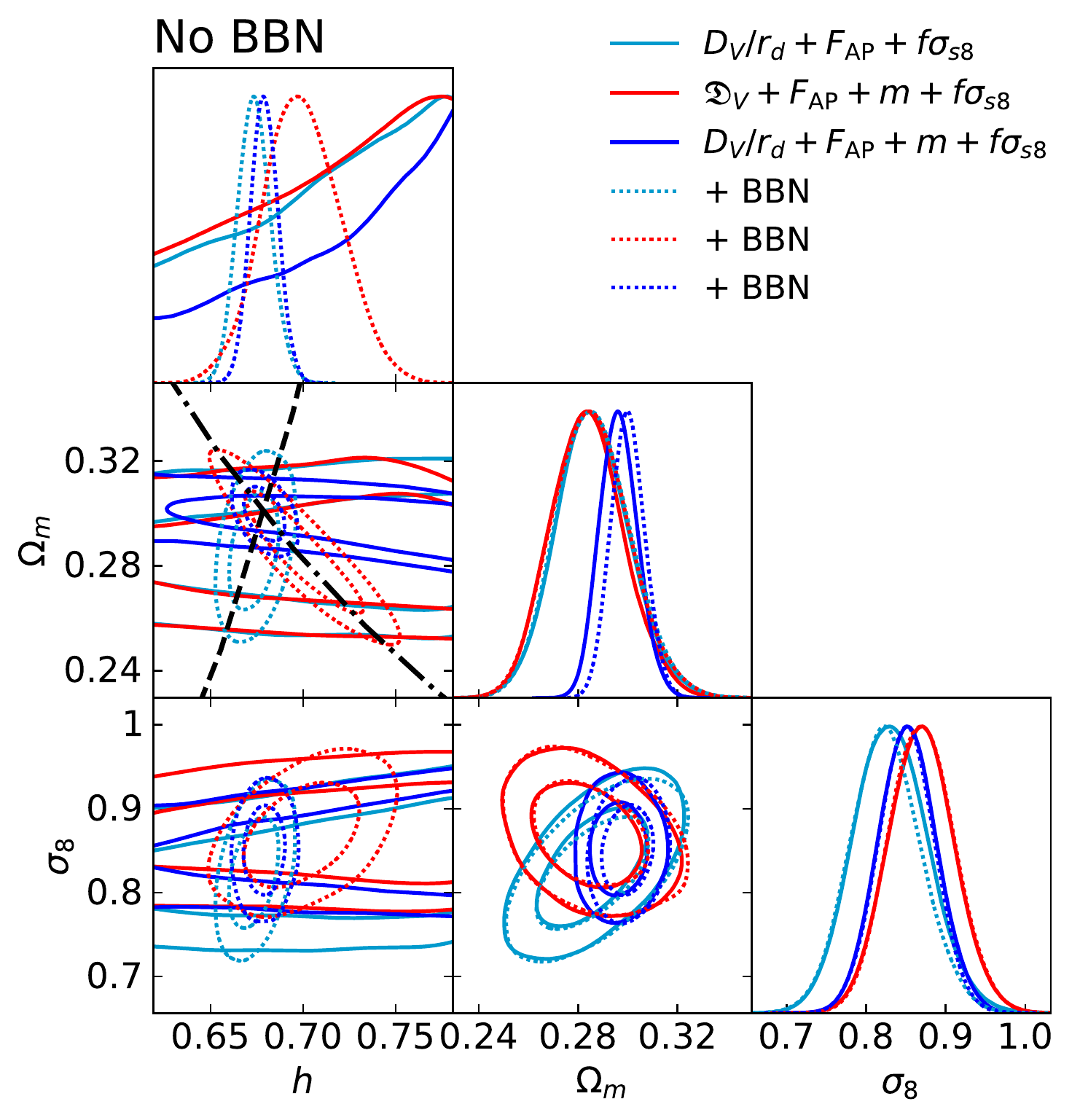}
    \includegraphics[width=0.49\textwidth]{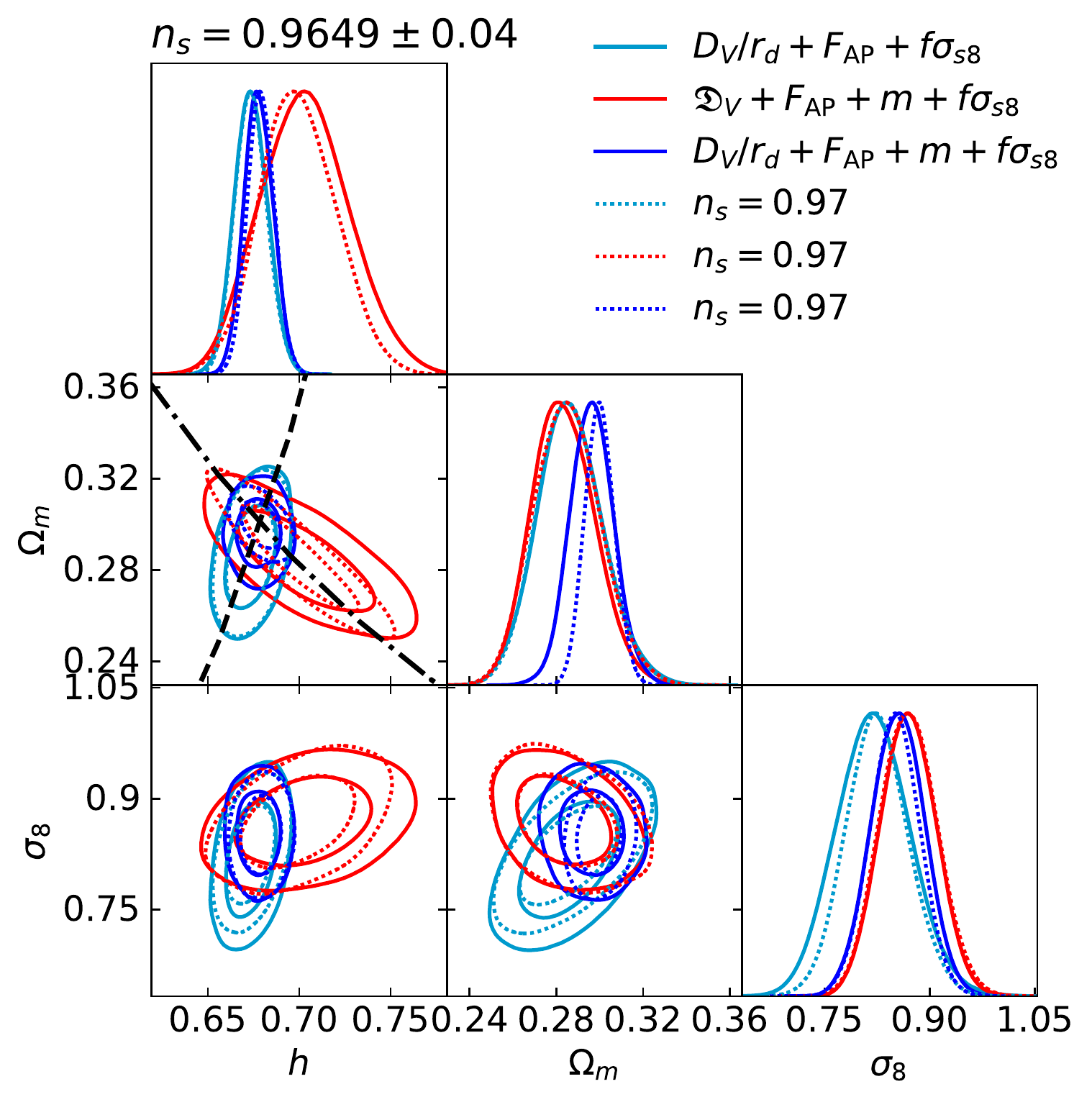}
\caption{{Left panel:} effect of removing the BBN prior on $\Omega_{\rm b}h^2$ in the parameters of a flat $\Lambda$CDM model, for the geometry+shape+growth case. The dotted lines display the most relevant cases of figure~\ref{fig:growth} (i.e., with the BBN prior on $\Omega_{\rm b}h^2)$, whereas the solid lines represent the effect of setting a wide and uninformative uniform prior on $\Omega_{\rm b}h^2$. As expected, $\Om$ is unaffected by the prior, as its constraints come from the relative BAO peak positions in different directions and at different redshifts ($\mathfrak{D}_V+F_{\rm AP}$ case); similarly $\sigma_8$ is also unaffected as it is constrained by the redshift space distortions, i.e., the relative amplitudes of the isotropic and anisotropic signals, mainly at large scales. Only $h$ is significantly affected: 1) a free $\Omega_{\rm b}h^2$ de-calibrates the horizon scale ruler size, which can take almost any value; 2) a free $\Omega_{\rm b}h^2$ also washes out the information on $\Om h^2$ from the shape, because of a `free' baryon suppression. As a consequence, without a precise $\Omega_{\rm b}h^2$ constraint the system looses its 'two anchors' and only $\Om$ and $\sigma_8$ can be efficiently determined. {Right panel:} effect of considering a Gaussian prior on $n_s$ centered around the Planck best-fit value $n_s=0.9649 \pm 0.04$ where the width corresponds to 10-$\sigma$ Planck sensitivity versus fixing $n_s=0.97$ in our baseline setup. As expected, this leads to a broadening of the  shape-derived constraints (red) on $h$, while the sound-horizon based constraint (cyan) is essentially unaffected. The $\Om$ constraints are completely unaffected, but there is a mild increase in error on $\sigma_8$.}
    \label{fig:varying-Ob}
\end{figure}

The effect of removing the BBN priors on $\Omega_{\rm b}h^2$ is displayed in the left panel of figure~\ref{fig:varying-Ob}, where the dashed (solid) lines represent the BBN prior (flat prior $\left[ 0.1<100\Ob h^2 <4.0 \right]$) results on the $\Lambda$CDM parameters for the different variable combinations using geometry, shape and growth, as described in the legend, and using the same colour notation as in figures~\ref{fig:growth},\ref{fig:varying-Mnu}. On the one hand,  the $\Omega_{\rm b}h^2$ prior  has no effect  on $\Omega_m$ and $\sigma_8$. $\Omega_m$  is essentially constrained from the uncalibrated BAO signal, which depends on relative $D_V(z_i)/r_{\rm d}$ measurements at different redshifts (noted as $\mathfrak{D}_V(z_i)$ when $r_{\rm d}$ is marginalized over), and by the Alcock-Paczynski parameter, $F_{\rm AP}$, both fully independent of $r_{\rm d}$, and therefore of $\Omega_{\rm b}h^2$. The $\sigma_8$ parameter is inferred from the redshift-space variable $f\sigma_{\rm s8}$ once $\Omega_m$ is known (from the uncalibrated BAO) and GR is assumed (for the $f\sim\Om^{0.56}$ relation), again fully independent of  $\Omega_{\rm b}h^2$. On the other hand, the results on $h$ are highly affected by the $\Omega_{\rm b}h^2$ prior. On the sound horizon-calibrated results (cyan lines), the $\Omega_{\rm b}h^2$ measurement is key to determine the absolute size of $r_{\rm d}$ (once the early-time $\Lambda$CDM physics are assumed) and therefore determine $h$. For results obtained via the shape variable $m$ (red lines), $\Omega_{\rm b}h^2$ plays a key role on determining the baryon suppression in the transfer function with sufficient precision to establish a connection between matter-radiation equality and the slope, and thus determine $h$.

Even when combining the BBN-free results taking the full $D_V/r_{\rm d}$ and shape, $m$, information, $h$ remains undetermined (dashed blue lines), although the effects of $\Omega_{\rm b}h^2$ on $r_{\rm d}$ on one hand and on the shape on the other hand are of very different nature. It appears that the physical CDM density $\Oc h^2$ is able to compensate for both effects in parallel while still fitting the \SF data. In order to break this degeneracy we would require additional information related to $\Omega_{\rm b}h^2$, for example the amplitude of the BAO peak, although it is not clear whether this would appreciably tighten our constraints in the case without BBN. We leave such an investigation for future work.

In the right panel of figure~\ref{fig:varying-Ob} we show the impact of employing a Gaussian prior on the spectral index $n_s=0.9649 \pm 0.04$ instead of fixing it to $n_s=0.97$ as in our baseline analysis. The prior width represents 10-$\sigma$ deviations given the Planck sensitivity \cite{Aghanim:2018eyx}. We can see that the impact of such a prior is mild. The error on $\sigma_8$ slightly increases (by 20\%) in the geometry+growth (cyan) case. The constraint on $h$ on the other hand barely changes in that case. Only upon including the shape $m$, varying $n_s$ appreciably changes the constraints. In particular, it inflates the $\Om-h$ degeneracy in the sound horizon-independent case (red), such that our fixed $n_s$ constraint $H_0 = 70.2^{+1.9}_{-2.1}\, {\rm km\,s}^{-1} {\rm Mpc}^{-1}$ changes to $H_0 = 70.4^{+2.2}_{-2.8}\, {\rm km\,s}^{-1} {\rm Mpc}^{-1}$ once we relax that assumption. As a consequence, the combined constraint (blue) mildly changes from $H_0 = 67.90_{-0.75}^{+0.76}\,{\rm km\,s}^{-1} {\rm Mpc}^{-1}$ to $H_0 = 67.80_{-0.84}^{+0.78}\,{\rm km\,s}^{-1} {\rm Mpc}^{-1}$, being still remarkably competitive with respect to the local distance determination \cite{Riess:2021jrx}. In parallel, the combined constraint on the matter density also changes from $\Om = 0.3000_{-0.0073}^{+0.0065}$ to $\Om = 0.2963_{-0.011}^{+0.0096}$, as an effect of the inflated $\Om-h$ degeneracy mentioned earlier. But note that this only affects the $\Om^\mathrm{early}$ part coming from the shape. The $\Om^\mathrm{late}$ part from geometry is not influenced by varying $n_s$ at all.

\section{Summary of all cosmological constraints} \label{app:s8_values}

Table~\ref{tab:full} displays the full results for different combinations presented in the paper, for the full LRG+QSO+Lyman-$\alpha$ full redshift range, $0.2<z<3.5$. 
Note that there are small differences in the error bar with respect to equivalent cases reported in table 9 of \cite{Brieden:2022lsd}. This is because the transformation of the $\left(D_M/\rd,  D_H/\rd\right)$ towards the $\left(D_V/\rd,  F_\mathrm{AP}\right)$ basis induces small numerical fluctuations.

\begin{table}[thb]
    \centering
    \begin{tabular}{|c|c|c|c|c|}
    \hline
    Model & Case & $H_0\scriptstyle{ [{\rm km\,s}^{-1} {\rm Mpc}^{-1}]}$ & $\Omega_m$ & $\sigma_8$  \\
        \hline
        \hline
        $\Lambda$CDM & $F_\mathrm{AP}+m$ & $69.5^{+4.2}_{-5.1}$ & $0.292_{-0.041}^{+0.034}$ & - \\
        $\Lambda$CDM & $\mathfrak{D}_V + F_\mathrm{AP}+m$ & $70.1^{+2.1}_{-2.1}$ & 
        $0.286_{-0.017}^{+0.013}$ & -   \\
        $\Lambda$CDM & $D_V/\rd+F_{\rm AP}$ & $67.42^{+0.88}_{-0.94}$ & $0.290_{-0.016}^{+0.015}$ & -  \\
        $\Lambda$CDM & $D_V/\rd+F_{\rm AP}+m$ & $67.90^{+0.76}_{-0.75}$ & $0.3019_{-0.0069}^{+0.0074}$ & -  \\
\hline
        $\Lambda$CDM & $F_\mathrm{AP}+m+f\sigma_{s8}$ & $72.8^{+4.0}_{-4.8}$ & $0.264_{-0.033}^{+0.029}$ & $0.897_{-0.055}^{+0.047}$  \\
        $\Lambda$CDM & $\mathfrak{D}_V + F_\mathrm{AP}+m+f\sigma_{s8}$ & $70.2^{+1.9}_{-2.1}$ & $0.284_{-0.016}^{+0.014}$ & $0.872^{+0.042}_{-0.042}$   \\
        $\Lambda$CDM & $D_V/\rd+F_{\rm AP}+f\sigma_{s8}$ & 
        $67.37_{-0.95}^{+0.86}$ & $0.286_{-0.016}^{+0.014}$ & $0.825^{+0.043}_{-0.047}$  \\
        $\Lambda$CDM & $D_V/\rd+F_{\rm AP}+m+f\sigma_{s8}$ & 
        $67.90_{-0.75}^{+0.76}$ & $0.3000_{-0.0073}^{+0.0065}$ & $0.850^{+0.035}_{-0.038}$  \\
\hline
        $\Lambda$CDM (free $n_s$) & $F_\mathrm{AP}+m+f\sigma_{s8}$ & $73.0^{+4.6}_{-4.7}$ & $0.265_{-0.034}^{+0.027}$ & $0.894_{-0.052}^{+0.051}$  \\
        $\Lambda$CDM (free $n_s$) & $\mathfrak{D}_V + F_\mathrm{AP}+m+f\sigma_{s8}$ & $70.4^{+2.2}_{-2.8}$ & $0.283_{-0.016}^{+0.014}$ & $0.870^{+0.042}_{-0.042}$   \\
        $\Lambda$CDM (free $n_s$) & $D_V/\rd+F_{\rm AP}+f\sigma_{s8}$ & 
        $67.36_{-0.94}^{+0.92}$ & $0.286_{-0.016}^{+0.014}$ & $0.820^{+0.052}_{-0.055}$  \\
        $\Lambda$CDM (free $n_s$) & $D_V/\rd+F_{\rm AP}+m+f\sigma_{s8}$ &
        $67.80_{-0.84}^{+0.78}$ & $0.2963_{-0.011}^{+0.0096}$ & $0.855^{+0.037}_{-0.039}$  \\
\hline
        $\Lambda$CDM \small{(no-BBN)} &  $F_\mathrm{AP}+m+f\sigma_{s8}$ & - &  $0.264_{-0.033}^{+0.028}$ & $0.892_{-0.055}^{+0.048}$  \\
        $\Lambda$CDM \small{(no-BBN)} & $\mathfrak{D}_V + F_\mathrm{AP}+m+f\sigma_{s8}$ & - & $0.284_{-0.016}^{+0.014}$ & $0.869_{-0.043}^{+0.040}$ \\
        $\Lambda$CDM \small{(no-BBN)} & $D_V/\rd+F_{\rm AP}+f\sigma_{s8}$ & - & $0.286_{-0.016}^{+0.014}$ & $0.831_{-0.048}^{+0.046}$   \\
        $\Lambda$CDM \small{(no-BBN)} & $D_V/\rd + F_\mathrm{AP}+m+f\sigma_{s8}$ & - & $0.2964_{-0.0083}^{+0.0073}$ & $0.852_{-0.038}^{+0.036}$   \\
        \hline
        $\nu\Lambda$CDM &  $F_\mathrm{AP}+m+f\sigma_{s8}$ & $76.2_{-5.5}^{+5.0}$ & $0.265_{-0.035}^{+0.029}$ & $0.935_{-0.065}^{+0.054}$   \\
        $\nu\Lambda$CDM & $\mathfrak{D}_V + F_\mathrm{AP}+m+f\sigma_{s8}$ & $73.3_{-3.3}^{+3.0}$ &  $0.284_{-0.015}^{+0.014}$ & $0.906_{-0.050}^{+0.048}$    \\
        $\nu\Lambda$CDM & $D_V/\rd+F_{\rm AP}+f\sigma_{s8}$ & $66.35_{-1.2}^{+1.1}$ & $0.287_{-0.016}^{+0.015}$ & $0.786_{-0.054}^{+0.054}$   \\
        $\nu\Lambda$CDM & $D_V/\rd + F_\mathrm{AP}+m+f\sigma_{s8}$ & $67.86_{-0.74}^{+0.78}$ & $0.304_{-0.011}^{+0.008}$ & $0.853_{-0.036}^{+0.035}$  \\
        \hline
        $N_{\rm eff}\Lambda$CDM &  $F_\mathrm{AP}+m+f\sigma_{s8}$ & $68.8_{-3.3}^{+3.2}$ & $0.291_{-0.025}^{+0.024}$ & $0.866_{-0.044}^{+0.041}$   \\
        $N_{\rm eff}\Lambda$CDM & $\mathfrak{D}_V + F_\mathrm{AP}+m+f\sigma_{s8}$ & $69.9_{-2.5}^{+2.5}$ & $0.283_{-0.016}^{+0.014}$ & $0.870_{-0.044}^{+0.039}$   \\
        $N_{\rm eff}\Lambda$CDM & $D_V/\rd+F_{\rm AP}+f\sigma_{s8}$ & $66.66_{-2.1}^{+2.1}$ & $0.286_{-0.015}^{+0.014}$ & $0.822_{-0.046}^{+0.045}$   \\
        $N_{\rm eff}\Lambda$CDM & $D_V/\rd + F_\mathrm{AP}+m+f\sigma_{s8}$ & $67.90_{-1.9}^{+1.9}$ & $0.2999_{-0.0087}^{+0.0075}$ & $0.850_{-0.035}^{+0.036}$  \\
        \hline
    \end{tabular}
    \caption{Results on $H_0=100h\,[{\rm km}\, {\rm s}^{-1}\,{\rm Mpc}^{-1}]$, $\Om$ and $\sigma_{8}$ when employing the full set of LRG+QSO+Lyman-$\alpha$ BOSS and eBOSS data in the range $0.2<z<3.5$. Different rows show different assumptions on the underlying model (first column) and different combination of compressed variables (second column, see section~\ref{sec:results} for notation). These results show the cases displayed in figure~\ref{fig:contibutions}-\ref{fig:varying-Mnu}.}
    \label{tab:full}
\end{table}

%
%
%


\def\jnl@style{\it}
\def\aaref@jnl#1{{\jnl@style#1}}

\def\aaref@jnl#1{{\jnl@style#1}}

\def\aj{\aaref@jnl{AJ}}                   
\def\araa{\aaref@jnl{ARA\&A}}             
\def\apj{\aaref@jnl{ApJ}}                 
\def\apjl{\aaref@jnl{ApJ}}                
\def\apjs{\aaref@jnl{ApJS}}               
\def\ao{\aaref@jnl{Appl.~Opt.}}           
\def\apss{\aaref@jnl{Ap\&SS}}             
\def\aap{\aaref@jnl{A\&A}}                
\def\aapr{\aaref@jnl{A\&A~Rev.}}          
\def\aaps{\aaref@jnl{A\&AS}}              
\def\azh{\aaref@jnl{AZh}}                 
\def\baas{\aaref@jnl{BAAS}}               
\def\jrasc{\aaref@jnl{JRASC}}             
\def\memras{\aaref@jnl{MmRAS}}            
\def\mnras{\aaref@jnl{MNRAS}}             
\def\pra{\aaref@jnl{Phys.~Rev.~A}}        
\def\prb{\aaref@jnl{Phys.~Rev.~B}}        
\def\prc{\aaref@jnl{Phys.~Rev.~C}}        
\def\prd{\aaref@jnl{Phys.~Rev.~D}}        
\def\pre{\aaref@jnl{Phys.~Rev.~E}}        
\def\prl{\aaref@jnl{Phys.~Rev.~Lett.}}    
\def\pasp{\aaref@jnl{PASP}}               
\def\pasj{\aaref@jnl{PASJ}}               
\def\qjras{\aaref@jnl{QJRAS}}             
\def\skytel{\aaref@jnl{S\&T}}             
\def\solphys{\aaref@jnl{Sol.~Phys.}}      
\def\sovast{\aaref@jnl{Soviet~Ast.}}      
\def\ssr{\aaref@jnl{Space~Sci.~Rev.}}     
\def\zap{\aaref@jnl{ZAp}}                 
\def\nat{\aaref@jnl{Nature}}              
\def\iaucirc{\aaref@jnl{IAU~Circ.}}       
\def\aplett{\aaref@jnl{Astrophys.~Lett.}} 
\def\apspr{\aaref@jnl{Astrophys.~Space~Phys.~Res.}}
\def\bain{\aaref@jnl{Bull.~Astron.~Inst.~Netherlands}} 
\def\fcp{\aaref@jnl{Fund.~Cosmic~Phys.}}  
\def\gca{\aaref@jnl{Geochim.~Cosmochim.~Acta}}   
\def\grl{\aaref@jnl{Geophys.~Res.~Lett.}} 
\def\jcp{\aaref@jnl{J.~Chem.~Phys.}}      
\def\jgr{\aaref@jnl{J.~Geophys.~Res.}}    
\def\jqsrt{\aaref@jnl{J.~Quant.~Spec.~Radiat.~Transf.}}
\def\memsai{\aaref@jnl{Mem.~Soc.~Astron.~Italiana}}
\def\nphysa{\aaref@jnl{Nucl.~Phys.~A}}   
\def\physrep{\aaref@jnl{Phys.~Rep.}}   
\def\physscr{\aaref@jnl{Phys.~Scr}}   
\def\planss{\aaref@jnl{Planet.~Space~Sci.}}   
\def\procspie{\aaref@jnl{Proc.~SPIE}}   
\def\jcap{\aaref@jnl{J. Cosmology Astropart. Phys.}}

\let\astap=\aap
\let\apjlett=\apjl
\let\apjsupp=\apjs
\let\applopt=\ao

\newcommand{\etal}{et al.\ }

\newcommand{\mpc}{\, {\rm Mpc}}
\newcommand{\kpc}{\, {\rm kpc}}
\newcommand{\hmpc}{\, h^{-1} \mpc}
\newcommand{\ihmpc}{\, h\, {\rm Mpc}^{-1}}
\newcommand{\ikms}{\, {\rm s\, km}^{-1}}
\newcommand{\kms}{\, {\rm km\, s}^{-1}}
\newcommand{\hkpc}{\, h^{-1} \kpc}
\newcommand{\lya}{Ly$\alpha$\ }
\newcommand{\lyb}{Lyman-$\beta$\ }
\newcommand{\lyaf}{Ly$\alpha$ forest}
\newcommand{\lr}{\lambda_{{\rm rest}}}
\newcommand{\bF}{\bar{F}}
\newcommand{\bS}{\bar{S}}
\newcommand{\bC}{\bar{C}}
\newcommand{\bB}{\bar{B}}
\newcommand{\vdF}{{\mathbf \delta_F}}
\newcommand{\vdS}{{\mathbf \delta_S}}
\newcommand{\vdf}{{\mathbf \delta_f}}
\newcommand{\vdn}{{\mathbf \delta_n}}
\newcommand{\vdC}{{\mathbf \delta_C}}
\newcommand{\vdX}{{\mathbf \delta_X}}
\newcommand{\xrei}{x_{rei}}
\newcommand{\lrmin}{\lambda_{{\rm rest, min}}}
\newcommand{\lrmax}{\lambda_{{\rm rest, max}}}
\newcommand{\lmin}{\lambda_{{\rm min}}}
\newcommand{\lmax}{\lambda_{{\rm max}}}
\newcommand{\hi}{\mbox{H\,{\scriptsize I}\ }}
\newcommand{\heii}{\mbox{He\,{\scriptsize II}\ }}
\newcommand{\vp}{\mathbf{p}}
\newcommand{\vq}{\mathbf{q}}
\newcommand{\vxperp}{\mathbf{x_\perp}}
\newcommand{\vkperp}{\mathbf{k_\perp}}
\newcommand{\vrperp}{\mathbf{r_\perp}}
\newcommand{\vx}{\mathbf{x}}
\newcommand{\vy}{\mathbf{y}}
\newcommand{\vk}{\mathbf{k}}
\newcommand{\vR}{\mathbf{r}}
\newcommand{\tdtwo}{\tilde{b}_{\delta^2}}
\newcommand{\tstwo}{\tilde{b}_{s^2}}
\newcommand{\tbthree}{\tilde{b}_3}
\newcommand{\tadtwo}{\tilde{a}_{\delta^2}}
\newcommand{\tastwo}{\tilde{a}_{s^2}}
\newcommand{\tabthree}{\tilde{a}_3}
\newcommand{\tpsi}{\tilde{\psi}}
\newcommand{\vv}{\mathbf{v}}
\newcommand{\fnl}{{f_{\rm NL}}}
\newcommand{\tfnl}{{\tilde{f}_{\rm NL}}}
\newcommand{\gnl}{g_{\rm NL}}
\newcommand{\orderfour}{\mathcal{O}\left(\delta_1^4\right)}
\newcommand{\SDSSPF}{\cite{2006ApJS..163...80M}}
\newcommand{\PF}{$P_F^{\rm 1D}(k_\parallel,z)$}
\newcommand\ionalt[2]{#1$\;${\scriptsize \uppercase\expandafter{\romannumeral #2}}}%
\newcommand{\vxone}{\mathbf{x_1}}
\newcommand{\vxtwo}{\mathbf{x_2}}
\newcommand{\vRot}{\mathbf{r_{12}}}
\newcommand{\cm}{\, {\rm cm}}

\bibliography{main.bib}{}

\providecommand{\href}[2]{#2}\begingroup\raggedright\begin{thebibliography}{10}

\bibitem{desi}
B.~{Abareshi}, J.~{Aguilar}, S.~{Ahlen}, S.~{Alam}, D.M.~{Alexander},
  R.~{Alfarsy} et~al., \emph{{Overview of the Instrumentation for the Dark
  Energy Spectroscopic Instrument}},
  \href{https://doi.org/10.3847/1538-3881/ac882b}{\emph{\aj} {\bfseries 164}
  (2022) 207} [\href{https://arxiv.org/abs/2205.10939}{{\ttfamily
  2205.10939}}].

\bibitem{euclid}
R.~{Laureijs}, J.~{Amiaux}, S.~{Arduini}, J.L.~{Augu{\`e}res}, J.~{Brinchmann},
  R.~{Cole} et~al., \emph{{Euclid Definition Study Report}}, {\emph{arXiv
  e-prints} (2011) arXiv:1110.3193}
  [\href{https://arxiv.org/abs/1110.3193}{{\ttfamily 1110.3193}}].

\bibitem{lsst}
{\v{Z}}.~{Ivezi{\'c}}, S.M.~{Kahn}, J.A.~{Tyson}, B.~{Abel}, E.~{Acosta},
  R.~{Allsman} et~al., \emph{{LSST: From Science Drivers to Reference Design
  and Anticipated Data Products}},
  \href{https://doi.org/10.3847/1538-4357/ab042c}{\emph{\apj} {\bfseries 873}
  (2019) 111} [\href{https://arxiv.org/abs/0805.2366}{{\ttfamily 0805.2366}}].

\bibitem{wfirst}
D.~{Spergel}, N.~{Gehrels}, C.~{Baltay}, D.~{Bennett}, J.~{Breckinridge},
  M.~{Donahue} et~al., \emph{{Wide-Field InfrarRed Survey
  Telescope-Astrophysics Focused Telescope Assets WFIRST-AFTA 2015 Report}},
  {\emph{arXiv e-prints} (2015) arXiv:1503.03757}
  [\href{https://arxiv.org/abs/1503.03757}{{\ttfamily 1503.03757}}].

\bibitem{SKA}
{\scshape SKA} collaboration, \emph{{Cosmology with Phase 1 of the Square
  Kilometre Array: Red Book 2018: Technical specifications and performance
  forecasts}}, \href{https://doi.org/10.1017/pasa.2019.51}{\emph{Publ. Astron.
  Soc. Austral.} {\bfseries 37} (2020) e007}
  [\href{https://arxiv.org/abs/1811.02743}{{\ttfamily 1811.02743}}].

\bibitem{SimonsObs}
P.~{Ade}, J.~{Aguirre}, Z.~{Ahmed}, S.~{Aiola}, A.~{Ali}, D.~{Alonso} et~al.,
  \emph{{The Simons Observatory: science goals and forecasts}},
  \href{https://doi.org/10.1088/1475-7516/2019/02/056}{\emph{\jcap} {\bfseries
  2019} (2019) 056} [\href{https://arxiv.org/abs/1808.07445}{{\ttfamily
  1808.07445}}].

\bibitem{Riess:2021jrx}
A.G.~Riess et~al., \emph{{A Comprehensive Measurement of the Local Value of the
  Hubble Constant with 1 km/s/Mpc Uncertainty from the Hubble Space Telescope
  and the SH0ES Team}},
  \href{https://doi.org/10.3847/2041-8213/ac5c5b}{\emph{Astrophys. J. Lett.}
  {\bfseries 934} (2022) L7}
  [\href{https://arxiv.org/abs/2112.04510}{{\ttfamily 2112.04510}}].

\bibitem{VTR_2019}
L.~{Verde}, T.~{Treu} and A.G.~{Riess}, \emph{{Tensions between the early and
  late Universe}},
  \href{https://doi.org/10.1038/s41550-019-0902-0}{\emph{Nature Astronomy}
  {\bfseries 3} (2019) 891} [\href{https://arxiv.org/abs/1907.10625}{{\ttfamily
  1907.10625}}].

\bibitem{Aghanim:2018eyx}
{Planck Collaboration}, N.~{Aghanim}, Y.~{Akrami}, M.~{Ashdown}, J.~{Aumont},
  C.~{Baccigalupi} et~al., \emph{{Planck 2018 results. VI. Cosmological
  parameters}}, \href{https://doi.org/10.1051/0004-6361/201833910}{\emph{\aap}
  {\bfseries 641} (2020) A6}
  [\href{https://arxiv.org/abs/1807.06209}{{\ttfamily 1807.06209}}].

\bibitem{wendy}
W.L.~{Freedman}, B.F.~{Madore}, T.~{Hoyt}, I.S.~{Jang}, R.~{Beaton}, M.G.~{Lee}
  et~al., \emph{{Calibration of the Tip of the Red Giant Branch}},
  \href{https://doi.org/10.3847/1538-4357/ab7339}{\emph{\apj} {\bfseries 891}
  (2020) 57} [\href{https://arxiv.org/abs/2002.01550}{{\ttfamily 2002.01550}}].

\bibitem{holicow}
K.C.~{Wong}, S.H.~{Suyu}, G.C.F.~{Chen}, C.E.~{Rusu}, M.~{Millon}, D.~{Sluse}
  et~al., \emph{{H0LiCOW - XIII. A 2.4 per cent measurement of H$_{0}$ from
  lensed quasars: 5.3{\ensuremath{\sigma}} tension between early- and
  late-Universe probes}},
  \href{https://doi.org/10.1093/mnras/stz3094}{\emph{\mnras} {\bfseries 498}
  (2020) 1420} [\href{https://arxiv.org/abs/1907.04869}{{\ttfamily
  1907.04869}}].

\bibitem{alamdr12}
S.~{Alam}, F.D.~{Albareti}, C.~{Allende Prieto}, F.~{Anders}, S.F.~{Anderson},
  T.~{Anderton} et~al., \emph{{The Eleventh and Twelfth Data Releases of the
  Sloan Digital Sky Survey: Final Data from SDSS-III}},
  \href{https://doi.org/10.1088/0067-0049/219/1/12}{\emph{ApJS} {\bfseries 219}
  (2015) 12} [\href{https://arxiv.org/abs/1501.00963}{{\ttfamily 1501.00963}}].

\bibitem{eboss_collaboration_dr16}
S.~{Alam}, M.~{Aubert}, S.~{Avila}, C.~{Balland}, J.E.~{Bautista},
  M.A.~{Bershady} et~al., \emph{{Completed SDSS-IV extended Baryon Oscillation
  Spectroscopic Survey: Cosmological implications from two decades of
  spectroscopic surveys at the Apache Point Observatory}},
  \href{https://doi.org/10.1103/PhysRevD.103.083533}{\emph{\prd} {\bfseries
  103} (2021) 083533} [\href{https://arxiv.org/abs/2007.08991}{{\ttfamily
  2007.08991}}].

\bibitem{wigglez}
C.~{Blake}, S.~{Brough}, M.~{Colless}, C.~{Contreras}, W.~{Couch}, S.~{Croom}
  et~al., \emph{{The WiggleZ Dark Energy Survey: joint measurements of the
  expansion and growth history at z < 1}},
  \href{https://doi.org/10.1111/j.1365-2966.2012.21473.x}{\emph{\mnras}
  {\bfseries 425} (2012) 405}
  [\href{https://arxiv.org/abs/1204.3674}{{\ttfamily 1204.3674}}].

\bibitem{DiValentino:2021izs}
E.~Di~Valentino, O.~Mena, S.~Pan, L.~Visinelli, W.~Yang, A.~Melchiorri et~al.,
  \emph{{In the realm of the Hubble tension\textemdash{}a review of
  solutions}}, \href{https://doi.org/10.1088/1361-6382/ac086d}{\emph{Class.
  Quant. Grav.} {\bfseries 38} (2021) 153001}
  [\href{https://arxiv.org/abs/2103.01183}{{\ttfamily 2103.01183}}].

\bibitem{Cuesta:2014asa}
A.J.~{Cuesta}, L.~{Verde}, A.~{Riess} and R.~{Jimenez}, \emph{{Calibrating the
  cosmic distance scale ladder: the role of the sound-horizon scale and the
  local expansion rate as distance anchors}},
  \href{https://doi.org/10.1093/mnras/stv261}{\emph{\mnras} {\bfseries 448}
  (2015) 3463} [\href{https://arxiv.org/abs/1411.1094}{{\ttfamily 1411.1094}}].

\bibitem{Aubourgetal15}
{\'E}.~{Aubourg}, S.~{Bailey}, J.E.~{Bautista}, F.~{Beutler}, V.~{Bhardwaj},
  D.~{Bizyaev} et~al., \emph{{Cosmological implications of baryon acoustic
  oscillation measurements}},
  \href{https://doi.org/10.1103/PhysRevD.92.123516}{\emph{\prd} {\bfseries 92}
  (2015) 123516} [\href{https://arxiv.org/abs/1411.1074}{{\ttfamily
  1411.1074}}].

\bibitem{H0olympics}
N.~{Sch{\"o}neberg}, G.F.~{Abell{\'a}n}, A.P.~{S{\'a}nchez}, S.J.~{Witte},
  V.~{Poulin} and J.~{Lesgourgues}, \emph{{The H$_{0}$ Olympics: A fair ranking
  of proposed models}},
  \href{https://doi.org/10.1016/j.physrep.2022.07.001}{\emph{\physrep}
  {\bfseries 984} (2022) 1} [\href{https://arxiv.org/abs/2107.10291}{{\ttfamily
  2107.10291}}].

\bibitem{Kamion_Riess22}
M.~{Kamionkowski} and A.G.~{Riess}, \emph{{The Hubble Tension and Early Dark
  Energy}}, {\emph{arXiv e-prints} (2022) arXiv:2211.04492}
  [\href{https://arxiv.org/abs/2211.04492}{{\ttfamily 2211.04492}}].

\bibitem{ShapeFit}
S.~{Brieden}, H.~{Gil-Mar{\'\i}n} and L.~{Verde}, \emph{{ShapeFit: extracting
  the power spectrum shape information in galaxy surveys beyond BAO and RSD}},
  \href{https://doi.org/10.1088/1475-7516/2021/12/054}{\emph{\jcap} {\bfseries
  2021} (2021) 054} [\href{https://arxiv.org/abs/2106.07641}{{\ttfamily
  2106.07641}}].

\bibitem{Brieden:2022lsd}
S.~{Brieden}, H.~{Gil-Mar{\'\i}n} and L.~{Verde}, \emph{{Model-agnostic
  interpretation of 10 billion years of cosmic evolution traced by BOSS and
  eBOSS data}},
  \href{https://doi.org/10.1088/1475-7516/2022/08/024}{\emph{\jcap} {\bfseries
  2022} (2022) 024} [\href{https://arxiv.org/abs/2204.11868}{{\ttfamily
  2204.11868}}].

\bibitem{ShapeFit:data}
S.~{Brieden}, H.~{Gil-Mar\'in} and L.~{Verde}, ``\textsc{BOSS} and
  e\textsc{BOSS} data-vectors and covariances for \textsc{ShapeFit}.''
  \url{https://www.ub.edu/bispectrum/sdss_shapefit.html}.

\bibitem{SDSS:2003tbn}
M.~{Tegmark}, M.R.~{Blanton}, M.A.~{Strauss}, F.~{Hoyle}, D.~{Schlegel},
  R.~{Scoccimarro} et~al., \emph{{The Three-Dimensional Power Spectrum of
  Galaxies from the Sloan Digital Sky Survey}},
  \href{https://doi.org/10.1086/382125}{\emph{\apj} {\bfseries 606} (2004) 702}
  [\href{https://arxiv.org/abs/astro-ph/0310725}{{\ttfamily
  astro-ph/0310725}}].

\bibitem{DAmico:2019fhj}
G.~{d'Amico}, J.~{Gleyzes}, N.~{Kokron}, K.~{Markovic}, L.~{Senatore},
  P.~{Zhang} et~al., \emph{{The cosmological analysis of the SDSS/BOSS data
  from the Effective Field Theory of Large-Scale Structure}},
  \href{https://doi.org/10.1088/1475-7516/2020/05/005}{\emph{\jcap} {\bfseries
  2020} (2020) 005} [\href{https://arxiv.org/abs/1909.05271}{{\ttfamily
  1909.05271}}].

\bibitem{Ivanov:2019pdj}
M.M.~{Ivanov}, M.~{Simonovi{\'c}} and M.~{Zaldarriaga}, \emph{{Cosmological
  parameters from the BOSS galaxy power spectrum}},
  \href{https://doi.org/10.1088/1475-7516/2020/05/042}{\emph{\jcap} {\bfseries
  2020} (2020) 042} [\href{https://arxiv.org/abs/1909.05277}{{\ttfamily
  1909.05277}}].

\bibitem{Trosteretal:2020}
T.~{Tr{\"o}ster}, A.G.~{S{\'a}nchez}, M.~{Asgari}, C.~{Blake}, M.~{Crocce},
  C.~{Heymans} et~al., \emph{{Cosmology from large-scale structure.
  Constraining {\ensuremath{\Lambda}}CDM with BOSS}},
  \href{https://doi.org/10.1051/0004-6361/201936772}{\emph{\aap} {\bfseries
  633} (2020) L10} [\href{https://arxiv.org/abs/1909.11006}{{\ttfamily
  1909.11006}}].

\bibitem{Hamann_shape}
J.~{Hamann}, S.~{Hannestad}, J.~{Lesgourgues}, C.~{Rampf} and Y.Y.Y.~{Wong},
  \emph{{Cosmological parameters from large scale structure - geometric versus
  shape information}},
  \href{https://doi.org/10.1088/1475-7516/2010/07/022}{\emph{\jcap} {\bfseries
  2010} (2010) 022} [\href{https://arxiv.org/abs/1003.3999}{{\ttfamily
  1003.3999}}].

\bibitem{Pisanti:2007hk}
O.~Pisanti, A.~Cirillo, S.~Esposito, F.~Iocco, G.~Mangano, G.~Miele et~al.,
  \emph{{PArthENoPE: Public Algorithm Evaluating the Nucleosynthesis of
  Primordial Elements}},
  \href{https://doi.org/10.1016/j.cpc.2008.02.015}{\emph{Comput. Phys. Commun.}
  {\bfseries 178} (2008) 956}
  [\href{https://arxiv.org/abs/0705.0290}{{\ttfamily 0705.0290}}].

\bibitem{Adelberger:2010qa}
E.G.~Adelberger et~al., \emph{{Solar fusion cross sections II: the pp chain and
  CNO cycles}}, \href{https://doi.org/10.1103/RevModPhys.83.195}{\emph{Rev.
  Mod. Phys.} {\bfseries 83} (2011) 195}
  [\href{https://arxiv.org/abs/1004.2318}{{\ttfamily 1004.2318}}].

\bibitem{BAOBBN}
G.E.~{Addison}, D.J.~{Watts}, C.L.~{Bennett}, M.~{Halpern}, G.~{Hinshaw} and
  J.L.~{Weiland}, \emph{{Elucidating {\ensuremath{\Lambda}}CDM: Impact of
  Baryon Acoustic Oscillation Measurements on the Hubble Constant
  Discrepancy}}, \href{https://doi.org/10.3847/1538-4357/aaa1ed}{\emph{\apj}
  {\bfseries 853} (2018) 119}
  [\href{https://arxiv.org/abs/1707.06547}{{\ttfamily 1707.06547}}].

\bibitem{Cuceu20}
A.~{Cuceu}, J.~{Farr}, P.~{Lemos} and A.~{Font-Ribera}, \emph{{Baryon Acoustic
  Oscillations and the Hubble constant: past, present and future}},
  \href{https://doi.org/10.1088/1475-7516/2019/10/044}{\emph{\jcap} {\bfseries
  2019} (2019) 044} [\href{https://arxiv.org/abs/1906.11628}{{\ttfamily
  1906.11628}}].

\bibitem{NilsBBN}
N.~Sch\"oneberg, J.~Lesgourgues and D.C.~Hooper, \emph{{The BAO+BBN take on the
  Hubble tension}},
  \href{https://doi.org/10.1088/1475-7516/2019/10/029}{\emph{JCAP} {\bfseries
  10} (2019) 029} [\href{https://arxiv.org/abs/1907.11594}{{\ttfamily
  1907.11594}}].

\bibitem{NilsBBN2}
N.~{Sch{\"o}neberg}, L.~{Verde}, H.~{Gil-Mar{\'\i}n} and S.~{Brieden},
  \emph{{BAO+BBN revisited - growing the Hubble tension with a 0.7 km/s/Mpc
  constraint}},
  \href{https://doi.org/10.1088/1475-7516/2022/11/039}{\emph{\jcap} {\bfseries
  2022} (2022) 039} [\href{https://arxiv.org/abs/2209.14330}{{\ttfamily
  2209.14330}}].

\bibitem{cunnington22}
S.~{Cunnington}, \emph{{Detecting the power spectrum turnover with H I
  intensity mapping}},
  \href{https://doi.org/10.1093/mnras/stac576}{\emph{\mnras} {\bfseries 512}
  (2022) 2408} [\href{https://arxiv.org/abs/2202.13828}{{\ttfamily
  2202.13828}}].

\bibitem{2005MNRAS.363.1329B}
C.~{Blake} and S.~{Bridle}, \emph{{Cosmology with photometric redshift
  surveys}},
  \href{https://doi.org/10.1111/j.1365-2966.2005.09526.x}{\emph{\mnras}
  {\bfseries 363} (2005) 1329}
  [\href{https://arxiv.org/abs/astro-ph/0411713}{{\ttfamily
  astro-ph/0411713}}].

\bibitem{Philcox_Sherwin_Farren_Baxter_21}
O.H.E.~{Philcox}, B.D.~{Sherwin}, G.S.~{Farren} and E.J.~{Baxter},
  \emph{{Determining the Hubble constant without the sound horizon:
  Measurements from galaxy surveys}},
  \href{https://doi.org/10.1103/PhysRevD.103.023538}{\emph{\prd} {\bfseries
  103} (2021) 023538} [\href{https://arxiv.org/abs/2008.08084}{{\ttfamily
  2008.08084}}].

\bibitem{Farren:2021grl}
G.S.~Farren, O.H.E.~Philcox and B.D.~Sherwin, \emph{{Determining the Hubble
  constant without the sound horizon: Perspectives with future galaxy
  surveys}}, \href{https://doi.org/10.1103/PhysRevD.105.063503}{\emph{Phys.
  Rev. D} {\bfseries 105} (2022) 063503}
  [\href{https://arxiv.org/abs/2112.10749}{{\ttfamily 2112.10749}}].

\bibitem{Philcox:2022sgj}
O.H.E.~{Philcox}, G.S.~{Farren}, B.D.~{Sherwin}, E.J.~{Baxter} and
  D.J.~{Brout}, \emph{{Determining the Hubble constant without the sound
  horizon: A 3.6 \% constraint on H$_{0}$ from galaxy surveys, CMB lensing, and
  supernovae}}, \href{https://doi.org/10.1103/PhysRevD.106.063530}{\emph{\prd}
  {\bfseries 106} (2022) 063530}
  [\href{https://arxiv.org/abs/2204.02984}{{\ttfamily 2204.02984}}].

\bibitem{Smith:2022iax}
T.L.~Smith, V.~Poulin and T.~Simon, \emph{{Assessing the robustness of sound
  horizon-free determinations of the Hubble constant}},
  \href{https://arxiv.org/abs/2208.12992}{{\ttfamily 2208.12992}}.

\bibitem{Amon_Efstathiou22}
A.~{Amon} and G.~{Efstathiou}, \emph{{A non-linear solution to the S$_{8}$
  tension?}}, \href{https://doi.org/10.1093/mnras/stac2429}{\emph{\mnras}
  {\bfseries 516} (2022) 5355}
  [\href{https://arxiv.org/abs/2206.11794}{{\ttfamily 2206.11794}}].

\bibitem{Maartens:2011yx}
R.~Maartens, \emph{{Is the Universe homogeneous?}},
  \href{https://doi.org/10.1098/rsta.2011.0289}{\emph{Phil. Trans. Roy. Soc.
  Lond. A} {\bfseries 369} (2011) 5115}
  [\href{https://arxiv.org/abs/1104.1300}{{\ttfamily 1104.1300}}].

\bibitem{Blas2016_BAOIR}
D.~{Blas}, M.~{Garny}, M.M.~{Ivanov} and S.~{Sibiryakov}, \emph{{Time-sliced
  perturbation theory II: baryon acoustic oscillations and infrared
  resummation}},
  \href{https://doi.org/10.1088/1475-7516/2016/07/028}{\emph{\jcap} {\bfseries
  2016} (2016) 028} [\href{https://arxiv.org/abs/1605.02149}{{\ttfamily
  1605.02149}}].

\bibitem{Eis2007}
D.J.~{Eisenstein}, H.-J.~{Seo}, E.~{Sirko} and D.N.~{Spergel}, \emph{{Improving
  Cosmological Distance Measurements by Reconstruction of the Baryon Acoustic
  Peak}}, \href{https://doi.org/10.1086/518712}{\emph{\apj} {\bfseries 664}
  (2007) 675} [\href{https://arxiv.org/abs/astro-ph/0604362}{{\ttfamily
  astro-ph/0604362}}].

\bibitem{burden_reconstruction_2015}
A.~Burden, W.J.~Percival and C.~Howlett, \emph{Reconstruction in {Fourier}
  space}, \href{https://doi.org/10.1093/mnras/stv1581}{\emph{Monthly Notices of
  the Royal Astronomical Society} {\bfseries 453} (2015) 456}
  [\href{https://arxiv.org/abs/1504.02591}{{\ttfamily 1504.02591}}].

\bibitem{Sherwin_2019}
B.D.~Sherwin and M.~White, \emph{The impact of wrong assumptions in {BAO}
  reconstruction},
  \href{https://doi.org/10.1088/1475-7516/2019/02/027}{\emph{Journal of
  Cosmology and Astroparticle Physics} {\bfseries 2019} (2019) 027}
  [\href{https://arxiv.org/abs/1808.04384}{{\ttfamily 1808.04384}}].

\bibitem{Blazek:2015ula}
J.~Blazek, J.E.~McEwen and C.M.~Hirata, \emph{{Streaming velocities and the
  baryon-acoustic oscillation scale}},
  \href{https://doi.org/10.1103/PhysRevLett.116.121303}{\emph{Phys. Rev. Lett.}
  {\bfseries 116} (2016) 121303}
  [\href{https://arxiv.org/abs/1510.03554}{{\ttfamily 1510.03554}}].

\bibitem{Slepian:2016nfb}
Z.~Slepian et~al., \emph{{Constraining the baryon\textendash{}dark matter
  relative velocity with the large-scale three-point correlation function of
  the SDSS BOSS DR12 CMASS galaxies}},
  \href{https://doi.org/10.1093/mnras/stx2723}{\emph{Mon. Not. Roy. Astron.
  Soc.} {\bfseries 474} (2018) 2109}
  [\href{https://arxiv.org/abs/1607.06098}{{\ttfamily 1607.06098}}].

\bibitem{Hirata:2017ivs}
C.M.~Hirata, \emph{{Small-scale structure and the Lyman-$\alpha$ forest baryon
  acoustic oscillation feature}},
  \href{https://doi.org/10.1093/mnras/stx2854}{\emph{Mon. Not. Roy. Astron.
  Soc.} {\bfseries 474} (2018) 2173}
  [\href{https://arxiv.org/abs/1707.03358}{{\ttfamily 1707.03358}}].

\bibitem{Nishimichi:2020tvu}
T.~Nishimichi, G.~D'Amico, M.M.~Ivanov, L.~Senatore, M.~Simonovi\'c, M.~Takada
  et~al., \emph{{Blinded challenge for precision cosmology with large-scale
  structure: results from effective field theory for the redshift-space galaxy
  power spectrum}},
  \href{https://doi.org/10.1103/PhysRevD.102.123541}{\emph{Phys. Rev. D}
  {\bfseries 102} (2020) 123541}
  [\href{https://arxiv.org/abs/2003.08277}{{\ttfamily 2003.08277}}].

\bibitem{PTchallenge:data}
T.~{Nishimichi}, ``Multipole moment data for \textsc{PT} challenges.''
  \url{https://www2.yukawa.kyoto-u.ac.jp/~takahiro.nishimichi/data/PTchallenge/}.

\bibitem{ShapeFitPT}
S.~{Brieden}, H.~{Gil-Mar{\'\i}n} and L.~{Verde}, \emph{{PT challenge:
  validation of ShapeFit on large-volume, high-resolution mocks}},
  \href{https://doi.org/10.1088/1475-7516/2022/06/005}{\emph{\jcap} {\bfseries
  2022} (2022) 005} [\href{https://arxiv.org/abs/2201.08400}{{\ttfamily
  2201.08400}}].

\bibitem{Reid:2015gra}
B.~{Reid}, S.~{Ho}, N.~{Padmanabhan}, W.J.~{Percival}, J.~{Tinker},
  R.~{Tojeiro} et~al., \emph{{SDSS-III Baryon Oscillation Spectroscopic Survey
  Data Release 12: galaxy target selection and large-scale structure
  catalogues}}, \href{https://doi.org/10.1093/mnras/stv2382}{\emph{\mnras}
  {\bfseries 455} (2016) 1553}
  [\href{https://arxiv.org/abs/1509.06529}{{\ttfamily 1509.06529}}].

\bibitem{ebossLRG_catalogue}
A.J.~{Ross}, J.~{Bautista}, R.~{Tojeiro}, S.~{Alam}, S.~{Bailey}, E.~{Burtin}
  et~al., \emph{{The Completed SDSS-IV extended Baryon Oscillation
  Spectroscopic Survey: Large-scale structure catalogues for cosmological
  analysis}}, \href{https://doi.org/10.1093/mnras/staa2416}{\emph{\mnras}
  {\bfseries 498} (2020) 2354}
  [\href{https://arxiv.org/abs/2007.09000}{{\ttfamily 2007.09000}}].

\bibitem{ebossQSO_catalogue}
B.W.~{Lyke}, A.N.~{Higley}, J.N.~{McLane}, D.P.~{Schurhammer}, A.D.~{Myers},
  A.J.~{Ross} et~al., \emph{{The Sloan Digital Sky Survey Quasar Catalog:
  Sixteenth Data Release}},
  \href{https://doi.org/10.3847/1538-4365/aba623}{\emph{\apjs} {\bfseries 250}
  (2020) 8} [\href{https://arxiv.org/abs/2007.09001}{{\ttfamily 2007.09001}}].

\bibitem{BriedenPRL21}
S.~{Brieden}, H.~{Gil-Mar{\'\i}n} and L.~{Verde}, \emph{{Model-independent
  versus model-dependent interpretation of the SDSS-III BOSS power spectrum:
  Bridging the divide}},
  \href{https://doi.org/10.1103/PhysRevD.104.L121301}{\emph{\prd} {\bfseries
  104} (2021) L121301} [\href{https://arxiv.org/abs/2106.11931}{{\ttfamily
  2106.11931}}].

\bibitem{Simon:2022lde}
T.~Simon, P.~Zhang, V.~Poulin and T.L.~Smith, \emph{{On the consistency of
  effective field theory analyses of BOSS power spectrum}},
  \href{https://arxiv.org/abs/2208.05929}{{\ttfamily 2208.05929}}.

\bibitem{Chenetal21}
S.-F.~{Chen}, Z.~{Vlah} and M.~{White}, \emph{{A new analysis of galaxy 2-point
  functions in the BOSS survey, including full-shape information and
  post-reconstruction BAO}},
  \href{https://doi.org/10.1088/1475-7516/2022/02/008}{\emph{\jcap} {\bfseries
  2022} (2022) 008} [\href{https://arxiv.org/abs/2110.05530}{{\ttfamily
  2110.05530}}].

\bibitem{bernal_trouble_2016}
J.L.~Bernal, L.~Verde and A.G.~Riess, \emph{The trouble with {H}0},
  \href{https://doi.org/10.1088/1475-7516/2016/10/019}{\emph{Journal of
  Cosmology and Astro-Particle Physics} {\bfseries 10} (2016) 019}
  [\href{https://arxiv.org/abs/1607.05617}{{\ttfamily 1607.05617}}].

\bibitem{2014PhRvL.113x1302H}
A.~{Heavens}, R.~{Jimenez} and L.~{Verde}, \emph{{Standard Rulers, Candles, and
  Clocks from the Low-Redshift Universe}},
  \href{https://doi.org/10.1103/PhysRevLett.113.241302}{\emph{\prl} {\bfseries
  113} (2014) 241302} [\href{https://arxiv.org/abs/1409.6217}{{\ttfamily
  1409.6217}}].

\bibitem{Standards2}
L.~{Verde}, J.L.~{Bernal}, A.F.~{Heavens} and R.~{Jimenez}, \emph{{The length
  of the low-redshift standard ruler}},
  \href{https://doi.org/10.1093/mnras/stx116}{\emph{\mnras} {\bfseries 467}
  (2017) 731} [\href{https://arxiv.org/abs/1607.05297}{{\ttfamily
  1607.05297}}].

\bibitem{Knox_Millea20}
L.~{Knox} and M.~{Millea}, \emph{{Hubble constant hunter's guide}},
  \href{https://doi.org/10.1103/PhysRevD.101.043533}{\emph{\prd} {\bfseries
  101} (2020) 043533} [\href{https://arxiv.org/abs/1908.03663}{{\ttfamily
  1908.03663}}].

\bibitem{SPT-3G:2021eoc}
{\scshape SPT-3G} collaboration, \emph{{Measurements of the E-mode polarization
  and temperature-E-mode correlation of the CMB from SPT-3G 2018 data}},
  \href{https://doi.org/10.1103/PhysRevD.104.022003}{\emph{Phys. Rev. D}
  {\bfseries 104} (2021) 022003}
  [\href{https://arxiv.org/abs/2101.01684}{{\ttfamily 2101.01684}}].

\bibitem{ACT:2020gnv}
{\scshape ACT} collaboration, \emph{{The Atacama Cosmology Telescope: DR4 Maps
  and Cosmological Parameters}},
  \href{https://doi.org/10.1088/1475-7516/2020/12/047}{\emph{JCAP} {\bfseries
  12} (2020) 047} [\href{https://arxiv.org/abs/2007.07288}{{\ttfamily
  2007.07288}}].

\bibitem{Cuceu:2021hlk}
A.~Cuceu, A.~Font-Ribera, B.~Joachimi and S.~Nadathur, \emph{{Cosmology beyond
  BAO from the 3D distribution of the Lyman-\ensuremath{\alpha} forest}},
  \href{https://doi.org/10.1093/mnras/stab1999}{\emph{Mon. Not. Roy. Astron.
  Soc.} {\bfseries 506} (2021) 5439}
  [\href{https://arxiv.org/abs/2103.14075}{{\ttfamily 2103.14075}}].

\bibitem{pantheonplus}
D.~{Scolnic}, D.~{Brout}, A.~{Carr}, A.G.~{Riess}, T.M.~{Davis}, A.~{Dwomoh}
  et~al., \emph{{The Pantheon+ Analysis: The Full Data Set and Light-curve
  Release}}, \href{https://doi.org/10.3847/1538-4357/ac8b7a}{\emph{\apj}
  {\bfseries 938} (2022) 113}
  [\href{https://arxiv.org/abs/2112.03863}{{\ttfamily 2112.03863}}].

\bibitem{brout22}
D.~{Brout}, D.~{Scolnic}, B.~{Popovic}, A.G.~{Riess}, A.~{Carr}, J.~{Zuntz}
  et~al., \emph{{The Pantheon+ Analysis: Cosmological Constraints}},
  \href{https://doi.org/10.3847/1538-4357/ac8e04}{\emph{\apj} {\bfseries 938}
  (2022) 110} [\href{https://arxiv.org/abs/2202.04077}{{\ttfamily
  2202.04077}}].

\bibitem{riess20}
A.G.~{Riess}, \emph{{The expansion of the Universe is faster than expected}},
  \href{https://doi.org/10.1038/s42254-019-0137-0}{\emph{Nature Reviews
  Physics} {\bfseries 2} (2020) 10}
  [\href{https://arxiv.org/abs/2001.03624}{{\ttfamily 2001.03624}}].

\bibitem{BernalTriangles}
J.L.~{Bernal}, L.~{Verde}, R.~{Jimenez}, M.~{Kamionkowski}, D.~{Valcin} and
  B.D.~{Wandelt}, \emph{{Trouble beyond H$_{0}$ and the new cosmic triangles}},
  \href{https://doi.org/10.1103/PhysRevD.103.103533}{\emph{\prd} {\bfseries
  103} (2021) 103533} [\href{https://arxiv.org/abs/2102.05066}{{\ttfamily
  2102.05066}}].

\bibitem{houetal2021}
J.~{Hou}, A.G.~{S{\'a}nchez}, A.J.~{Ross}, A.~{Smith}, R.~{Neveux},
  J.~{Bautista} et~al., \emph{{The completed SDSS-IV extended Baryon
  Oscillation Spectroscopic Survey: BAO and RSD measurements from anisotropic
  clustering analysis of the quasar sample in configuration space between
  redshift 0.8 and 2.2}},
  \href{https://doi.org/10.1093/mnras/staa3234}{\emph{\mnras} {\bfseries 500}
  (2021) 1201} [\href{https://arxiv.org/abs/2007.08998}{{\ttfamily
  2007.08998}}].

\bibitem{neveuxetal2020}
R.~{Neveux}, E.~{Burtin}, A.~{de Mattia}, A.~{Smith}, A.J.~{Ross}, J.~{Hou}
  et~al., \emph{{The completed SDSS-IV extended Baryon Oscillation
  Spectroscopic Survey: BAO and RSD measurements from the anisotropic power
  spectrum of the quasar sample between redshift 0.8 and 2.2}},
  \href{https://doi.org/10.1093/mnras/staa2780}{\emph{\mnras} {\bfseries 499}
  (2020) 210} [\href{https://arxiv.org/abs/2007.08999}{{\ttfamily
  2007.08999}}].

\bibitem{dumasdesBorboux2020}
H.~du~Mas~des Bourboux et~al., \emph{{The Completed SDSS-IV Extended Baryon
  Oscillation Spectroscopic Survey: Baryon Acoustic Oscillations with
  Ly\ensuremath{\alpha} Forests}},
  \href{https://doi.org/10.3847/1538-4357/abb085}{\emph{Astrophys. J.}
  {\bfseries 901} (2020) 153}
  [\href{https://arxiv.org/abs/2007.08995}{{\ttfamily 2007.08995}}].

\bibitem{Cooke:2018}
R.J.~{Cooke}, M.~{Pettini} and C.C.~{Steidel}, \emph{{One Percent Determination
  of the Primordial Deuterium Abundance}},
  \href{https://doi.org/10.3847/1538-4357/aaab53}{\emph{\apj} {\bfseries 855}
  (2018) 102} [\href{https://arxiv.org/abs/1710.11129}{{\ttfamily
  1710.11129}}].

\bibitem{2011JCAP...07..034B}
D.~{Blas}, J.~{Lesgourgues} and T.~{Tram}, \emph{{The Cosmic Linear Anisotropy
  Solving System (CLASS). Part II: Approximation schemes}},
  \href{https://doi.org/10.1088/1475-7516/2011/07/034}{\emph{Journal of
  Cosmology and Astro-Particle Physics} {\bfseries 2011} (2011) 034}
  [\href{https://arxiv.org/abs/1104.2933}{{\ttfamily 1104.2933}}].

\bibitem{Brinckmann:2018cvx}
T.~Brinckmann and J.~Lesgourgues, \emph{{MontePython 3: boosted MCMC sampler
  and other features}},
  \href{https://doi.org/10.1016/j.dark.2018.100260}{\emph{Phys. Dark Univ.}
  {\bfseries 24} (2019) 100260}
  [\href{https://arxiv.org/abs/1804.07261}{{\ttfamily 1804.07261}}].

\bibitem{gil-marin_bispectrum_dr12}
H.~{Gil-Mar{\'\i}n}, W.J.~{Percival}, L.~{Verde}, J.R.~{Brownstein},
  C.-H.~{Chuang}, F.-S.~{Kitaura} et~al., \emph{{The clustering of galaxies in
  the SDSS-III Baryon Oscillation Spectroscopic Survey: RSD measurement from
  the power spectrum and bispectrum of the DR12 BOSS galaxies}},
  \href{https://doi.org/10.1093/mnras/stw2679}{\emph{\mnras} {\bfseries 465}
  (2017) 1757} [\href{https://arxiv.org/abs/1606.00439}{{\ttfamily
  1606.00439}}].

\bibitem{des_collaboration_dark_2017}
T.M.C.~{Abbott}, F.B.~{Abdalla}, A.~{Alarcon}, J.~{Aleksi{\'c}}, S.~{Allam},
  S.~{Allen} et~al., \emph{{Dark Energy Survey year 1 results: Cosmological
  constraints from galaxy clustering and weak lensing}},
  \href{https://doi.org/10.1103/PhysRevD.98.043526}{\emph{\prd} {\bfseries 98}
  (2018) 043526} [\href{https://arxiv.org/abs/1708.01530}{{\ttfamily
  1708.01530}}].

\bibitem{Aver:2015iza}
E.~Aver, K.A.~Olive and E.D.~Skillman, \emph{{The effects of He I
  \ensuremath{\lambda}10830 on helium abundance determinations}},
  \href{https://doi.org/10.1088/1475-7516/2015/07/011}{\emph{JCAP} {\bfseries
  07} (2015) 011} [\href{https://arxiv.org/abs/1503.08146}{{\ttfamily
  1503.08146}}].

\bibitem{Gil-Marin:2020bct}
H.~{Gil-Mar{\'\i}n}, J.E.~{Bautista}, R.~{Paviot}, M.~{Vargas-Maga{\~n}a},
  S.~{de la Torre}, S.~{Fromenteau} et~al., \emph{{The Completed SDSS-IV
  extended Baryon Oscillation Spectroscopic Survey: measurement of the BAO and
  growth rate of structure of the luminous red galaxy sample from the
  anisotropic power spectrum between redshifts 0.6 and 1.0}},
  \href{https://doi.org/10.1093/mnras/staa2455}{\emph{\mnras} {\bfseries 498}
  (2020) 2492} [\href{https://arxiv.org/abs/2007.08994}{{\ttfamily
  2007.08994}}].

\bibitem{McDonald_2009}
P.~McDonald and A.~Roy, \emph{Clustering of dark matter tracers: generalizing
  bias for the coming era of precision {LSS}},
  \href{https://doi.org/10.1088/1475-7516/2009/08/020}{\emph{Journal of
  Cosmology and Astroparticle Physics} {\bfseries 2009} (2009) 020}
  [\href{https://arxiv.org/abs/0902.0991}{{\ttfamily 0902.0991}}].

\bibitem{Saito:2014qha}
S.~Saito, T.~Baldauf, Z.~Vlah, U.~Seljak, T.~Okumura and P.~McDonald,
  \emph{{Understanding higher-order nonlocal halo bias at large scales by
  combining the power spectrum with the bispectrum}},
  \href{https://doi.org/10.1103/PhysRevD.90.123522}{\emph{Phys. Rev.}
  {\bfseries D90} (2014) 123522}
  [\href{https://arxiv.org/abs/1405.1447}{{\ttfamily 1405.1447}}].

\bibitem{Taruya:2010mx}
A.~Taruya, T.~Nishimichi and S.~Saito, \emph{{Baryon Acoustic Oscillations in
  2D: Modeling Redshift-space Power Spectrum from Perturbation Theory}},
  \href{https://doi.org/10.1103/PhysRevD.82.063522}{\emph{Phys. Rev.}
  {\bfseries D82} (2010) 063522}
  [\href{https://arxiv.org/abs/1006.0699}{{\ttfamily 1006.0699}}].

\bibitem{smithetal2020}
T.L.~{Smith}, V.~{Poulin} and M.A.~{Amin}, \emph{{Oscillating scalar fields and
  the Hubble tension: A resolution with novel signatures}},
  \href{https://doi.org/10.1103/PhysRevD.101.063523}{\emph{\prd} {\bfseries
  101} (2020) 063523} [\href{https://arxiv.org/abs/1908.06995}{{\ttfamily
  1908.06995}}].

\bibitem{poulinetal2018}
V.~{Poulin}, T.L.~{Smith}, D.~{Grin}, T.~{Karwal} and M.~{Kamionkowski},
  \emph{{Cosmological implications of ultralight axionlike fields}},
  \href{https://doi.org/10.1103/PhysRevD.98.083525}{\emph{\prd} {\bfseries 98}
  (2018) 083525} [\href{https://arxiv.org/abs/1806.10608}{{\ttfamily
  1806.10608}}].

\bibitem{poulinetal2019}
V.~{Poulin}, T.L.~{Smith}, T.~{Karwal} and M.~{Kamionkowski}, \emph{{Early Dark
  Energy can Resolve the Hubble Tension}},
  \href{https://doi.org/10.1103/PhysRevLett.122.221301}{\emph{\prl} {\bfseries
  122} (2019) 221301} [\href{https://arxiv.org/abs/1811.04083}{{\ttfamily
  1811.04083}}].

\bibitem{Simon:2022adh}
T.~Simon, P.~Zhang, V.~Poulin and T.L.~Smith, \emph{{Updated constraints from
  the effective field theory analysis of BOSS power spectrum on Early Dark
  Energy}},  \href{https://arxiv.org/abs/2208.05930}{{\ttfamily 2208.05930}}.

\bibitem{herold_ferreira22}
L.~{Herold} and E.G.M.~{Ferreira}, \emph{{Resolving the Hubble tension with
  Early Dark Energy}}, {\emph{arXiv e-prints} (2022) arXiv:2210.16296}
  [\href{https://arxiv.org/abs/2210.16296}{{\ttfamily 2210.16296}}].

\end{thebibliography}\endgroup
\bibliographystyle{JHEP}

\end{document}